\documentclass[preprint,12pt,showpacs,aps,amsmath,amssymb, 
nofootinbib,superscriptaddress,showkeys]{revtex4} 
 
\usepackage{epsfig} 
\usepackage{amsmath} 
\usepackage{amssymb} 
\usepackage{bbm} 
\usepackage{wasysym} 
\usepackage{mathrsfs} 
\usepackage{graphicx} 
\usepackage{amsfonts} 
\usepackage{amsbsy} 
\usepackage{amscd} 
\usepackage{bbm}  
\usepackage{accents} 
\usepackage{axodraw} 
\usepackage{pstricks} 
\newcommand{\beq}{\begin{equation}} 
\newcommand{\eeq}{\end{equation}} 
\newcommand{\beqa}{\begin{eqnarray}} 
\newcommand{\eeqa}{\end{eqnarray}} 
\newcommand{\barr}{\begin{array}} 
\newcommand{\earr}{\end{array}} 
 
\newcommand{\nn}{\nonumber} 
\newcommand{\Sb}{\overline{\Sigma}} 
\newcommand{\dmsq}{\Delta m^2} 
\newcommand{\dma}{{\Delta m}^2_{\rm atm}} 
 
\newcommand{\D}{D\hspace{-8pt}\slash} 
\newcommand{\p}{p\hspace{-4pt}\slash}

\newcommand{\calG}{\mathcal{G}} 
\newcommand{\Ysigman}{\accentset{(n)}{{Y}_\Sigma}} 
 
\newcommand{\Qn}{\accentset{(n)}{Q}} 
 
\def\lsim{\ \raisebox{-.45ex}{\rlap{$\sim$}} \raisebox{.45ex}{$<$}\ } 
 
\DeclareMathOperator{\re}{Re} 
\DeclareMathOperator{\im}{Im} 
 
\DeclareMathOperator{\Tr}{Tr} 
 
\begin{document} 
 
\author{Joydeep Chakrabortty} 
\email{joydeep@hri.res.in} 
\affiliation{Harish-Chandra Research Institute, Chhatnag Road,  
Jhunsi, Allahabad  211019, India}

\author{Amol Dighe} 
\email{amol@theory.tifr.res.in} 
\affiliation{Tata Institute of Fundamental Research,  
Homi Bhabha Road, Colaba, Mumbai 400005, India}

\author{Srubabati Goswami} 
\email{sruba@prl.res.in} 
\affiliation{Physical Research Laboratory, Ahmedabad 380009, India}

\author{Shamayita Ray} 
\email{shamayitar@theory.tifr.res.in} 
\affiliation{Tata Institute of Fundamental Research,  
Homi Bhabha Road, Colaba, Mumbai 400005, India} 
 
\begin{flushright} 
TIFR/TH/08-56 \\ 
HRI-P-08-12-001 
\end{flushright}

\title{Renormalization group evolution  of neutrino masses and mixing 
in the Type-III seesaw mechanism} 
 
\begin{abstract} 
 
We consider the standard model extended by heavy right handed  
fermions transforming as triplets under SU(2)$_L$,  
which generate neutrino masses through the Type-III seesaw mechanism.  
At energies below their respective mass scales, the heavy fields  
get sequentially decoupled to give an effective dimension-5 operator. 
Above their mass thresholds, these fields also participate in the  
renormalization of the wavefunctions, masses and coupling constants. 
We compute the renormalization group evolution of the  
effective neutrino mass matrix in this model, with particular 
emphasis on the threshold effects. 
The evolution equations are obtained in a basis of neutrino  
parameters where all the quantities are well-defined everywhere, 
including at $\theta_{13} = 0$. 
We also point out the important role of the threshold effects 
and Majorana phases in the evolution of mixing angles through  
illustrative examples. 
\end{abstract} 
 
 

\maketitle 
 
\section{Introduction} 
\label{sec-intro} 
 
In the last decade, results from solar, atmospheric, accelerator  
and reactor experiments looking for neutrino flavour oscillations  
have succeeded in establishing that atleast two of the neutrinos are  
massive and there is mixing between different flavors \cite{maltonireview}.  
The present best-fit values of the mass squared differences and 
mixing angles determined from analyses of global data on neutrino 
oscillation are \cite{Schwetz:2008er} 
\begin{center} 
$\Delta m^2_{21} = 7.65^{+0.69}_{-0.60} \times 10^{-5}$ eV$^2$, $\quad$ 
$|\Delta m^2_{31}| = 2.40^{+0.35}_{-0.33} \times 10^{-3}$ eV$^2$ , 
\\ 
$\sin^2\theta_{12} = 0.30^{+0.07}_{-0.05}$ 
$\; , \quad$  
$\sin^2\theta_{23} = 0.50^{+0.17}_{-0.14}$ 
$\; , \quad$  
$\sin^2\theta_{13} = 0.01^{+0.046}_{-0.01}$ $\; ,$  
\end{center}
where $\dmsq_{ij} \equiv m_i^2 - m_j^2$ are the mass squared differences 
and $\theta_{ij}$ the mixing angles. 
The relative position of the third mass eigenstate $m_3$ with respect to  
the other two is unknown, though the solar neutrino data give  
$\Delta m^2_{21} > 0$. This results in two possible  
orderings of the neutrino masses: normal ($m_1 < m_2 < m_3$)  
and inverted ($m_3 < m_1 < m_2$).

One of the most distinctive features emerging out of the above results  
is the occurrence of two large and one small mixing angles which is  
rather different from the quark sector where all three mixing angles are 
small.  
The absolute masses of neutrinos are also orders of magnitude 
smaller than those of quarks and charged leptons,  
the current bound from cosmology on the sum of neutrino masses 
being $\sum{m_i} \lsim 1.5$ eV \cite{hannestad}.  
The most favored mechanisms to generate such small neutrino masses 
and nontrivial mixings 
are the so called seesaw mechanisms which need the introduction of 
one or more heavy fields. At energies below their mass scales,  
the heavy fields get integrated out giving rise to an effective  
dimension-5 operator \cite{dimension5} 
\beq 
{\cal{L}}_5 = \kappa_5 l_L l_L \phi \phi \; , 
\label{kappa-intro} 
\eeq 
where $l_L$ and $\phi$ are respectively the lepton and Higgs doublets 
belonging to the standard model (SM).  
Here $\kappa_5$ is the effective coupling which  
has inverse mass dimension and can be expressed in terms of a dimensionless  
coupling $a_5$ as $\kappa_5 = a_5/\Lambda$, with $\Lambda$ some  
high energy scale. In this picture the SM serves as an effective theory  
valid upto the mass scale $\Lambda$, which can be taken to be the  
mass of the lightest of the heavy fields. Such an operator violates  
lepton number by two units and hence gives rise to Majorana masses  
for neutrinos: ${\mathbbm m}_\nu \sim \kappa_5 v^2$, where $v$ is the  
vacuum expectation value of the Higgs field $\phi$ after spontaneous symmetry  
breaking. Taking $v \sim 246$ GeV, a neutrino mass of $\sim 0.05$ eV  
implies $\Lambda \sim 10^{15}$ GeV if $a_5 \sim 1$.   
 
There are four possible ways to form a dimension-5 gauge singlet term  
out of the two lepton doublets and two Higgs doublets: 
(i) each $l_L$-$\phi$ pair forms a fermion singlet,  
(ii) each of the $l_L$-$l_L$ and $\phi$-$\phi$ pair forms a scalar triplet,  
(iii) each $l_L$-$\phi$ pair forms a fermion triplet, and 
(iv) each of the $l_L$-$l_L$ and $\phi$-$\phi$ pair forms a scalar singlet. 
Case (i) can arise from the tree level exchange of a right  
handed fermion singlet and this corresponds to the Type-I seesaw  
mechanism \cite{seesaw1}. 
Case (ii) arises when the heavy particle is a Higgs triplet 
giving rise to the Type-II seesaw mechanism \cite{seesaw2,Schechter:1980gr}.  
For case (iii) the exchanged particle should be a right-handed  
fermion triplet, which corresponds to generating neutrino mass  
through the Type-III seesaw mechanism \cite{seesaw3}. 
The last scenario gives terms of the form $\overline{\nu_L^C} e_L$ 
which cannot generate a neutrino mass.

Type-III seesaw mechanism mediated by heavy fermion triplets  
transforming in the adjoint representation has been considered earlier in  
\cite{seesaw3,dpma}. Very recently there has been a renewed interest in  
these type of models. 
The smallness of neutrino masses usually implies the mass of the 
heavy particle to be high $\sim 10^{11 - 15}$ GeV. 
However, it is also possible to assume that one or more of the triplets 
have masses near the TeV scale, making it possible to search for their 
signatures at the LHC \cite{goran-lhc,goran-triplet,gavela-2,triplet-lhc}. 
In such models, the Yukawa couplings need to be small to suppress the 
neutrino mass.  
Lepton flavour violating decays in the context of Type-III seesaw  
models have also been considered in \cite{gavela}.   
Recently it has also been suggested that the neutral member of the triplet  
can serve as the dark matter and can be instrumental in generating  
small neutrino mass radiatively \cite{ma-dm}.  

The possibility of being able to add one triplet fermion per family without
creating anomalies was one of the consequences of a general analysis in 
\cite{barr86} which discussed adding an extra U(1) gauge group to the SM. 
Possible ways of adding fermion triplets in an anomaly-free 
manner have been explored \cite{ma-anomaly}, with
some specific models studied in \cite{ma-rathin}. A possible origin of such an extra U(1)
gauge group has been proposed in \cite{ilja}. Fermions in the adjoint representation
fit naturally into the 24-dimensional representation of SU(5), and can rectify the
two main problems encountered in SU(5) Grand Unified Theory (GUT) models,
viz. generation of neutrino masses and gauge coupling unification 
\cite{goran-lhc,goran-triplet,perez-su5,okada-su5}.
The latter requirement constrains the fermionic triplets to be of mass below
TeV for $M_{\rm GUT} \sim 10^{16}$ GeV, making the model testable at the LHC. 
Leptogenesis mediated by triplet fermions has been explored in 
\cite{triplet-leptogenesis}. 
Additional fermions transforming as triplet representations in the 
context of left-right symmetric model have been studied in \cite{triplet-lr}.
Minimal supersymmetric standard model extended by triplet fermions has 
recently been considered in \cite{temssm}.

Whether the exchanged particle at the high scale is a singlet fermion  
(Type-I seesaw) or a triplet fermion (Type-III seesaw), the  
light neutrino mass matrix  
is given as ${\mathbbm m}_D^T {\mathbbm M}_R^{-1} {\mathbbm m}_D$.  
Here ${\mathbbm m}_D$ is the Dirac mass  
matrix coupling the left handed neutrinos with the right handed heavy fields, 
and ${\mathbbm M}_R$ is the Majorana mass matrix for the right  
handed fields. Thus the generation of the light neutrino mass matrix 
is similar in the Type-I and Type-III seesaw mechanisms, both of which  
are fermion mediated.  
Since the neutrino mass is generated at the high scale while the neutrino  
masses and mixings are  measured experimentally at a low scale, the  
renormalization group (RG) evolution effects need to be included.  
These radiative corrections in Type-I and Type-III seesaw are different, 
since the heavy fermions couple differently to the other particles 
in the theory. 
We note that below the mass scale of the lightest of the heavy particles,  
the effect of all heavy degrees of freedom are integrated out and  
the effective mass operators in these scenarios become identical.

The effect of RG induced quantum corrections on leptonic masses and 
mixings have been studied extensively in the literature 
\cite{chankowski-pokorski,babu-pantaleone,ellis-lola,Carena:1999xz, antusch,antusch-CP, 
antusch-HDM-MSSM,Fukuyama:2002ch}.  
These effects can have interesting consequences such as 
the generation of large mixing angles 
\cite{tanimoto,haba-LA,balaji-dighe, Balaji:2000ma,mpr,Agarwalla:2006dj},  
small mass splittings for degenerate neutrinos 
\cite{vissani-deg,branco-deg,Casas:1999tp, casas-deg,haba-deg,adhikari-deg, 
Joshipura:2002xa,Joshipura:2003fy,xing-deg}, or 
radiative generation of $\theta_{13}$ starting from a zero value 
at the high scale 
\cite{Joshipura:2002kj,Joshipura:2002gr,Mei:2004rn,jcppaper}. 
RG induced deviations from various high scale  symmetries like 
tri-bimaximal mixing scenario 
\cite{Plentinger:2005kx,tbm-planck,qlctbm} or quark-lepton 
complimentarity \cite{qlctbm,rgqlc,Schmidt:2006rb,Hirsch:2006je}  
and correlations with low scale observables have been explored. 
Such effects can have significant contributions from the  
threshold corrections 
\cite{King:2000hk,antusch-LMA,Mohapatra:2005gs}.  
The RG evolution of  the neutrino mass operator 
in the SM  and the Minimal Supersymmetric Standard Model 
(MSSM) in the context of Type-I seesaw 
\cite{antusch-LMA,seesaw1rg, Mei:2005qp} 
and Type -II seesaw \cite{seesaw2rg,Chao:2006ye} 
have been studied in the literature. 
In the context of Type-III seesaw with degenerate heavy fermions,  
the impact of the RG evolution 
on the vacuum stability and perturbativity bounds of the Higgs Boson  
has been explored in \cite{shafi-PLB}. 
 
In this work we study the RG evolution in the SM in the context of 
the Type-III seesaw model with nondegenerate heavy fermions.
Our model consists  of the SM with additional massive 
fermion triplets $\Sigma$ with masses $\sim M_{i}$, ($i =1,2,\cdots,M_r$) 
such that $M_1 < M_2 < \cdots < M_r$.  
Below the mass scale $M_1$ all the triplets will be decoupled from the  
model and  the RG evolution will be according to the SM.  
The triplets will manifest themselves at this low scale in the form of  
an effective operator $\kappa$ obtained by integrating out the heavy fields.  
For energy scales above $M_1$, the effect of the heavy fermions will come  
into play successively and above $M_r$ all the three triplets will  
contribute to the RG running.  
We evaluate the contributions of these fermion triplets to the wavefunction,  
mass and coupling constant renormalization of the SM fields and 
of the triplet fields themselves. 
We obtain the $\beta$-functions for RG evolution of 
the Yukawa couplings, the Higgs self-coupling, 
the Majorana mass matrix of the fermion triplets, 
the effective vertex $\kappa$ and the gauge couplings, 
including the extra contribution due to the additional triplets 
wherever applicable. We obtain 
analytic expressions for the runnings of the masses, mixing angles and 
phases in a basis where all the quantities are well-defined 
at every point in the parameter 
space including $\theta_{13} = 0$ \cite{jcppaper}.  
We also solve the RG equations numerically and  
present some illustrative examples of running of masses and mixing angles.  
We analyze the effect of the seesaw thresholds and Majorana phases and  
check if such a scheme can generate masses and mixing angles consistent  
with the current bounds.

The plan of the paper is as follows. In Sec.~\ref{model} we outline the 
basic features of the Type-III seesaw model including extra SU(2)$_L$-triplet 
fermions, and  describe how the effective neutrino mass  operator can arise 
by this mechanism. In Sec.~\ref{beta-function},  
we describe how to include the varying  
mass thresholds of the heavy particles in the analysis, discuss the  
renormalization of the SM extended with heavy triplets, and give the  
expressions of the $\beta$ functions including the effect of the extra  
triplets.  
In Sec.~\ref{running}, we detail the changes in the RG equations of 
the effective neutrino mass operator due to the inclusion of the 
extra fermion triplets.  
In Sec.~\ref{numerical}, we numerically demonstrate the modifications 
in the RG equations of the neutrino masses and mixing angles.  
We summarize our results in Sec.~\ref{concl}.

\section{The Type-III Seesaw Model} 
\label{model} 
 
We consider the Type-III seesaw model where three heavy fermions are added  
to each family of the SM. 
These fermions have zero weak hypercharge, i.e. they are 
singlets of the gauge group U(1)$_Y$ of the SM. 
However, under the SU(2)$_L$ gauge, they transform as a triplet 
in the adjoint representation. 
In the basis of the Pauli matrices $\{\sigma^1,\;\sigma^2,\;\sigma^3 \}$, 
this triplet can be represented as 
\beq 
\Sigma_R = \left( \barr{cc} 
\Sigma_R^0/\sqrt{2} & \Sigma_R^+ \\ 
\Sigma_R^- & -\Sigma_R^0/\sqrt{2} 
\earr \right)  \equiv  \frac{\Sigma_R^i \sigma^i}{\sqrt{2}} \; , 
\label{Sigma} 
\eeq 
where $\Sigma_R^\pm = {(\Sigma_R^1 \mp i \Sigma_R^2)}{\sqrt{2}} $. 
For the sake of simplicity of further calculations, we combine 
$\Sigma_R$ with its charge conjugate 
\beqa 
\Sigma_R^C = \left( 
\barr{cc} 
\Sigma_R^{0C}/\sqrt{2} & \Sigma_R^{-C} \\ 
\Sigma_R^{+C} & -\Sigma_R^{0C}/\sqrt{2} 
\earr \right) \equiv  \frac{\Sigma_R^{Ci} \sigma^i}{\sqrt{2}} \; , 
\label{SigmaC} 
\eeqa 
and use the quantity $\Sigma$, defined as 
\beq 
\Sigma \equiv \Sigma_R + {(\Sigma_R)}^C \, . 
\label{sigma_maj} 
\eeq 
Clearly, $\Sigma$ also transforms in the adjoint representation  
of SU(2)$_L$. 
Note that though formally $\Sigma = \Sigma^C$,  the 
individual elements of $\Sigma$ are not all Majorana particles. 
While the diagonal elements of $\Sigma$ are indeed 
Majorana spinors which represent the neutral component of $\Sigma$, the 
off-diagonal elements are charged Dirac spinors.

\subsection{The Lagrangian} 
\label{sec-lagrangian} 
 
Introduction of the fermionic triplets  
$\Sigma$ will introduce new terms in the Lagrangian. The net Lagrangian is  
\beqa 
{\cal L} = {\cal L}_{SM} +  {\cal{L}}_{\Sigma} \; ,  
\eeqa 
where 
\beqa 
{\cal{L}}_{\Sigma} = {\cal{L}}_{\Sigma,kin} + {\cal{L}}_{\Sigma,mass} 
                      + {\cal{L}}_{\Sigma,Yukawa} \; . 
\eeqa 
Here, 
\beqa 
{\cal{L}}_{\Sigma,kin} & = & \Tr[\overline{\Sigma} i \D  \Sigma] \; , \\ 
{\cal{L}}_{\Sigma,mass} &  = & \frac{1}{2} \Tr[\Sb {\mathbbm M}_\Sigma \Sigma] \; , \\ 
{\cal{L}}_{\Sigma,Yukawa} & = & 
-\overline{l_L} \sqrt{2} Y_\Sigma^\dagger 
\Sigma \widetilde{\phi} - 
\phi^T \varepsilon ^T \Sb \sqrt{2} Y_\Sigma l_L \; , 
\label{L-Sigma} 
\eeqa 
where  
\beqa 
\varepsilon = \left( \barr{cc} 0 & 1 \\ -1 & 0 \earr\right) 
\eeqa 
is the completely anti-symmetric tensor in the SU(2)$_L$ space.  
Here we have not written the generation indices explicitly.  
${\mathbbm M}_\Sigma$ is the Majorana mass matrix of the heavy  
fermion triplets and $Y_\Sigma$ is the Yukawa coupling.  
The SM fields $l_L$, $\phi$ and $\widetilde \phi$ are SU(2)$_L$ doublets  
and can be written as 
\beqa 
l_L = {\left( \barr{c} \nu_L \\ e_L^- \earr \right)}_{Y=-1} \; , \quad 
\phi = {\left( \barr{c} \phi^+ \\ \phi^0 \earr \right)}_{Y=1} \; , \quad 
\widetilde{\phi} =  
\varepsilon \phi^* = 
{\left( \barr{c} {\phi^0} \\ -\phi^- \earr \right)}_{Y=-1} \; .  
\eeqa 
Each member of the SU(2)$_L$ doublet $l_L$ is a 4-component Dirac spinor. 
Since the fermion triplet $\Sigma$ is in the adjoint  
representation of SU(2)$_L$, the covariant derivative  
of $\Sigma$ is defined as 
\beqa 
D_{\mu} \Sigma   = \partial_\mu \Sigma + i g_2 [W_\mu, \Sigma] \; , 
\eeqa 
where $g_2$ is the SU(2)$_L$ gauge coupling. 
 
All the Feynman diagrams for the new vertices involving the triplet  
fermionic field $\Sigma$ are given in the Appendix~\ref{App-feyn-Sigma}. 
The Feynman diagrams for the SM particles are shown in the  
Appendix~\ref{App-feyn-SM}.

\subsection{The effective vertex} 
\label{sec-eff-vertex} 
  
In the low energy limit of the extended standard model, we  
have an effective theory which will be described by the  
SM Lagrangian with the additional operators obtained 
by integrating out the heavy fermion triplets added to it.  
The lowest dimensional one of such operators is the dimension-5  
operator\footnote{ 
We use this form to emphasize the triplet nature of the $l_L$-$\phi$ 
pairs. Since all the dimension-5 operators are equivalent, 
we choose the normalization such that $\kappa_{fg}$ defined here 
matches that in \cite{kersten-thesis}. 
} 
\beqa 
{\cal L}_\kappa & = & \kappa_{fg}  
\left( {\overline{l_L^C}}^f \sigma^i \varepsilon  \phi \right)   
\left( \phi^T \sigma^i \varepsilon l_L^g \right) + {\rm h.c.} ,   
\\ 
&=& - \kappa_{fg} \left( {\overline{l_L^C}}^f_c  
\phi_a  l_{Lb}^g \phi_d \right)  
\frac{1}{2} \left( \varepsilon_{ac} \varepsilon_{bd} +  
\varepsilon_{ab} \varepsilon_{cd} \right) + {\rm h.c.} \; \; , 
\label{L-kappa} 
\eeqa 
where $\kappa$ is a symmetric complex matrix with mass dimension  
$(-1)$. Generation indices $f,g \in \{ 1,2,3\}$ are shown  
explicitly and $a, b, c, d \in \{1,2\}$ are the SU(2)$_L$ indices.  
In writing the last line we have used 
\begin{eqnarray} 
(\sigma^i)_{ab} (\sigma^i)_{cd} &=& 2 \delta_{ad} \delta_{bc} - \delta_{ab} \delta_{cd} \nn \\ 
\Rightarrow \quad  
(\sigma^i \varepsilon)_{ba} (\sigma^i \varepsilon)_{dc} &=& 
2 \varepsilon_{da} \varepsilon_{bc} - \varepsilon_{ba} \varepsilon_{dc} 
\label{su2-algebra} 
\end{eqnarray} 
and utilizing the $\phi_d \leftrightarrow \phi_a$ symmetry, we can write 
\begin{equation} 
2 \varepsilon_{da} \varepsilon_{bc} - \varepsilon_{ba} \varepsilon_{dc} 
= \frac{1}{2} \left( \varepsilon_{ab} \varepsilon_{dc} +  
\varepsilon_{db} \varepsilon_{ac} \right) \; . 
\label{symmetrize} 
\end{equation} 

\begin{figure} 
\begin{center} 
\includegraphics[scale=0.8]{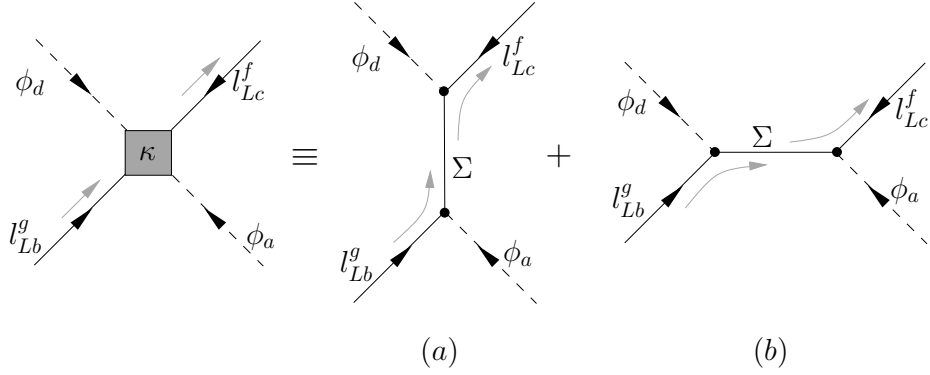} 
\caption{The effective vertex $\kappa$ at an energy $\mu \ll M_1$,  
after all the heavy fermions have been decoupled from the theory.  
$f,g \in \{ 1,2,3 \}$ are the generation indices. The SU(2)$_L$ and
generation indices for $\Sigma$ are not shown explicitly since they
are summed over.
\label{effective-vertex}} 
\end{center} 
\end{figure} 

The relevant diagrams in the complete theory giving rise to  
the effective operators in the low energy limit are shown in   
Fig~\ref{effective-vertex}. The ``shaded box'' on the left hand side  
represents the effective low energy vertex $\kappa$, while  
${\cal A}_{(a)}$ and ${\cal A}_{(b)}$ are the amplitudes of the  
diagrams labeled as $(a)$ and $(b)$ on the right hand side. 
The amplitudes are given by 
\beqa 
{\cal A}_{(a)} &=& i \mu^{\epsilon} \left( Y_\Sigma^T{\mathbbm M}_\Sigma^{-1}  
Y_\Sigma \right)_{fg} \left[ (\varepsilon^T \sigma^i)_{ab}  
(\varepsilon^T \sigma^i)_{cd} \right] P_L \; , \\ 
{\cal A}_{(b)} &=& i \mu^{\epsilon} \left( Y_\Sigma^T {\mathbbm M}_\Sigma^{-1}  
Y_\Sigma \right)_{fg} \left[ (\varepsilon^T \sigma^i)_{db}  
(\varepsilon^T \sigma^i)_{ca} \right] P_L \; ,  
\eeqa 
with $\epsilon = 4 - D$ where $D$ is the dimensionality 
that we introduce in order to use dimensional regularization. 
Note that ${\cal A}_{(b)}$ is obtained from ${\cal A}_{(a)}$ just by  
$d \leftrightarrow a$ interchange. 
Using Eq.~(\ref{su2-algebra}) one  
finally gets 
\beqa 
{\cal A}_{(a)} + {\cal A}_{(b)} &=& -i \mu^{\epsilon}  
\left(Y_\Sigma^T {\mathbbm M}_\Sigma^{-1} Y_\Sigma \right)_{fg} 
\left( \varepsilon_{ab}\varepsilon_{cd}  
+ \varepsilon_{ac}\varepsilon_{bd} \right) P_L \; . 
\label{kappa-rhs} 
\eeqa 
This is equal to the left hand side of Fig.~\ref{effective-vertex}  
with the identification  
\beqa 
\kappa = 2 Y_\Sigma^T {\mathbbm M}_\Sigma^{-1} Y_\Sigma \; . 
\label{kappa} 
\eeqa 
Equation~(\ref{kappa}) gives the Feynman rule for the low  
energy effective vertex $\kappa$, as shown in the  
Appendix~\ref{App-feyn-Sigma}. 
From Eqs.~(\ref{kappa}) and (\ref{L-kappa}), one gets  
the neutrino mass after spontaneous symmetry breaking to be  
\beq 
{\mathbbm m}_\nu = -\frac{v^2}{2}  Y_\Sigma^T {\mathbbm M}_\Sigma^{-1} Y_\Sigma 
\eeq  
which is the Type-III seesaw relation.  
Here, $v$ denotes the vacuum expectation value of the  
Higgs  field.

\section{Radiative corrections in Type-III seesaw} 
\label{beta-function} 
 
\subsection{Sequential decoupling of heavy fermions} 
 
Let us consider the most general case when there are $r$ triplets  
having masses $M_1 < M_2 < \cdots < M_{r-1} < M_r$. 
Above the heaviest  mass $M_r$,  all  
the $r$-triplets are coupled to the theory  
and will contribute to the neutrino mass through seesaw mechanism as 
\beqa 
\accentset{(r+1)}{{\mathbbm m}_\nu} &=& -\frac{v^2}{2} \;   
\accentset{(r+1)}{{Y}_\Sigma^T} \; ~ 
\accentset{(r+1)}  
{\mathbbm M}_\Sigma^{-1} \; ~ 
\accentset{(r+1)} 
{Y_\Sigma}~~~~~(\mu > M_r) \; . 
\label{mnup} 
\eeqa 
Here, $\accentset{(r+1)}Y_\Sigma$ is a $[r \times n_F]$ dimensional matrix  
($n_F$ is the number of flavors, which is 3 in our case),  
$\accentset{(r+1)}{{\mathbbm M}_\Sigma}$   is a $[r \times r]$ matrix  and  
$\accentset{(r+1)}{{\mathbbm m}_\nu}$ is a $[n_F \times n_F]$  
dimensional matrix.  
Below the scale $M_r$, the heaviest triplet decouples from the  
theory. Integrating out this degree of freedom gives rise to an effective 
operator $\accentset{(r)}\kappa$.   
The matching condition at $\mu = M_r$ is  
\beqa 
\left. \accentset{(r)}\kappa 
\right\arrowvert_{M_r} &=& 
\left. 2 \accentset{(r+1)}{Y_\Sigma^T}\; 
(M_r)^{-1} \;  
\accentset{(r+1)} 
{Y_\Sigma} 
\right\arrowvert_{M_r} \; . 
\label{mnum} 
\eeqa 
This condition ensures the continuity of ${\mathbbm m}_\nu$ at 
$\mu=M_r$. 
In order to get the value of the threshold $M_r$, we need to  
write the above matching condition in the basis where  
${\mathbbm M}_{\Sigma} = {\rm diag}( M_1, M_2, \cdots , M_r)$. 
Here it is worth mentioning that the matching scale has to be found  
carefully since ${{\mathbbm M}_\Sigma}$ itself runs with the energy scale, 
i.e. $M_i = M_i(\mu)$. The threshold scale $M_i$ is therefore to 
be understood as $M_i(\mu=M_i)$.

In the energy range $M_{r-1} < \mu < M_r$, the  
effective mass of the neutrinos will be given as 
\beqa 
\accentset{(r)}{{\mathbbm m}_\nu} &=& -\frac{v^2}{4}  
\left( \accentset{(r)}{\kappa} +  
2 \accentset{(r)}{Y}_\Sigma^{\, T}\; 
\accentset{(r)} 
{\mathbbm M}_\Sigma^{-1}\; 
\accentset{(r)} 
{Y_\Sigma}\; 
\right) \; . 
\label{mnum-1} 
\eeqa 
The first term in Eq.~(\ref{mnum-1}) is the contribution  
of the  integrated out triplet of mass $M_r$ through the  
effective operator $\accentset{(r)}{\kappa}$. The second term 
represents  
the contribution of the remaining $(r-1)$ heavy fermion triplets, which  
are still coupled to the theory,  through the seesaw mechanism.  
$\accentset{(r)}{{\mathbbm M}}_\Sigma$ is now a $[(r-1) \times (r-1)]$ matrix  
while $\accentset{(r)}{Y_\Sigma}$ is a $[(r-1) \times n_F]$ dimensional  
matrix.

The matching condition at $\mu = M_{r-1}$ is  
\beqa 
\left. \accentset{(r-1)}\kappa \; \;\right\arrowvert_{M_{r-1}} & = &  
\left. {\accentset{(r)}\kappa} \; \right\arrowvert_{M_{r-1}} +  
\left. 2 {\accentset{(r)}{Y_\Sigma}^{\; T}} \;  
(M_{r-1})^{-1} \; 
{\accentset{(r)}{Y}_\Sigma} \right\arrowvert_{M_{r-1}} \; . 
\label{matching} 
\eeqa

Generalizing the above sequence, we can say that if we consider  
the intermediate energy region between the $(n-1)^{\rm th}$ and the  
$n^{\rm th}$ threshold, i.e.  
$M_n > \mu > M_{n-1}$, then  all the heavy triplets from masses $M_r$   
down to $M_n$ have been decoupled.   
In this region the Yukawa matrix $\accentset{(n)}Y_\Sigma$  
will be $[(n-1) \times n_F]$ dimensional and will be given as 
\beqa 
{Y}_\Sigma \rightarrow  
\left( \begin{array}{ccc} 
{(y_\Sigma)}_{1,1} & \cdots & {(y_\Sigma)}_{1,n_F} \\ 
\vdots & & \vdots \\ 
{(y_\Sigma)}_{n-1,1} & \cdots & {(y_\Sigma)}_{n-1,n_F}\\ 
\hline 
0 & \cdots & 0 \\ 
\vdots &  & \vdots \\ 
0 & \cdots & 0 
\end{array} \right)  \: 
\begin{array}{cl} 
\left.\begin{array}{c}\\[1.7cm]\end{array}\right\} 
& =\Ysigman\;, 
 \\ 
\left.\begin{array}{c}\\[1.7cm]\end{array}\right\} 
& \begin{array}{l}\text{heavy triplets with masses }\\[-0.3cm] 
M_n \mbox{---} M_r  \; \text{integrated out}\;.\end{array} 
\end{array} \; 
\label{Y-sigma-n} 
\eeqa 
$\accentset{(n)}{\mathbbm M}_\Sigma$ will be $[(n-1) \times (n-1)]$ dimensional.  
In this energy range the effective neutrino mass matrix will be  
\beqa 
\accentset{(n)}{{\mathbbm m}}_\nu &=& -\frac{v^2}{4} 
\left( \;\accentset{(n)}{\kappa} + 
2 \accentset{(n)}{Q} \right) \; , 
\label{mnu} 
\eeqa 
with 
\beq 
\Qn \equiv \Ysigman^{\; T} \; \accentset{(n)}{{\mathbbm M}}_\Sigma^{-1} \Ysigman \; , 
\label{Qn} 
\eeq 
while the matching condition  
at $\mu = M_{n}$ is given by Eq.~(\ref{matching}) with  
$r$ replaced by $(n+1)$. For $\mu < M_1$, all the heavy triplets  
will get decoupled and thus only ${\accentset{(1)}\kappa}$  
will contribute, which is the low energy effective neutrino 
mass operator.

\subsection{Dimensional regularization and renormalization} 
\label{sec-renormSM} 
 
Now we consider the radiative corrections to the fields,  
masses and couplings in our model, on the lines of that  
performed in \cite{antusch,kersten-thesis}  
in the context of Type-I seesaw. The  
wavefunction renormalizations are defined as 
\beq 
\psi_B^f = \left( Z_\psi^{\frac{1}{2}} \right)_{fg}\psi^g \; , 
\label{Zphi} 
\eeq 
where $\psi \in \{ l_L, q_L,e_R,u_R,d_R \}$. We denote the  
renormalized quantities as $X$ and the corresponding bare fields as $X_B$. 
For the fermion triplets 
\beq 
\Sigma^{fi}_B = \left( Z_\Sigma^{\frac{1}{2}} \right)_{fg} \Sigma^{gi} \; . 
\eeq 
For the doublet Higgs 
\beq 
\phi_B = Z_\phi^{\frac{1}{2}} \phi \; , 
\eeq 
whereas 
\beq 
A_B = Z_A^{\frac{1}{2}} A \;  
\eeq 
for the gauge bosons where $A \in \{B, W^i, G^A \}$. For the  
Faddeev-Popov ghosts one has  
\beq 
c_B = Z_c^{\frac{1}{2}} c \; , 
\label{Zc} 
\eeq 
however the ghosts will not appear in the RG evolution of the  
relevant quantities at one loop level. 
We introduce the abbreviation 
\beq 
\delta Z_X = Z_X - 1\; , 
\eeq 
where $Z_X$ denotes the renormalization constant of any  
of the relevant quantities $X$. 
 
We will use the dimensional regularization  
and the minimal subtraction scheme  
for renormalization. In this renormalization formalism,  
the counter terms are  
defined such that they only cancel out the divergent parts.  
Thus the renormalization constants are of the form 
\beqa 
Z_X = 1 + \displaystyle\sum_{k \geq 1} \delta Z_{X,k} \frac{1}{\epsilon^k} \; , 
\eeqa 
where the $\delta Z_{X,k}$ are independent of $\epsilon$.  
In our scenario, at the one loop level, the renormalization  
constants are proportional to $1/\epsilon$. 
The final results of course will be independent of the particular  
regularization as well as the renormalization scheme  
used for the calculations.   
 
The diagrams contributing to the  
renormalization constants of the different quantities are all  
shown explicitly in Appendix~\ref{App-renorm}.  
The renormalization constants of different quantities are given by 
\beqa 
\delta Z_{\phi} &=& - \frac{1}{16 \pi^2}  
\Bigl( 2 T - \frac{3}{10} (3 - \xi_1) g_1^2  
- \frac{3}{2} (3 - \xi_2) g_2^2 \Bigr) \frac{1}{\epsilon} \; , \\ 
\delta Z_{l_L} &=& - \frac{1}{16 \pi^2} \Bigl( Y_e^\dagger Y_e  
+ 3 \Ysigman^{\; \dagger} \Ysigman +  
\frac{3}{10} \xi_1 g_1^2 + \frac{3}{2} \xi_2 g_2^2  \Bigr)  
\frac{1}{\epsilon} \; , \\ 
\delta Z_{e_R} &=& - \frac{1}{16 \pi^2} \Bigl( 2 Y_e Y_e^\dagger  
+ \frac{6}{5} \xi_1  g_1^2  \Bigr) \frac{1}{\epsilon} \; ,\\ 
\delta Z_{\Sigma} &=&  - \frac{1}{16 \pi^2}  
\Biggl[ \Bigl(2 \accentset{(n)}{Y}_\Sigma \accentset{(n)}{Y}_\Sigma^\dagger 
+ 4 \xi_2 g_2^2  \Bigr) P_R +  
\Bigl(2 (\accentset{(n)}{Y}_\Sigma \accentset{(n)}{Y}_\Sigma^\dagger)^\ast  
+ 4 \xi_2 g_2^2  \Bigr) P_L \Biggr] \frac{1}{\epsilon} \; .
\eeqa 
where we have used the $R_\xi$ gauge, and the 
GUT normalization of the gauge couplings \cite{chankowski-pokorski}.

The Yukawa couplings are renormalized as\footnote{ 
In \cite{shafi-PLB}  
the contributions of fermion triplets to some 
of the above renormalization constants 
are calculated in the context of  SM extended  
with these fields. Their conventions  
of field normalizations are different and hence the results may differ  
upto numerical constants in certain cases.  
However, their Eq.~(19)    
for $\delta Y_\nu$, which is the same quantity as our  
$\delta Z_{Y_\Sigma}$ in Eq.~(\ref{ZYsigma}), is missing the  
$Y_e^\dagger Y_e$ term.  
The source of this term is the diagram labelled as (F2) 
in Appendix~\ref{App-renorm}.  
The extra contribution to $\delta Z_{Ye}$ from the fermion triplets 
has also not been calculated in \cite{shafi-PLB}.  
} 
\beqa 
\delta Z_{Y_e} &=& - \frac{1}{16 \pi^2}  
\Bigl( -6 \Ysigman^{\; \dagger} \Ysigman 
+ \frac{9}{10} \left( 2 + \xi_1 \right) g_1^2  
+ \frac{3}{2} \xi_2 g_2^2  \Bigr) \frac{1}{\epsilon} \; , 
\label{ZYe} \\ 
\delta Z_{Y_\Sigma} &=& - \frac{1}{16 \pi^2} \Bigl( 2 Y_e^\dagger Y_e  
- \frac{3}{10} \xi_1 g_1^2 - \frac{1}{2}  
\left( 12 + 7 \xi_2  \right) g_2^2\Bigr) \frac{1}{\epsilon} \; , 
\label{ZYsigma} 
\eeqa 
while the Majorana neutrino mass matrix gets renormalized as 
\beq 
\delta Z_{{\mathbbm M}_\Sigma} 
= -\frac{1}{16 \pi^2} \left( 12 + 4 \xi_2 \right) g_2^2 \; \frac{1}{\epsilon} \; . 
\eeq 
The addition of the right handed fermion triplets to the SM will  
contribute one extra diagram to the renormalization of the  
Higgs self-coupling $\lambda$, as shown in the diagram (G1) 
of the Appendix~\ref{App-renorm}.  
This contribution will be\footnote{
Note that Ref.~\cite{shafi-PLB} gives this quantity
($\delta \lambda$ in their Eq.~(20)) to be of the form
$\Tr (Y_\Sigma^\dagger Y_\Sigma)$.
However, the additional contribution to the  
Higgs quartic coupling  $\delta Z_\lambda$ should be of the form
$\Tr ( Y_\Sigma^\dagger Y_\Sigma  Y_\Sigma^\dagger Y_\Sigma)$,
since it comes from the diagram (G1) in Appendix~\ref{App-renorm}. 
} 
\beq 
\left. \delta Z_\lambda \right\arrowvert_{\rm new} =-\frac{5 i}{4 \pi^2}  
\Tr\left[ \Ysigman^{\; \dagger} \Ysigman \Ysigman^{\; \dagger} \Ysigman \right] 
\left( \delta_{ab} \delta_{cd} +\delta_{ac} \delta_{bd}  \right)  
\frac{1}{\epsilon} \; . 
\label{Zlambda-new} 
\eeq 

Finally for the effective vertex $\accentset{(n)}{\kappa}$, the renormalization  
constant is 
\beq 
\delta \accentset{(n)}{\kappa} = - \frac{1}{16 \pi^2}  
\Bigl[ 2 \; \accentset{(n)}{\kappa} \left( Y_e^\dagger Y_e \right) 
+ 2 \left( Y_e^\dagger Y_e \right)^T \accentset{(n)}{\kappa}  
- \lambda \accentset{(n)}{\kappa}  
- \left( \frac{3}{2} - \xi_1 \right) g_1^2 \; \accentset{(n)}{\kappa}  
- \left( \frac{3}{2} - 3 \xi_2 \right) g_2^2 \; \accentset{(n)}{\kappa} \; \Bigr]  
\frac{1}{\epsilon} \; . 
\eeq 
We observe  that there is  
no contribution from the fermion triplet $\Sigma$ in the loop,  
which means that $\delta \accentset{(n)}{\kappa}$ will not directly depend  
on the fermion triplets still coupled to the theory. However,  
during RG evolution  
an indirect dependence will creep in via the  other couplings.  
 
 
\subsection{Calculation of the $\beta$ functions} 
\label{sec-beta-func} 
%
To calculate the $\beta$ functions for the RG evolution of the  
Yukawa couplings, Majorana mass matrix, the effective vertex $\kappa$  
and other relevant quantities, we consider the relations between the  
bare ($X_B$) and the corresponding renormalized ($X$) quantities  
given by 
\beqa 
{Z_{\Sigma}^T}^{\frac{1}{2}} {\mathbbm M}_{\Sigma B} Z_\Sigma^{\frac{1}{2}} &=& 
Z_{{\mathbbm M}_\Sigma} {\mathbbm M}_\Sigma \; , \\
Z_{\Sigma_R}^{\frac{1}{2}} Y_{\Sigma B} Z_\phi^{\frac{1}{2}} Z_{l_L}^{\frac{1}{2}} &=& 
\mu^{\frac{\epsilon}{2}} Y_\Sigma Z_{Y_\Sigma} \; , \\
Z_{e_R}^{\frac{1}{2}} Y_{e B} Z_\phi^{\frac{1}{2}} Z_{l_L}^{\frac{1}{2}} &=& 
\mu^{\frac{\epsilon}{2}} Y_e Z_{Y_e} \; , \\
{Z_{l_L}^T}^{\frac{1}{2}} Z_\phi^{\frac{1}{2}} \kappa_B Z_\phi^{\frac{1}{2}} 
Z_{l_L}^{\frac{1}{2}} &=& \mu^\epsilon (\kappa + \delta \kappa) \; ,  
\eeqa 
where $Z_{\Sigma_R} = P_R Z_\Sigma$. We further 
use the functional differentiation method as in  
\cite{kersten-thesis} to find the $\beta$ functions for the  
Yukawa couplings as 
\beqa 
16 \pi^2 \beta_{Y_e} &=& Y_e \left( \frac{3}{2} Y_e^\dagger Y_e  
+ \frac{15}{2} \Ysigman^{\; \dagger} \Ysigman + T - \frac{9}{4} g_1^2  
- \frac{9}{4} g_2^2 \right) \; ,  
\label{beta-Ye} \\ 
16 \pi^2 \beta_{Y_\Sigma} &=& \Ysigman \left( \frac{5}{2} Y_e^\dagger Y_e  
+ \frac{5}{2} \Ysigman^{\; \dagger} \Ysigman + T - \frac{9}{20} g_1^2  
- \frac{33}{4} g_2^2 \right) \; , \label{beta-YSigma} \\ 
16 \pi^2 \beta_{Y_u} &=& Y_u \left( \frac{3}{2} Y_u^\dagger Y_u  
- \frac{3}{2} Y_d^\dagger Y_d + T  
-\frac{17}{20} g_1^2 - \frac{9}{4} g_2^2 - 8 g_3^2 \right) \; ,  
\label{beta-Yu} \\ 
16 \pi^2 \beta_{Y_d} &=& Y_d \left( \frac{3}{2} Y_d^\dagger Y_d  
- \frac{3}{2} Y_u^\dagger Y_u + T  
-\frac{1}{4} g_1^2 - \frac{9}{4} g_2^2 - 8 g_3^2 \right) \; .  
\label{beta-Yd} 
\eeqa 
Here 
\beqa 
T &=& {\rm Tr} \left[ Y_e^\dagger Y_e + 3 \Ysigman^{\; \dagger} \Ysigman  
+ 3 Y_u^\dagger Y_u + 3 Y_d^\dagger Y_d  \right] \; , 
\label{trace} 
\eeqa 
and $\beta_X \equiv \mu (dX / d \mu)$. 
Note that $\Ysigman$ is given in Eq.~(\ref{Y-sigma-n}),  
with $(n-1)$ the number of heavy fermion triplets still coupled to the theory.

Since the fermion triplets have non-zero SU(2)$_L$ charge, they  
couple to the $W$ bosons and hence will affect the RG evolution of  
the gauge coupling $g_2$ via 
\beq 
16 \pi^2 \beta_{g_2} = b_2 g_2^3  
\label{beta-g2} \; , 
\eeq 
where  
\beq 
b_2 = -\frac{19}{6} + \frac{4 (n-1)}{3} \; . 
\label{b2} 
\eeq 
\begin{figure} 
\begin{center} 
\includegraphics[scale=0.9]{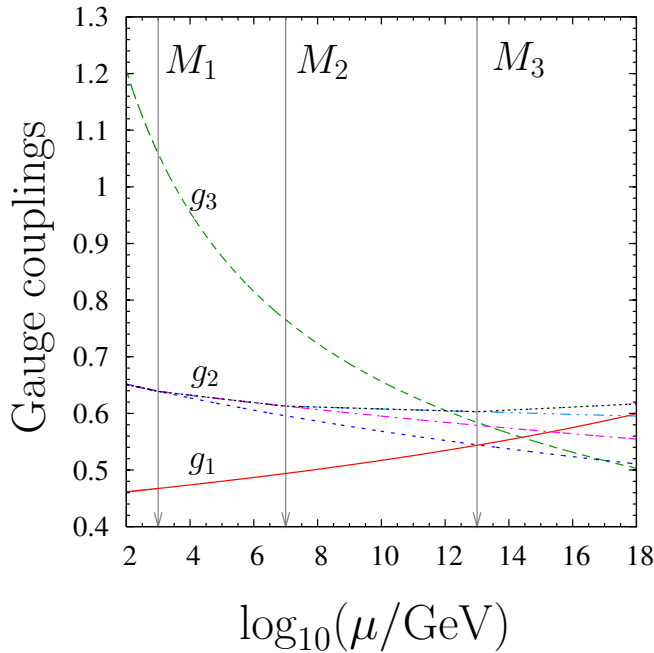} 
\caption{The solid (red) line and the dashed (green) lines  
show the energy scale variations of $g_1$ and $g_3$  
respectively in the SM, which is unaffected in Type-III seesaw. 
The dotted (blue) line gives the SM  
running of $g_2$, while dot-dashed (magenta),  
dot-dot-dashed (sky) and densely dotted (black) lines show  
the running if there were one, two or three fermion triplets  
respectively.  
\label{gauge-run}} 
\end{center} 
\end{figure} 
Note that if the number of heavy fermion triplets is $\leq 2$,  
the value of $b_2$ is always negative. On the other hand, if the  
number is $\geq 3$, then $b_2$ becomes positive above the mass  
scale $M_3$. Adding fermion triplets shifts the $g_1$-$g_2$  
intersection to higher energy scales, and the $g_2$-$g_3$ 
intersection to lower energy scales, as can be seen from  
Fig.~\ref{gauge-run}. 
The exact situation would depend on the values of $M_i$.

The RG evolution of $\lambda$ is given by 
\beqa 
16 \pi^2 \beta_\lambda &=& 6 \lambda^2 - 3 \lambda  
\left( \frac{3}{5} g_1^2 + 3 g_2^2 \right) 
+ 3 g_2^4 + \frac{3}{2} \left( \frac{3}{5} g_1^2 + g_2^2 \right)^2  
+ 4 \lambda T \nn \\ 
&& - 8 \; {\rm Tr} [Y_e^\dagger Y_e Y_e^\dagger Y_e  
+ 3 Y_u^\dagger Y_u Y_u^\dagger Y_u  
+ 3 Y_d^\dagger Y_d Y_d^\dagger Y_d] - 20 \;  
{\rm Tr}[\Ysigman^{\; \dagger} \Ysigman \Ysigman^{\; \dagger} \Ysigman] \; .  
\label{beta-lambda} 
\eeqa 
As it is evident from Eq.~(\ref{beta-lambda}), the last term is the  
new contribution to the $\beta$-function from the heavy triplets  
still coupled to the theory.

The RG evolution of the Majorana mass matrix of the heavy triplet  
fermions is given by 
\beqa 
16 \pi^2 \beta_{{\mathbbm M}_\Sigma} &=&  
\left[ \left( \Ysigman \Ysigman^{\; \dagger} \right) P_L +  
{ \left( \Ysigman \Ysigman^{\; \dagger} \right)}^\ast P_R \right] {\mathbbm M}_\Sigma \nn \\ 
&+& {\mathbbm M}_\Sigma \left[  {\left( \Ysigman \Ysigman^{\; \dagger} \right)}^\ast P_L + 
\left( \Ysigman \Ysigman^{\; \dagger} \right) P_R \right] - 12 g_2^2 {\mathbbm M}_\Sigma \; , 
\label{beta-M}  
\eeqa 
where it is always possible to separate the components of  
different chirality to get the left-chiral part as 
\beqa 
16 \pi^2 \beta_{{\mathbbm M}_\Sigma} &=& \left(  
\Ysigman \Ysigman^{\; \dagger} \right){\mathbbm M}_\Sigma  
+ {\mathbbm M}_\Sigma {\left( \Ysigman \Ysigman^{\; \dagger} \right)}^T  
- 12 g_2^2 {\mathbbm M}_\Sigma \; , 
\label{beta-M-L} 
\eeqa 
since $P_L + P_R = {\mathbbm I}$. Thus all the $\beta$-functions are  
gauge-independent, as they should be. The anomalous dimension of  
${\mathbbm M}_\Sigma$ is 
\beqa 
-16 \pi^2 \; {\accentset{(0)}{\gamma}}_{{\mathbbm M}_\Sigma} &=&  
{\mathbbm M}_\Sigma^{-1} \left[ \left( \Ysigman \Ysigman^{\; \dagger} \right) P_L +  
{ \left( \Ysigman \Ysigman^{\; \dagger} \right)}^\ast P_R \right] {\mathbbm M}_\Sigma \nn \\ 
&+& \left[  {\left( \Ysigman \Ysigman^{\; \dagger} \right)}^\ast P_L + 
\left( \Ysigman \Ysigman^{\; \dagger} \right) P_R \right] - 12 g_2^2 \; . 
\label{gamma}  
\eeqa 
Similar to the left-chiral component of $\beta_{{\mathbbm M}_\Sigma}$ in  
Eq.~(\ref{beta-M-L}), the left-chiral component of 
$\accentset{(0)}{\gamma}_{{\mathbbm M}_\Sigma}$ is  
\beqa 
-16 \pi^2 \; {\accentset{(0)}{\gamma}}_{{\mathbbm M}_\Sigma} =  
{\mathbbm M}_\Sigma^{-1} \left( \Ysigman \Ysigman^{\; \dagger} \right) {\mathbbm M}_\Sigma 
+ {\left( \Ysigman \Ysigman^{\; \dagger} \right)}^\ast - 12 g_2^2 \; . 
\eeqa 

As seen from Eq.~(\ref{mnu}), the RG evolution of the light neutrino  
mass matrix $\accentset{(n)}{\mathbbm m}_\nu$ is controlled by the  
evolutions of both $\accentset{(n)}{\kappa}$ and $\Qn$, which are  
given by 
\beqa 
16 \pi^2 \beta_\kappa &=& \alpha_\kappa \accentset{(n)}{\kappa} 
+ P_\kappa^T \, \accentset{(n)}{\kappa}  
+ \accentset{(n)}{\kappa} P_\kappa \; ,  
\label{beta-kappa} 
\\ 
16 \pi^2 \beta_{Q} &=& \alpha_{Q} \accentset{(n)}{Q}  
+ P_Q^T \accentset{(n)}{Q}  
+ \accentset{(n)}{Q} P_Q \; , 
\label{beta-Q} 
\eeqa 
with 
\beqa 
P_\kappa &=& \frac{3}{2} \Ysigman^{\; \dagger} \Ysigman 
- \frac{3}{2} Y_e^\dagger Y_e \; ; \quad 
\alpha_\kappa = 2 T + \lambda - 3 g_2^2 \; ,  
\label{P-Kappa} 
\eeqa 
\beqa 
P_Q &=& \frac{3}{2} \Ysigman^{\; \dagger} \Ysigman  
+ \frac{5}{2} Y_e^\dagger Y_e \; ; \quad 
\alpha_Q = 2 T - \frac{9}{10} g_1^2 - \frac{9}{2} g_2^2 \; . 
\label{P-Q} 
\eeqa 
%
 
\section{RG running of neutrino masses and mixing angles} 
\label{running} 
 
To derive the RG evolution for the neutrino masses and mixings  
we follow the standard procedure \cite{antusch,seesaw1rg}.  
At any energy scale $\mu$, the neutrino mass matrix  
${\mathbbm m}_\nu$ can be diagonalized by a unitary transformation via  
\beq 
U_\nu(\mu)^T {\mathbbm m}_\nu(\mu) U_\nu(\mu) 
= {\rm diag}(m_1(\mu), m_2(\mu), m_3(\mu)) \;.  
\eeq  
In a basis where $Y_e$ is diagonal,  the neutrino mixing matrix is given as  
\beq 
U_{\rm PMNS} = U_\nu \; , 
\eeq 
where $U_{\rm PMNS}$ is the Pontecorvo-Maki-Nakagawa-Sakata  
neutrino mixing matrix \cite{pontecorvo,mns}. 
From Eqs.~(\ref{beta-Ye}) it is seen that  
above and between the thresholds, off-diagonal  
terms will be generated in  
$Y_e$ even if we start with a diagonal $Y_e$ at the high scale, 
due to the $Y_\Sigma^\dagger Y_\Sigma$ terms.  
These terms will give additional contributions to the evolution of   
different parameters. In the presence of $Y_e$ with off-diagonal entries,  
the neutrino mixing matrix will be given as 
\beq 
U_{\rm PMNS} = U_e^\dagger U_\nu \; , 
\eeq 
where $U_e$ is the unitary matrix that diagonalizes $Y_e^\dagger Y_e$  
by a unitary transformation. $U_{\rm PMNS}$ is parameterized as 
\cite{mns,Schechter:1980gr}
\beq 
U_{\rm PMNS} = {\rm diag}(e^{i \delta_e}, \, e^{i \delta_\mu}, \, e^{i \delta_\tau} ) \; .   
\; {\cal U} \; . \;   
{\rm diag}(e^{-i \phi_1}, \,  e^{-i \phi_2}, 1) \; , 
\eeq 
with 
\beqa 
{\cal U} = \left( 
 \barr{ccc} 
 c_{12}c_{13} & s_{12}c_{13} & s_{13}e^{-i \delta}\\ 
 -c_{23}s_{12}-s_{23}s_{13}c_{12}e^{i \delta} & 
 c_{23}c_{12}-s_{23}s_{13}s_{12}e^{i \delta} & s_{23}c_{13}\\ 
 s_{23}s_{12}-c_{23}s_{13}c_{12}e^{i \delta} & 
 -s_{23}c_{12}-c_{23}s_{13}s_{12}e^{i \delta} & c_{23}c_{13} 
 \earr 
 \right) \; . 
\eeqa 
Here $c_{ij}$ and $s_{ij}$ are the cosines and sines respectively 
of the mixing angle $\theta_{ij}$,  
$\delta$ is the  
Dirac CP violating phase, $\phi_i$ are the Majorana phases.  
The ``flavor'' phases $\delta_e$, $\delta_\mu$ and $\delta_\tau$  
do not play any role in the phenomenology of neutrino mixing.

In this work,  we consider $r = 3$ heavy fermion triplets, one for each  
generation. Then $Y_\Sigma$ is a $3 \times 3$ matrix at high scale and  
is identically zero for $\mu < M_1$.  
The RG evolution of the neutrino parameters is then controlled by 
\beqa 
16 \pi^2 \beta_{Y_e} &=& Y_e F + \alpha_e Y_e \; ,  
\label{beta-Ye-1} \\ 
16 \pi^2 \beta_{{\mathbbm m}_\nu} &=& P^T {\mathbbm m}_\nu  
+ {\mathbbm m}_\nu P + \alpha_\nu {\mathbbm m}_\nu \; ,  
\label{beta-mnu}  
\eeqa 
where 
\beqa 
P &=& C_e Y_e^\dagger Y_e + C_\Sigma Y_\Sigma^{\dagger} Y_\Sigma \; ,  
\label{P}\\ 
F &=& D_e Y_e^\dagger Y_e + D_\Sigma Y_\Sigma^{\dagger} Y_\Sigma \; .  
\label{F} 
\eeqa 
Eqs.~(\ref{beta-Ye-1}) and (\ref{beta-mnu}) are essentially the  
same as the $\beta$-functions given in Eqs.~(\ref{beta-Ye}),  
(\ref{beta-kappa}) and  (\ref{beta-Q}), which we rewrite in the  
above form for later discussions. For $\mu > M_3$ and $\mu < M_1$,  
the evolutions of $Y_e$ and ${\mathbbm m}_\nu$ can be written in simple  
analytic forms, using Table~\ref{tab-1}. Note that for $\mu > M_3$  
the running of the neutrino masses will be governed by $\beta_Q$  
and so $P$ in Eq.~(\ref{P}) is the same as $P_Q$ as defined in  
Eq.~(\ref{P-Q}). On the other hand, for $\mu < M_1$, we have  
$P = P_\kappa$ as given in Eq.~(\ref{P-Kappa}). 
\begin{table}[t!] 
 \begin{center} 
 \begin{tabular}{ccccccccccccc} 
 \hline 
 &  
& $C_e$ &  $\phantom{sp}$  
& $C_\Sigma$ &  $\phantom{sp}$  
& $D_e$ & $\phantom{sp}$  
& $D_\Sigma$ & $\phantom{spc}$  
&$\alpha_e$ & $\phantom{spc}$ &$\alpha_\nu$  
\\ 
\hline 
 $\mu > M_3$ & $\phantom{spc}$  
&$\frac{5}{2}$ & $\phantom{sp}$  
& $\frac{3}{2}$ &  $\phantom{sp}$  
& $\frac{3}{2}$ & $\phantom{sp}$  
& $\frac{15}{2}$ & $\phantom{spc}$ & 
$T - \frac{9}{4} g_1^2 - \frac{9}{4} g_2^2$ & $\phantom{spc}$  
& $2 T - \frac{9}{10} g_1^2 - \frac{9}{2} g_2^2$  
\\ 
\hline 
 $\mu < M_1$ & $\phantom{spc}$  
&-$\frac{3}{2}$ &  $\phantom{sp}$  
& $0$ & $\phantom{sp}$  
& $\frac{3}{2}$ &  $\phantom{sp}$  
& $0$ & $\phantom{spc}$  
&$T - \frac{9}{4} g_1^2 - \frac{9}{4} g_2^2$ & $\phantom{spc}$ & 
 $2 T + \lambda - 3 g_2^2$  
\\ 
 \hline 
 \end{tabular} 
 \end{center} 
\caption{Coefficients of the $\beta$-functions governing the  
running of neutrino masses and mixings in the energy regimes  
$\mu > M_3$ and $\mu < M_1$. The quantity $T$ is defined in  
Eq.~(\ref{trace}). 
 \label{tab-1} } 
 \end{table} 
$P$ and $F$ are $3 \times 3$ matrices, with the rows and columns  
representing generations. We denote the elements of $P$ and $F$  
by $P_{fg}$ and $F_{fg}$. The coefficient of $P_{fg}$ and $F_{fg}$  
in the running of $Y_e$ and ${\mathbbm m}_\nu$ can be read off directly from  
\cite{seesaw1rg}, since the structure of Eqs.~(\ref{beta-Ye-1})  
and (\ref{beta-mnu}) remain the same both in Type-I and Type-III  
seesaw. The values of $P_{fg}$ and $F_{fg}$ themselves will however  
be different because of different underlying theories. 
The values of the relevant coefficients in Type-III seesaw are shown
in Table~\ref{tab-1}.  
 
If we consider the running equations in the basis  
${\cal P}_\delta = \{ m_i; \theta_{12}, \theta_{13}, \theta_{23}; \phi_i; \delta \}$,  
then both $\delta$ and $\dot{\delta}$ become ill-defined at  
$\theta_{13} = 0$ \cite{antusch,antusch-CP} and as a consequence,  
$\dot{\theta}_{13}$ also becomes ill-defined  
because of its $\delta$ dependence. 
This is only an apparent singularity. One can get rid of 
it by imposing a particular value of $\cot \delta$ at $\theta_{13} = 0$ 
\cite{antusch,antusch-CP} or by using the basis 
${\cal P}_J = \{ m_i; \theta_{12}, \theta_{23}, \theta_{13}^2; \phi_i; 
J_{\rm CP}, J'_{\rm CP} \}$, where the singularity does not appear at 
all \cite{jcppaper}. 
Here $J_{\rm CP}$ and $J'_{\rm CP}$ are defined as 
\beqa 
J_{\rm CP} &\equiv& \frac{1}{2} s_{12} c_{12} s_{23} c_{23} s_{13} c_{13}^2  
\sin{\delta} \; , \\ 
J'_{\rm CP} &\equiv& \frac{1}{2} s_{12} c_{12} s_{23} c_{23} s_{13} c_{13}^2  
\cos{\delta} \;.  
\eeqa 
In the limit $\theta_{13} \to 0$,   
$J_{\rm CP}, \, J'_{\rm CP} \to 0$.    
 From the point of view of the experiments also,  
the Jarlskog invariant $J_{\rm CP}$ is  
the quantity which appears in the probability expressions 
for CP violation in neutrino oscillation experiments, and is 
therefore directly measurable. $J'_{\rm CP}$ is  
needed in order to have complete information on $\delta$, since  
$J_{\rm CP}$ has no information on the sign of $\cos\delta$. 
We also choose to write the RG evolution for $\theta_{13}^2$ instead  
of $\theta_{13}$ as is traditionally done. This quantity turns out 
to have a smooth behaviour at $\theta_{13} = 0$. 
Moreover, since $\theta_{13} \geq 0$ by convention, the complete  
information about $\theta_{13}$ lies within $\theta_{13}^2$.  
The information about the Dirac phase will be present  
in $J_{\rm CP}$, $J'_{\rm CP}$. 

\begin{table}[t!] 
\centering 
  \begin{tabular}{ccccccc} 
\hline 
  &  
 & $32\pi^2 \, \Dot\theta_{12}$ & $\phantom{spc}$ 
 & $64\pi^2 \, \Dot{\overline{\theta_{13}^2}}$ & $\phantom{spc}$ 
 & $32\pi^2 \, \Dot\theta_{23}$ \\ 
\hline 
 
 $ P_{11}$ & $\phantom{spc}$  
 & $\mathcal{Q}^+_{12}\sin 2\theta_{12}$ & $\phantom{spc}$ 
 & $0$  & $\phantom{spc}$ 
 & $0$ \\ 
 
   $P_{22}$ & $\phantom{spc}$ 
 & $-\mathcal{Q}^+_{12}\sin 2\theta_{12}c_{23}^2$ & $\phantom{spc}$ 
 & $\left( \mathcal{\widetilde{A}}^+_{23} -  
\mathcal{\widetilde{A}}^+_{13} \right)  
\sin 2\theta_{12}\sin 2\theta_{23}$  & $\phantom{spc}$ 
 &  $\left( \mathcal{Q}^+_{23}c_{12}^2  
+ \mathcal{Q}^+_{13}s_{12}^2 \right) \sin 2\theta_{23}$  
\\ 
 
 $P_{33}$ & $\phantom{spc}$ 
 & $-\mathcal{Q}^+_{12} \sin 2\theta_{12}s_{23}^2$ & $\phantom{spc}$ 
 & $-\!\left( \mathcal{\widetilde{A}}^+_{23}  
- \mathcal{\widetilde{A}}^+_{13} \right)  
\sin 2\theta_{12} \sin 
 2\theta_{23}$ & $\phantom{spc}$ 
 & $-\!\left( \mathcal{Q}^+_{23}c_{12}^2 + \mathcal{Q}^+_{13}s_{12}^2 \right)  
\sin 2\theta_{23}$ \\ 
 
 $\re P_{21}$ & $\phantom{spc}$  
 & $2\mathcal{Q}^+_{12}\cos 2\theta_{12} c_{23}$ & $\phantom{spc}$  
 & $4 \left( \mathcal{\widetilde{A}}^+_{13}c_{12}^2  
+ \mathcal{\widetilde{A}}^+_{23}s_{12}^2 \right) 
 s_{23}$ & $\phantom{spc}$  
 & $\left( \mathcal{Q}^+_{23} - \mathcal{Q}^+_{13} \right)  
\sin 2\theta_{12}s_{23}$ \\ 
 
 $\re P_{31}$ & $\phantom{spc}$  
 & $-2\mathcal{Q}^+_{12} \cos 2\theta_{12} s_{23}$ & $\phantom{spc}$  
 & $4 \left( \mathcal{\widetilde{A}}^+_{13}c_{12}^2  
+ \mathcal{\widetilde{A}}^+_{23}s_{12}^2 
 \right)c_{23}$ & $\phantom{spc}$  
 & $\left( \mathcal{Q}^+_{23} - \mathcal{Q}^+_{13} \right)\sin 2\theta_{12} 
 c_{23}$ \\ 
  
 $\re P_{32}$ & $\phantom{spc}$ 
 & $\mathcal{Q}^+_{12} \sin 2\theta_{12}\sin 2\theta_{23}$ & $\phantom{spc}$ 
 & $2\! \left( \mathcal{\widetilde{A}}^+_{23}  
- \mathcal{\widetilde{A}}^+_{13} \right)  
\sin 2\theta_{12}\cos 2\theta_{23}$ & $\phantom{spc}$ 
 & $2\! \left( \mathcal{Q}^+_{23}c_{12}^2  
 + \mathcal{Q}^+_{13}s_{12}^2 \right) \cos 
 2\theta_{23}$ \\ 
  
 $\im P_{21}$ & $\phantom{spc}$  
 & $4 \mathcal{S}_{12}c_{23}$ & $\phantom{spc}$  
 & $4\left( \mathcal{\widetilde{B}}^-_{13}c_{12}^2  
+ \mathcal{\widetilde{B}}^-_{23}s_{12}^2 \right) s_{23}$& $\phantom{spc}$  
 & $2 \left( \mathcal{S}_{23} - \mathcal{S}_{13} \right) \sin 2\theta_{12}s_{23} $ 
 \\ 
  
 $\im P_{31}$ & $\phantom{spc}$ 
 & $-4\mathcal{S}_{12} s_{23}$ & $\phantom{spc}$ 
 & $4 \left( \mathcal{\widetilde{B}}^-_{13}c_{12}^2  
+ \mathcal{\widetilde{B}}^-_{23}s_{12}^2 
 \right)c_{23}$ & $\phantom{spc}$ 
 & $2 \left( \mathcal{S}_{23} - \mathcal{S}_{13} \right)  
\sin 2\theta_{12}c_{23} $ 
 \\ 
  
 $\im P_{32}$ & $\phantom{spc}$  
 & $0$ & $\phantom{spc}$ 
 &  $2 \left( \mathcal{\widetilde{B}}^-_{23}  
- \mathcal{\widetilde{B}}^-_{13} \right)  
\sin 2\theta_{12}$ & $\phantom{spc}$ 
 & $4 \left( \mathcal{S}_{23}c_{12}^2  
+ \mathcal{S}_{13}s_{12}^2 \right)$\\\hline 
\end{tabular} 
\caption{Coefficients of $P_{fg}$ in the RG evolution  
equations of the mixing angles 
$\theta_{12}$, $\theta_{13}^2$ and $\theta_{23}$,   
in the limit $\theta_{13}\to 0$.  
\label{tab-theta12-theta23} } 
\end{table} 
 
The expressions for the running of masses and Majorana phases are
the same as the ones obtained in \cite{seesaw1rg} for the Type-I
seesaw mechanism. (See Tables 5, 6, and 14 therein. Note that 
$\phi_i$ in our paper corresponds to $\varphi_i/2$ in 
\cite{seesaw1rg}.)
The running of masses and  
the Majorana phases does not depend on the Dirac phase to the  
lowest order in $\theta_{13}$. Hence the RG evolution equations  
do not change with the change in basis  
${\cal P}_\delta \rightarrow {\cal P}_J$. 
Running of the two large mixing angles  
$\theta_{12}$ and $\theta_{23}$, as given in  
Table~\ref{tab-theta12-theta23}, is also the same as that in the  
${\cal P}_\delta$ basis since the quantities  
$\mathcal{S}_{ij}$ and $\mathcal{Q}_{ij}^\pm$, defined as 
\beqa 
\mathcal{Q}^\pm_{13} = \frac{|m_3 \pm m_1 e^{2 i \phi_1}|^2}{\Delta 
  m^2_\mathrm{atm}\left(1+\zeta\right)} \; , &\quad& 
\mathcal{Q}^\pm_{23} = \frac{|m_3 \pm m_2 e^{2 i \phi_2}|^2}{\Delta 
  m^2_\mathrm{atm}}\; , \quad 
\mathcal{Q}^\pm_{12} = \frac{|m_2 e^{ 2 i \phi_2} \pm m_1 
  e^{2 i \phi_1}|^2}{\Delta m^2_\mathrm{sol}} \; , \label{Qij} \\ 
 \mathcal{S}_{13} = \frac{m_1 m_3 \sin{2 \phi_1}}{\Delta m^2_\mathrm{atm} 
\left( 1+\zeta \right)} \; , &\quad& 
\mathcal{S}_{23} =  
\frac{m_2 m_3 \sin{ 2 \phi_2} }{\Delta m^2_\mathrm{atm}} \; , \quad 
\mathcal{S}_{12} =  
\frac{m_1 m_2 \sin{( 2 \phi_1- 2 \phi_2)} }{\Delta  m^2_\mathrm{sol}} \; , 
\label{Sij} 
\eeqa 
depend on the mass eigenvalues and Majorana phases only. 
However the running of $\theta_{13}^2$, as seen from the  
Table~\ref{tab-theta12-theta23},  
depends on the quantities $\mathcal{\widetilde{A}}^\pm_{ij}$, 
$\mathcal{\widetilde{B}}^\pm_{ij}$ defined as
\beqa 
\mathcal{\widetilde{A}}^\pm_{13} =  
\frac{ 4 \left(m_1^2+m_3^2\right) J'_{\rm CP} \pm 8 m_1m_3  
( J'_{\rm CP} \cos{2 \phi_1} +  
 J_{\rm CP} \sin{2 \phi_1})}{  a \Delta m_\mathrm{atm}^2 
\left(1+\zeta\right)} \; , \label{Atilde13} \\ 
\mathcal{\widetilde{A}}^\pm_{23} =  
\frac{ 4 \left(m_2^2+m_3^2\right) J'_{\rm CP} \pm 8 m_2m_3  
( J'_{\rm CP} \cos{2 \phi_2} +  
 J_{\rm CP} \sin{2 \phi_2})}{ a \Delta m_\mathrm{atm}^2} \; ,  
\label{Atilde23} 
\eeqa 
\beqa 
\mathcal{\widetilde{B}}^\pm_{13} =  
\frac{ 4 \left(m_1^2+m_3^2\right) J_{\rm CP} \pm 8 m_1m_3  
( J_{\rm CP} \cos{2 \phi_1} -  
 J'_{\rm CP} \sin{2 \phi_1})}{  a \Delta m_\mathrm{atm}^2 
\left(1+\zeta\right)} \; , \label{Btilde13}\\ 
\mathcal{\widetilde{B}}^\pm_{23} =  
\frac{ 4 \left(m_2^2+m_3^2\right) J_{\rm CP}  
\pm 8 m_2m_3 ( J_{\rm CP} \cos{2 \phi_2} -  
 J'_{\rm CP} \sin{2 \phi_2})}{  a \Delta m_\mathrm{atm}^2} \; , 
\label{Btilde23} 
\eeqa 
where $a \equiv s_{12} c_{12} s_{23} c_{23}$.
Clearly these quantities depend on $J_{\rm CP}$, $J'_{\rm CP}$ 
in addition to the masses and Majorana phases. The coefficients for 
the RG evolution of $J_{CP}$ and $J'_{CP}$  
are presented in Table~\ref{tab-J}, where the quantities  
$\calG_{0,c,s}^{\pm}$ are given by 
\begin{table}[t!] 
\centering 
\begin{tabular}{ccccccc}
\hline 
& $\phantom{spc}$
& $ 64\pi^2 \,  \Dot J_{\rm CP} / a$ & $\phantom{spc}$ 
& $ 64\pi^2 \,  \Dot J'_{\rm CP} / a $  
& $\phantom{spc}$ &
$32 \pi^2 (\dot\phi_1 - \dot\phi_2)$ 
\\\hline
  
$P_{11}$ & $\phantom{spc}$  
& $0$  & $\phantom{spc}$ 
& $0$  
& $\phantom{spc}$ &
$-4\mathcal{S}_{12}\cos 2\theta_{12}$ 
\\ 
 
$P_{22}$ & $\phantom{spc}$ 
& $-4 a \calG_s^+$ & $\phantom{spc}$ 
& $2 a ( \calG_0^- - 2 \calG_c^-)$ 
& $\phantom{spc}$ &
$4\mathcal{S}_{12}c_{23}^2 \cos 2\theta_{12}$ 
\\ 
  
$P_{33}$ & $\phantom{spc}$ 
& $4 a \calG_s^-$ & $\phantom{spc}$ 
& $ - 2 a ( \calG_0^- -  2 \calG_c^-)$  
& $\phantom{spc}$ &
 $4 \mathcal{S}_{12}s_{23}^2\cos 2\theta_{12}$
\\ 
 
$\re P_{21}$ & $\phantom{spc}$ 
& $4 s_{23} \calG_s^+$ & $\phantom{spc}$ 
& $2 s_{23} (\calG_0^+ + 2 \calG_c^+)$  
& $\phantom{spc}$ &
 $-8\mathcal{S}_{12}c_{23}\cos 2\theta_{12}\cot 2\theta_{12}$ 

\\ 
 
$\re P_{31}$ & $\phantom{spc}$ 
& $4 c_{23} \calG_s^+$ & $\phantom{spc}$ 
& $ 2 c_{23} (\calG_0^+ + 2 \calG_c^+)$  
& $\phantom{spc}$ &
 $8\mathcal{S}_{12}s_{23}\cos 2\theta_{12} \cot 2\theta_{12}$ 
\\ 
 
$\re P_{32}$ & $\phantom{spc}$ 
& $ - 2 \sin{2 \theta_{12}} \cos{2 \theta_{23}} \, \calG_s^- \, $  
& $\phantom{spc}$ 
& $ \sin{2 \theta_{12}} \cos{2 \theta_{23}} (\calG_0^- - 2 \calG_c^-)$  
& $\phantom{spc}$ &
 $-4\mathcal{S}_{12} \cos 2\theta_{12}\sin 2\theta_{23}$ 
\\ 
 
$\im P_{21}$ & $\phantom{spc}$ 
& $ 2 s_{23} (\calG_0^+ - 2 \calG_c^+)$ & $\phantom{spc}$ 
& $ 4 s_{23} \calG_s^+ $  
& $\phantom{spc}$ &
 $-4\mathcal{Q}^-_{12}c_{23} \cot 2\theta_{12}$ 
\\ 
 
$\im P_{31}$ & $\phantom{spc}$ 
&  $ 2 c_{23} (\calG_0^+ - 2 \calG_c^+)$ & $\phantom{spc}$ 
& $ 4 c_{23} \calG_s^+ $  
& $\phantom{spc}$ &
 $4\mathcal{Q}^-_{12} s_{23}\cot 2\theta_{12}$ 
\\ 
 
$\im P_{32}$ & $\phantom{spc}$ 
& $  \sin{2 \theta_{12}} (\calG_0^- + 2 \calG_c^-)$ & $\phantom{spc}$ 
& $ - 2 \sin{2 \theta_{12}} \calG_s^- $  
& $\phantom{spc}$ & 0
\\\hline 
\end{tabular} 
\caption{
Coefficients of $P_{fg}$ in the RG evolution equations of  
the Jarlskog invariant $J_{\rm CP}$, the quantity $J'_{\rm CP} \equiv
J_{\rm CP} \cot \delta$, and the Majorana phase difference 
$(\phi_1-\phi_2)$, in the limit $\theta_{13}\to 0$. The convention
used here is $a \equiv s_{12} c_{12} s_{23} c_{23}$, and 
$J_{\rm CP} \equiv (a/2) s_{13} c_{13}^2 \sin \delta$.  
\label{tab-J} } 
\end{table} 
\beqa 
\calG_0^\pm &=& \frac{m_2^2 + m_3^2}{\dmsq_\mathrm{atm}} \pm   
\frac{m_1^2 + m_3^2}{\dmsq_\mathrm{atm} (1+\zeta)} \; , \label{G0}\\ 
\calG_s^\pm &=& \frac{m_1 m_3 \sin{2 \phi_1}}{\dmsq_\mathrm{atm} (1+\zeta)} \pm   
\frac{m_2 m_3 \sin{2 \phi_2}}{\dmsq_\mathrm{atm}} \; , \label{Gs}\\ 
\calG_c^\pm &=& \frac{m_1 m_3 \cos{2 \phi_1}}{\dmsq_\mathrm{atm} (1+\zeta)} \pm   
\frac{m_2 m_3 \cos{2 \phi_2}}{\dmsq_\mathrm{atm}} \;\label{Gc} . 
\eeqa 
Thus all the the quantities appearing in the  
evolution equations (\ref{Atilde13}) -- (\ref{Gc})  
have finite well-defined limits for $\theta_{13} \to 0$  
in the ${\cal P}_J$ basis. 
 
Even if one starts with diagonal $Y_e$ (i.e. $Y_e = {\rm diag}(y_e,y_\mu,y_\tau)$) 
at the high scale,  
non-zero off-diagonal elements of $Y_e$ will be generated  
through Eqs.~(\ref{beta-Ye-1}) -- (\ref{F}) since  
$\Ysigman^{\;\dagger} \Ysigman$ is not diagonal. These  
off-diagonal elements will give additional contributions to the  
running of masses and mixing above and between the thresholds 
through $F$ and $\alpha_e$.  
Since $\alpha_e$ is flavor diagonal, it will contribute to  
the running of $y_e$, $y_\mu$ and $y_\tau$,  
while off-diagonal conponents of $F$ will  
contribute additional terms in the $\beta$-functions of angles  
and phases, as tabulated in Table~\ref{tab-F}. 
These contributions will just get added  
to the $P_{fg}$ contribution for the evolution of the quantities  
in Tables~\ref{tab-theta12-theta23}, \ref{tab-J}, \ref{tab-F}. 
Note that the $F_{fg}$ coefficients are $\lesssim {\cal O}(1)$,  
whereas the $P_{fg}$ coefficients are $\gtrsim {\cal O}(m_i^2/\dma)$.  
Since the running is significant only when $m_i^2 \gg \dma$, in  
almost all the region of interest $P_{fg}$ contributions  
dominate over the $F_{fg}$ contribution. 
 
Note that the analytical expressions obtained in Eq.~(\ref{Qij}) onwards, 
and those given in the tables,  
are valid only in the two extreme regions $\mu>M_3$ and $\mu < M_1$. 
For the intermediate energy scales, 
${\mathbbm m}_\nu$ will receive contributions from both $\accentset{(n)}{\kappa}$  
and $\accentset{(n)}{Q}$. In  
the SM these two quantities have non-identical evolutions, as seen from  
Eqs.~(\ref{beta-kappa}) and (\ref{beta-Q}), and therefore the net  
evolution of $Y_e$ and ${\mathbbm m}_\nu$ is rather complicated. 
We perform it numerically in the next section. 
 
\begin{table}[t!] 
\centering 
  \begin{tabular}{ccccccccc} 
\hline 
& $ \; 16 \pi^2 \, \Dot \theta_{12}^{U_e} \; $  
& $ \; 16 \pi^2 \, {\Dot {\theta^2}_{13}}^{U_e} \; $ 
& $ \; 16 \pi^2 \, \Dot \theta_{23}^{U_e} \; $  
& $ \; 16 \pi^2 \, \Dot J_{\rm CP}^{U_e} \; $ 
& $ \; 16 \pi^2 \, \Dot J_{\rm CP}^{' \; U_e} \; $ 
& $ \; 16 \pi^2 \, \Dot \phi_1^{U_e} \; $  
&  \; $16 \pi^2 \, \Dot \phi_2^{U_e} \; $ 
\\\hline 
 
$F_{11}$ 
& $0$ 
& $0$ 
& $0$ 
& $0$ 
& $0$ 
& $0$ 
& $0$ 
\\ 
 
$F_{22}$ 
& $0$ 
& $0$ 
& $0$ 
& $0$ 
& $0$ 
& $0$ 
& $0$ 
\\ 
 
$F_{33}$ 
& $0$ 
& $0$ 
& $0$ 
& $0$ 
& $0$ 
& $0$ 
& $0$ 
\\ 
 
$\re F_{21}$ 
& $-c_{23}$ 
& $ - 4 s_{23} J'_{\rm CP} / a $ 
& $0$ 
& $0$ 
& $- s_{23} a /2$ 
& $0$ 
& $0$ 
\\ 
 
$\re F_{31}$ 
& $s_{23}$ 
& $- 4 c_{23} J'_{\rm CP} / a $ 
& $0$ 
& $0$ 
& $- c_{23} a / 2$ 
& $0$ 
& $0$  
\\ 
 
$\re F_{32}$ 
& $0$ 
& $0$ 
& $1$ 
& $0$ 
& $0$ 
& $0$ 
& $0$  
\\ 
 
$\im F_{21}$ 
& $0$ 
& $ - 4 s_{23} J_{\rm CP} / a $ 
& $0$ 
& $ - s_{23} a/2 \; $ 
& $0$ 
& $ \; c_{23} c_{12} / s_{12} \; \; $ 
& $ \; -c_{23} s_{12} / c_{12} \; \; $ 
\\ 
 
$\im F_{31}$ 
& $0$ 
& $- 4 c_{23} J_{\rm CP} / a$ 
& $0$ 
& $- c_{23} a / 2$ 
& $0$ 
& $ \; -s_{23} c_{12} / s_{12} \; \; $ 
& $ \; s_{23} s_{12} / c_{12} \; \; $ 
\\ 
 
$\im F_{32}$ 
& $0$ 
& $0$ 
& $0$ 
& $0$ 
& $0$ 
&$ \; -1/(c_{23} s_{23}) \; \; $ 
&$ \; -1/(c_{23} s_{23}) \; \; $ 
\\\hline 
 
\end{tabular} 
\caption{Coefficients of $F_{fg}$ in the RG evolution equations of all 
the angles ($\theta_{12}$, $\theta_{13}^2$, $\theta_{23}$),  
$J_{\rm CP}, J'_{\rm CP}$ and the Majorana phases $\phi_i$  
in the limit $\theta_{13}\to 0$.  
The convention
used here is $a \equiv s_{12} c_{12} s_{23} c_{23}$, and 
$J_{\rm CP} \equiv (a/2) s_{13} c_{13}^2 \sin \delta$.  
We neglect $y_e$ and $y_\mu$ compared to $y_\tau$, 
and take vanishing flavor phases.} 
\label{tab-F} 
\end{table} 
%
 
\section{Illustrative examples of RG running of masses and mixing} 
\label{numerical} 
 
In this section we numerically calculate the RG evolution 
of the masses and mixing parameters   
within the Type-III seesaw model 
including the impact of running between the thresholds.  
This analysis is done by imposing suitable matching conditions  
(\ref{matching}) at the thresholds. For illustration, we  
start at $\mu_0 = 10^{16}$ GeV and choose the basis  
in which $Y_e$ is diagonal, so that $U_{\rm PMNS} = U_\nu$. 
We further choose $U_\nu$ at this high scale to be the bimaximal  
mixing matrix $U_{\nu,{\rm bimax}}$ \cite{vissani-deg,bimax}, 
i.e. $\theta_{12} = \theta_{23} = \pi/4$ and $\theta_{13} =0$. 
This scenario is clearly inconsistent with the current data in 
the absence of RG evolution. 
We shall check if the radiative corrections to the masses and 
mixing angles can make it consistent with the data at the low scale.

If the low energy theory in the complete energy range $\mu < \mu_0$ 
is the SM, then $\theta_{12}$ decreases as the energy scale decreases, 
however the running is not sufficient to achieve compatibility with 
the low energy data. 
If the low energy theory is the MSSM, then $\theta_{12}$ increases 
with decreasing energy scale \cite{rgqlc}, so that compatibility with  
the data is not possible. 
However, it has been shown in \cite{antusch-LMA,Miura:2003if,Shindou:2004tv} 
in the context of Type-I seesaw mechanism, that the  
inclusion of threshold effects can make the mixing angle $\theta_{12}$  
decrease substantially as we go to lower energy scale and can give the  
correct values consistent with the Large Mixing Angle (LMA) solution.   
In this section we study the evolution from bi-maximal  
mixing at high scale in the context of Type-III seesaw scenario, 
including the seesaw threshold effects.

\begin{figure}[t!] 
\parbox{3in}{ 
\epsfig{file=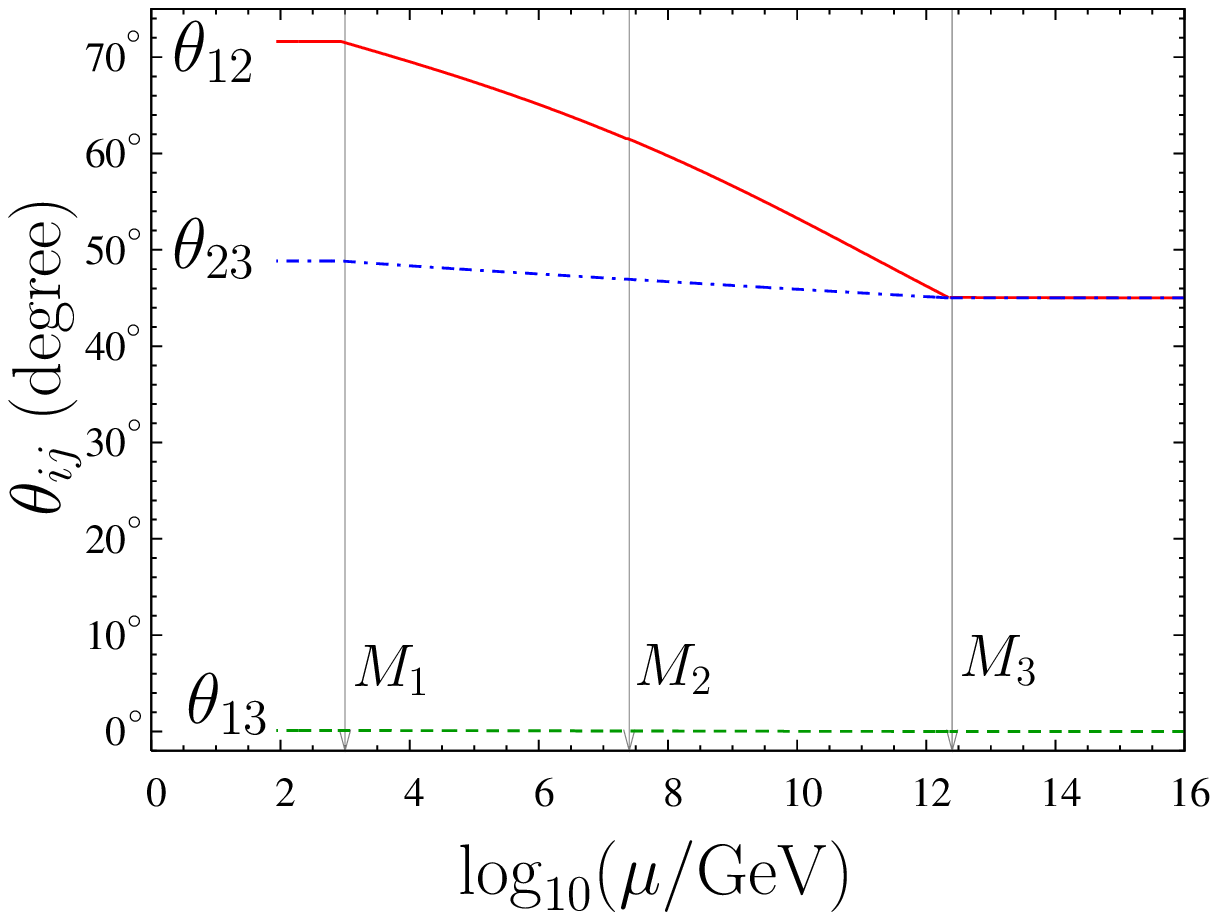,width=3in,height=2.4in} 
} 
\parbox{3in}{ 
\epsfig{file=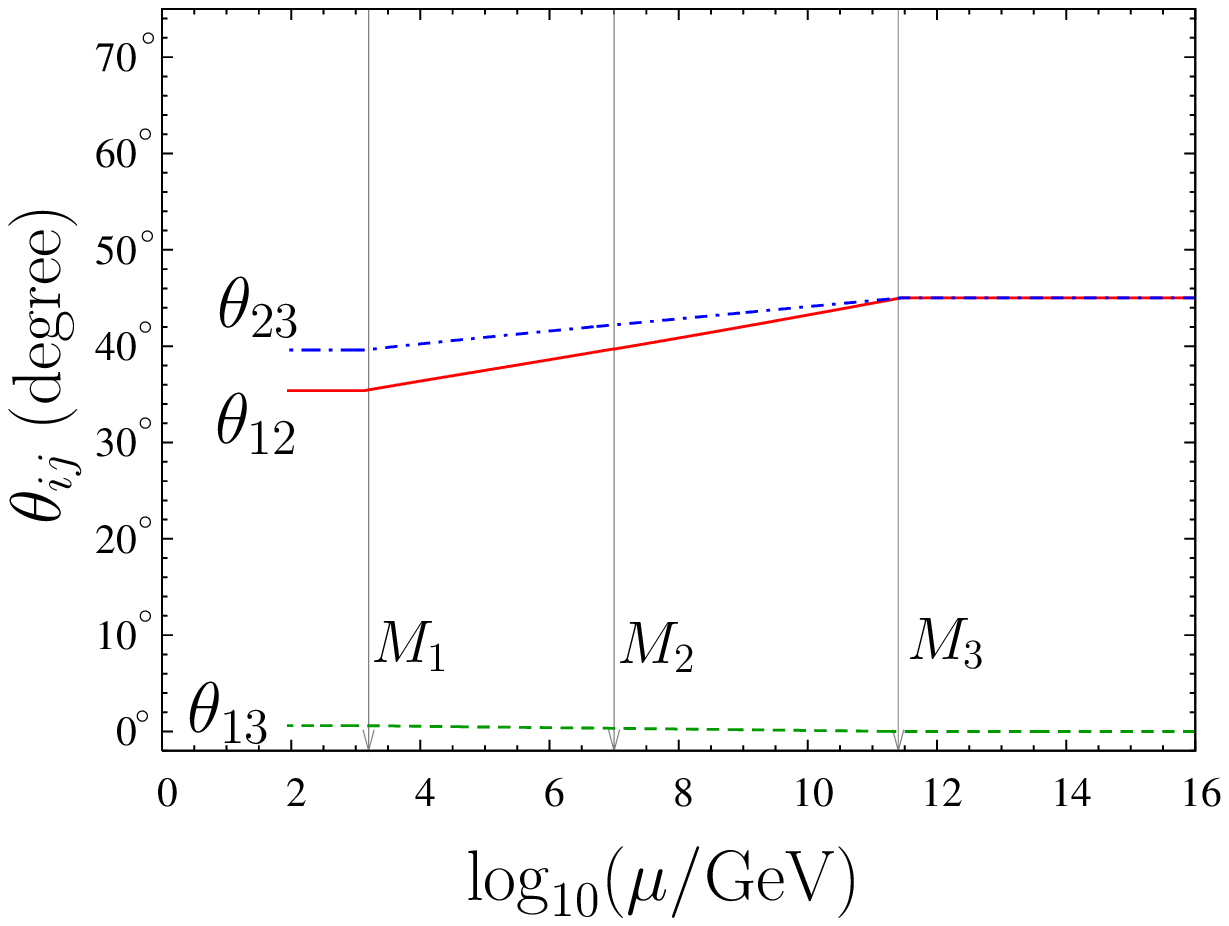,width=3in,height=2.4in} 
} 
\parbox{3in}{ 
\epsfig{file=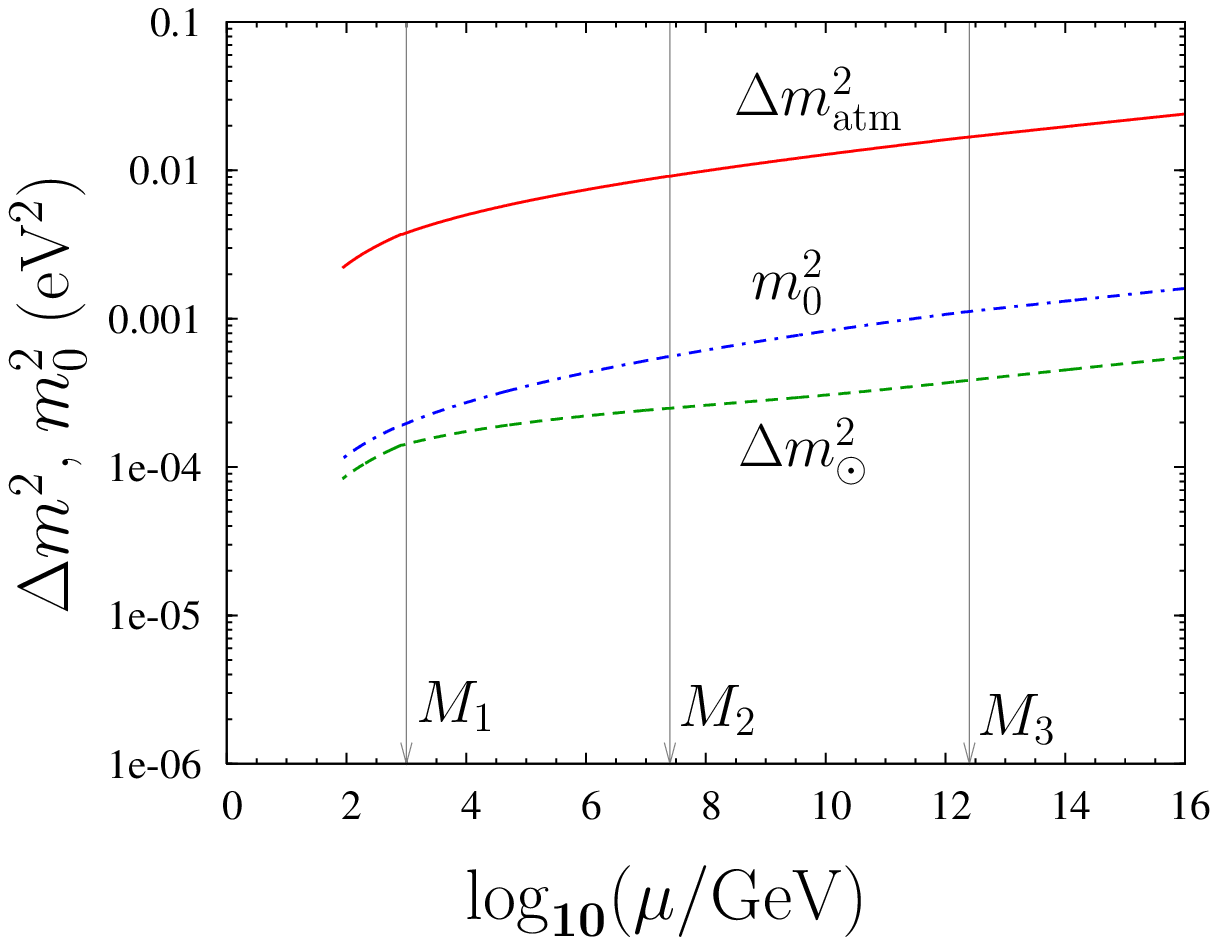,width=3in} 
} 
\parbox{3in}{ 
\epsfig{file=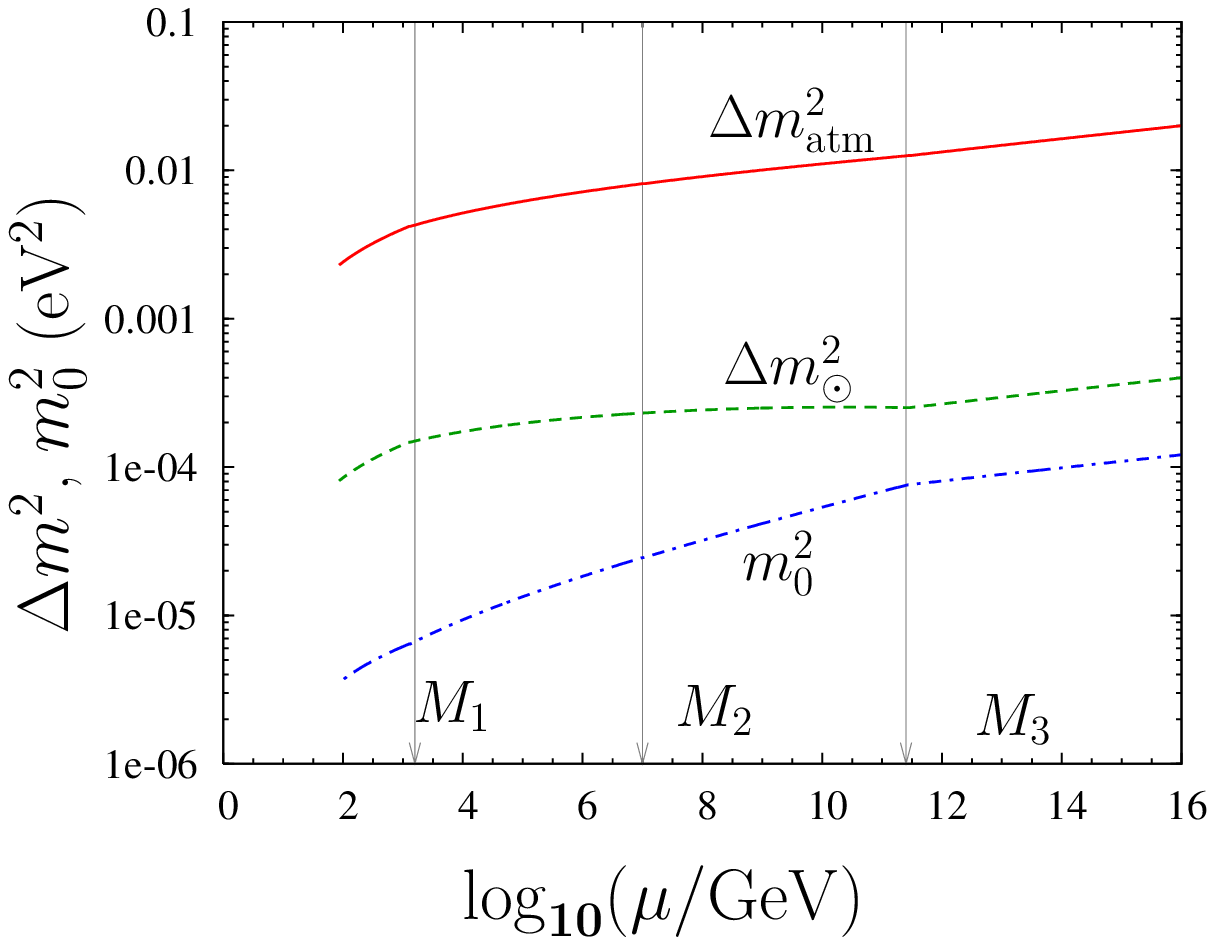,width=3in} 
} 
\caption{RG evolution of mixing angles and mass squared differences, 
starting from bimaximal mixing at $\mu_0 = 10^{16}$ GeV, for normal 
mass ordering and hierarchical neutrino masses. 
The left panels represent the scenario where the  
Majorana phases vanish at $\mu_0$. 
The right panel shows a representative case of nonzero Majorana phases
($\phi_1=89.0^\circ,\phi_2=0.4^\circ$) at $\mu_0$. 
The values of parameters at the high scale have been chosen such that 
the $\dmsq$'s and $g_2$ at the low scale are reproduced. 
\label{nophase-fig} 
} 
\end{figure} 

We write the neutrino mass matrix as 
\beqa 
{\mathbbm m}_\nu = U^{*}_{\nu,\mathrm bimax}  
{\rm diag}(m_1,m_2,m_3) U^\dagger_{\nu,\mathrm bimax} \; , 
\eeqa 
with $\delta_e = \delta_\mu = \delta_\tau=0$ at the high scale. 
Given the masses of the three fermion triplets and the  
light neutrino masses at the high scale, one can determine a  
$Y_\Sigma$ at the high scale\footnote{ 
The solution for $Y_\Sigma$ need not be unique, however any one of the  
solutions would suffice for the illustration. For practicality, we first  
choose an ``trial'' $Y_\Sigma$, calculate the corresponding  
${\mathbbm M}_\Sigma$ from the seesaw relation, and then apply the  
basis transformation that makes ${\mathbbm M}_\Sigma$ diagonal and  
takes the ``trial'' $Y_\Sigma$ to its final form.} 
that satisfies the seesaw relation  
${\mathbbm m}_\nu = -(v^2/2) Y_\Sigma^T {\mathbbm M}_\Sigma^{-1} Y_\Sigma$.  
We then evolve the parameters using the analysis of  
Sec~\ref{running}. 
 
Among the neutrino mixing angles, $\theta_{12}$ is expected to be  
the most sensitive to RG effects. Table~\ref{tab-theta12-theta23} 
shows that $\dot{\theta}_{12}$ is proportional to $\mathcal{Q}^+_{12}$ 
and $\mathcal{S}_{12}$, which are in turn proportional 
to $\left( m_i^2/\dmsq_{\rm sol} \right)$ 
as can be seen from Eqs.~(\ref{Qij}) and (\ref{Sij}). 
For the other angles $\theta_{ij}$, the corresponding quantities 
 $\mathcal{Q}^+_{ij}$ and $\mathcal{S}_{ij}$ are proportional to  
$\left( m_i^2/\dmsq_{\rm atm} \right)$, 
so the evolution of these angles is smaller.
The direction of $\theta_{12}$ evolution depends on the details of 
the Yukawa coupling matrix and masses of the heavy fermions. 
  
Since the values of Majorana phases at the low scale are  
completely unknown, we first consider the case where  
$\phi_1 = \phi_2 = 0$. 
In this case the CP violation will remain zero   
at all energy scales. The left panels of Fig.~\ref{nophase-fig} show 
the running of mixing angles and mass squared differences for the normal   
mass ordering in this scenario.  
It is observed that $\theta_{12}$ in the intermediate energy region 
changes more rapidly than in the extreme regions, however this 
change is in the opposite direction to what is required. 
As a result, bimaximal mixing at the high scale is not compatible with 
the low energy data in our model when the Majorana phases vanish. 
With nonzero Majorana phases, however, it is possible to achieve 
compatibility with the low scale data, as can be seen from the  
right panels of the figure. 
 
The lower panels of Fig.~\ref{nophase-fig} show the evolution of $m_0$, 
the lowest mass scale, and the two mass squared differences. 
As can be observed, the running of masses is quite substantial 
in Type-III seesaw, as compared to the SM, the MSSM \cite{antusch-CP},  
or the Type-I seesaw \cite{antusch-LMA}. 
Most of this running occurs in the intermediate energy range 
$M_1 < \mu < M_3$, where threshold effects play a crucial role 
in enhancing the running. 
Note that the values of $m_0$ required to cause substantial 
running of mixing angles is quite small: in the case of 
vanishing (non vanishing)  Majorana phases, we have taken 
$m_0 = 0.04 (0.01)$ eV at $\mu=\mu_0$. 
Thus, even at extremely small $m_0$, substantial running 
of neutrino parameters can be present in the Type-III seesaw.  

The example of the bimaximal mixing discussed above was just for 
illustration. However, it brings out certain salient features 
of the RG running in Type-III seesaw scenario. 
The running of neutrino masses can be quite substantial here 
in the intermediate energy range.  
Moreover, threshold effects can enhance the extent of running of  
mixing angles, as well as the direction of the evolution, 
similar to the Type-I seesaw scenario \cite{antusch-LMA}. 
Majorana phases are also seen to play an important role 
in determining the extent and the direction of RG running 
of neutrino mixing parameters.

\begin{figure}[t!] 
\parbox{3.2in}{ 
\epsfig{file=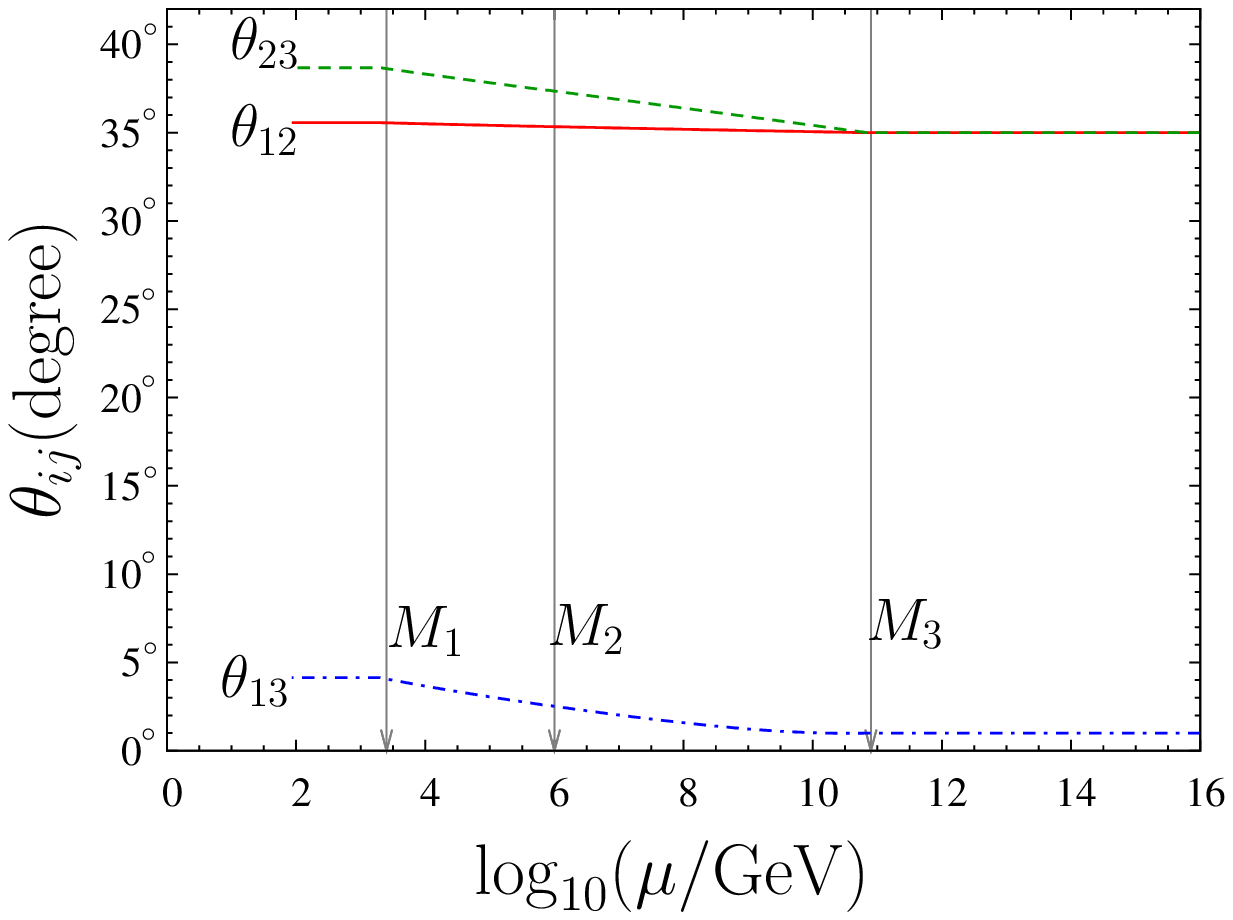,width=3.2in,height=2.3in} 
} 
\parbox{3in}{ 
\epsfig{file=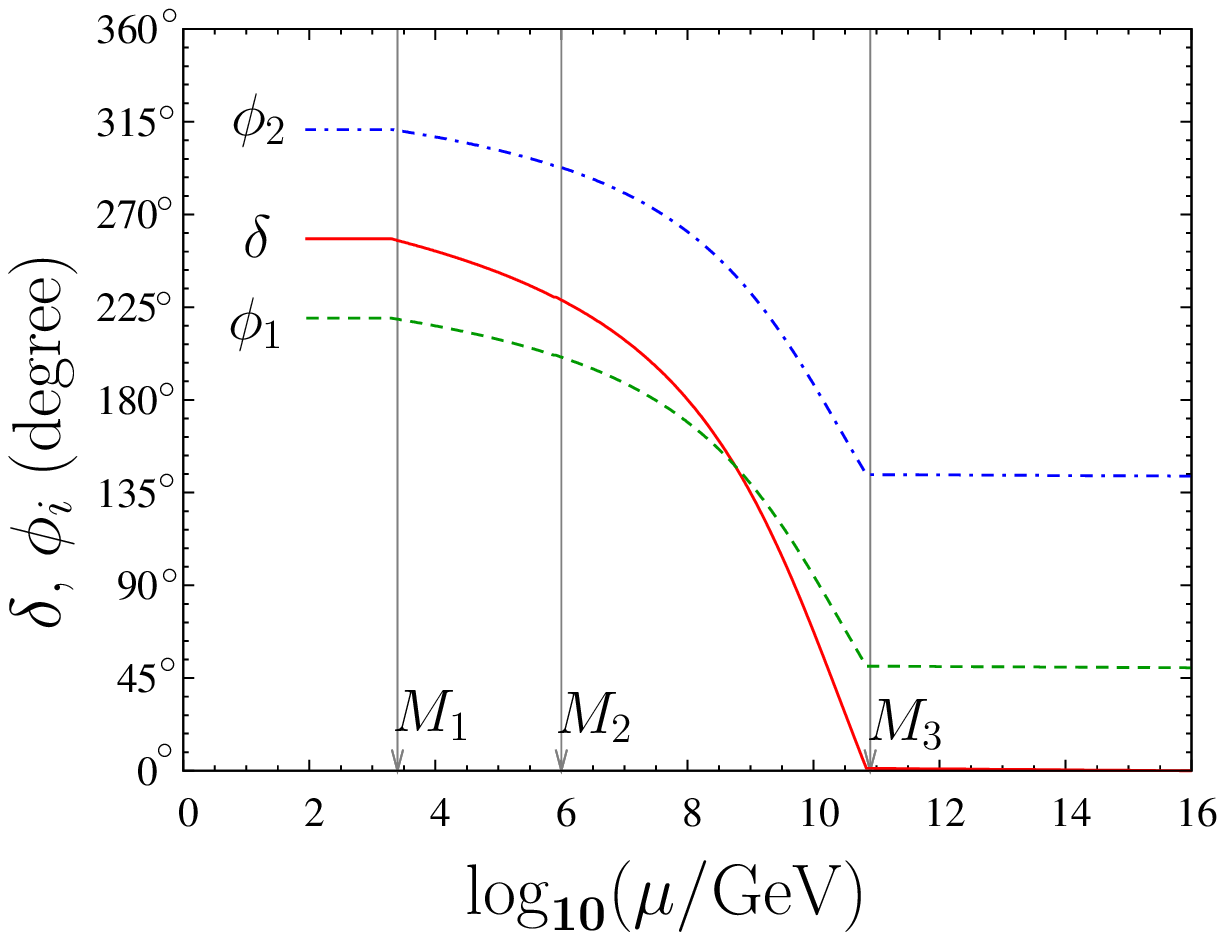,width=3in} 
} 
\parbox{3.2in}{ 
\epsfig{file=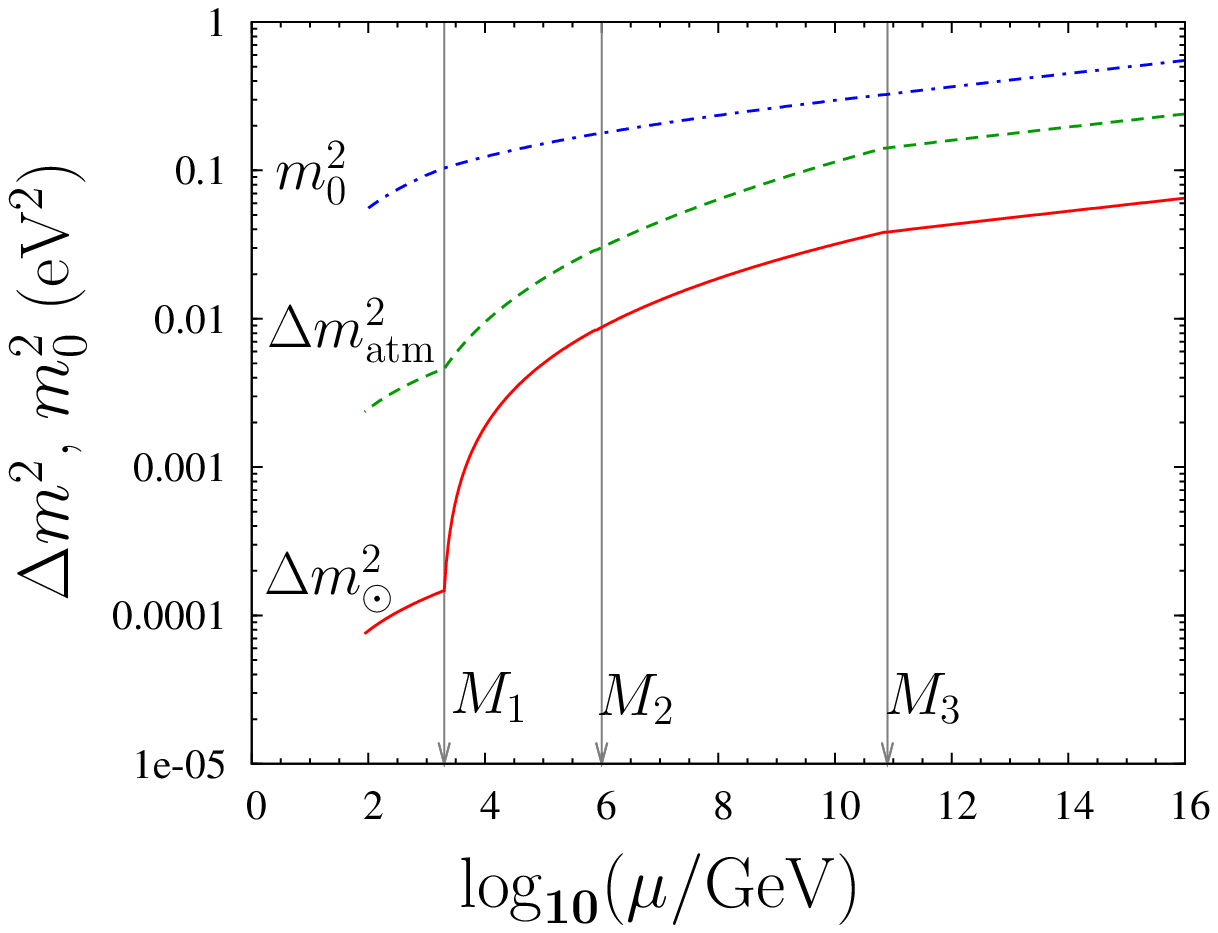,width=3.2in,height=2.3in} 
} 
\parbox{3in}{ 
\epsfig{file=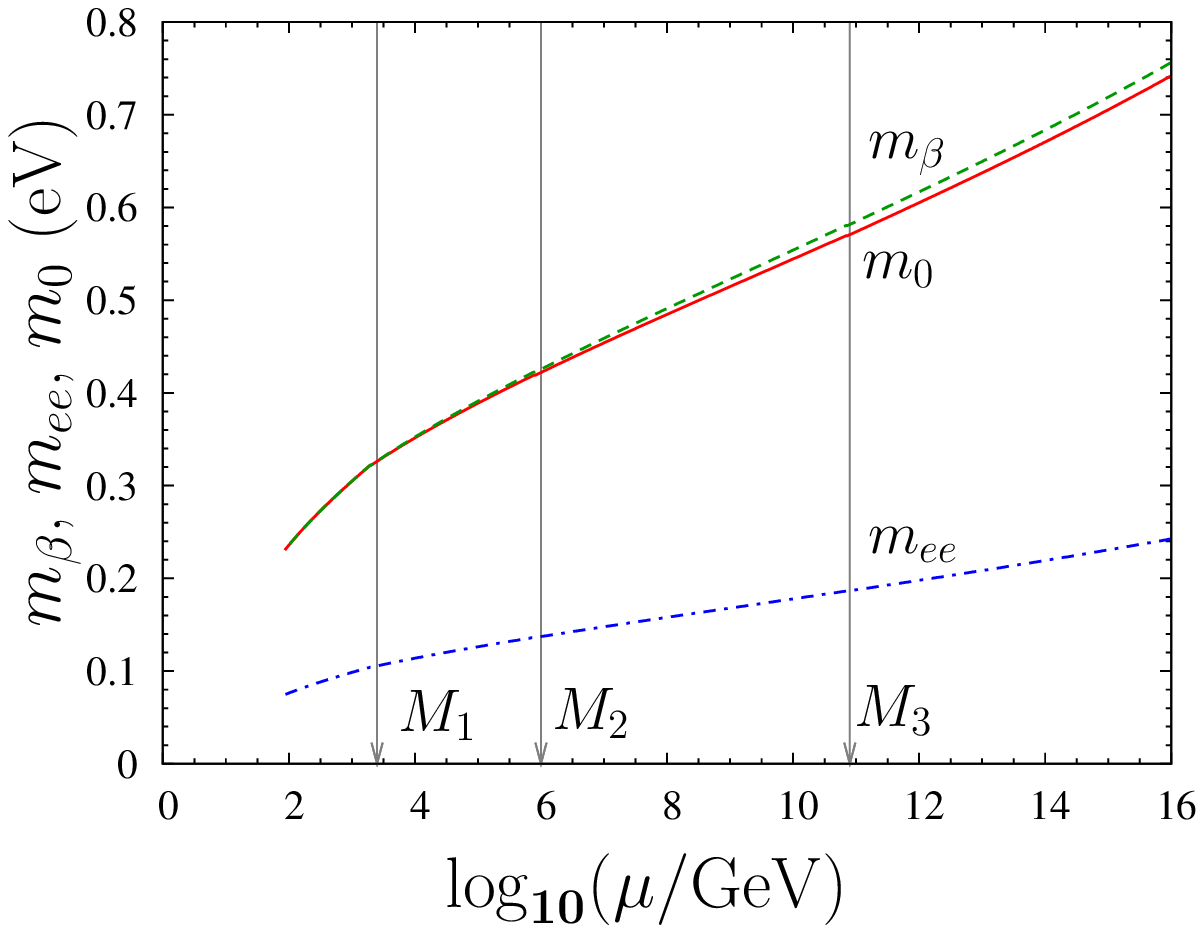,width=3in} 
} 
\caption{RG evolution of mixing angles, mass squared differences,
and CP violating phases, for quasi-degenerate neutrino masses and 
normal mass ordering.
The values of parameters at the high scale have been chosen such that 
the $\dmsq$'s and $g_2$ at the low scale are reproduced. 
Note that for the Majorana phases $\phi_i$, the regions $(0^\circ-180^\circ)$
and $(180^\circ-360^\circ)$ should be identified with each other. 
\label{degenerate-evol} 
} 
\end{figure} 

In Fig.~\ref{degenerate-evol}, we illustrate the RG evolution of 
parameters when the neutrino masses are quasi-degenerate.
We have taken the parameter values at the high scale to achieve
compatibility with the low scale data, without imposing any
special symmetry.
However in order to bring out certain salient features of the 
RG evolution that are independent of the threshold effects.
we have chosen a small $\theta_{13}$ value, 
$|\phi_1-\phi_2| \approx \pi/2$, and $Y_\Sigma^\dagger Y_\Sigma$
to be almost diagonal in the charged lepton basis, with
hierarchical eigenvalues.
These conditions ensure that $P_{21}$ and $P_{31}$ are small,
and $\mathcal{S}_{12}$ vanishes, so that from Table~\ref{tab-J},
the evolution of $(\phi_1 - \phi_2)$ is extremely small.
Thus $|\phi_1 - \phi_2|$ is expected to stay close to $\pi/2$ 
even after evolution, which is verified by the figure.
Moreover, combined with $m_1 \approx m_2$, the choice 
$|\phi_1 - \phi_2| \approx \pi/2$ makes ${\cal Q}_{12}^+$
extremely small, thus restricting the $\theta_{12}$ evolution.

It is observed that the running of $\theta_{23}$ is now large,
owing to $m_0^2/\dma \sim 1$. This makes it possible to mimic
maximal mixing accidentally, even if the mixing generated at
the high scale is arbitrary. 
The value of $\theta_{13}$ also quadruples from its high 
scale value.
The Dirac phase, which was chosen to vanish at $\mu_0$, is generated 
by the RG evolution.
The running of Dirac as well as Majorana phases is substantial
between the thresholds.

The right hand bottom panel of Fig.~\ref{degenerate-evol} shows the
evolution of $m_\beta \equiv \sqrt{\sum_i |U_{ei}|^2 m_i^2}$, 
the effective neutrino mass measured in the Tritium beta decay experiments
\cite{katrin},
as well as $m_{ee} \equiv | \sum_i U_{ei}^2 m_i |$, 
the effective neutrino Majorana mass in the neutrinoless double 
beta decay.
Note that since $\theta_{13}$ is small, $m_1 \approx m_2 \approx m_0$,
and since $|\phi_1 - \phi_2| \approx \pi/2$ in addition, we have 
$m_{ee} \approx m_0 \cos 2\theta_{12}$.  
Also in the quasi-degenerate case, the sum of neutrino masses that
is resticted by cosmology is $\sum m_i \approx 3 m_0$. 
The large running of these masses suggests that, even if the 
beta decay experiments were to bound $m_\beta$ 
to $\leq 0.3$ eV, or the
neutrinoless double beta decay experiments were to bound
$m_{ee}$ to $\leq 0.1$ eV,
or the cosmological observations were to restrict 
$m_0$ at the low scale to $\leq 0.3$ eV, 
the value of $m_0$ generated at the high scale 
can still be substantially larger.

It is thus observed that in Type-III seesaw, the RG evolution of
masses, angles as well as CP violating phases can be significant
between the thresholds even at low $m_0$ values.
The reason behind this, as well as the exact dependence of the
evolution on the mass thresholds and Majorana phases, needs to be 
studied in further detail for a better understanding 
of the allowed neutrino parameter space at high energies.

\section{Summary and Conclusions} 
\label{concl} 
 
In this paper we have studied the RG evolution  
of neutrino masses and mixing angles in the context of  
Type-III seesaw mechanism  mediated by heavy fermions $\Sigma$  
transforming as triplets under SU(2)$_L$.  
Tree level exchange of such particles gives rise to an effective  
operator $\kappa_5 l_L l_L \phi \phi$ below their lowest mass threshold.  
If one or more such triplets are present in the model, 
they affect the RG evolution of wavefunctions, masses and couplings. 
We compute these extra contributions using dimensional regularization  
and minimal subtraction scheme.  
We calculate the beta functions for the Yukawa couplings  
$Y_e$, $Y_u$, $Y_d$ and $Y_\Sigma$, the SU(2)$_L$ gauge coupling $g_2$, 
the Higgs self-coupling $\lambda$, the heavy fermion triplet mass matrix  
${\mathbbm M}_\Sigma$, and finally the light neutrino mass matrix  
${\mathbbm m}_\nu$. 
We do our calculation in the $R_\xi$ gauge and show the gauge invariance  
explicitly  by demonstrating that the terms containing $\xi$ are not  
present in the $\beta$-functions.

It is found that the presence of the triplets does not  
give rise to any additional diagram for the effective vertex  
$\kappa$. However, the presence of these fields 
is felt indirectly in the running of $\kappa$  through their  
contribution to the evolution of the other quantities. Since  
the fermion triplets couple to W bosons, the evolution of the  
SU(2)$_L$ gauge coupling $g_2$ is significantly affected, with more  
than two $\Sigma$ triplets changing the sign of the $\beta$  
function for $g_2$. This may also have implications for the  
unification of gauge couplings. In turn, the masses of the  
$\Sigma$'s are also affected substantially due to the coupling  
with $g_2$. 
 
We give the analytic expressions for the  
RG evolutions of the neutrino masses and mixing   
above the highest mass threshold and below the lowest one.  
We use a basis   
${\cal P}_J = \{m_i, \theta_{12}, \theta_{23}, \theta_{13}^2, \phi_i,  
J_{\rm CP}, J'_{\rm CP}\}$ instead of the commonly used basis  
${\cal P}_\delta = \{ m_i,\theta_{12},\theta_{23}, \theta_{13}, 
 \phi_i, \delta \}$.  
The advantage of the ${\cal P}_J$ basis is that 
all the evolution equations are explicitly non-singular at all  
points in the parameter space including at $\theta_{13} =0$ \cite{jcppaper}.

We consider the scenario with three triplets having non-degenerate  
masses and include the effect of successive  
decoupling of the heavy triplets at their respective mass thresholds 
by imposing suitable matching conditions at each threshold.  
We present illustrative examples of running of masses and mixings  
by numerical diagonalization of the effective neutrino mass matrix.  
Although the running of neutrino parameters is  
not very large in the SM, in our model the running can be large  
due to threshold effects of the heavy triplets.  
In particular we find that starting from bi-maximal mixing at a high scale  
it is possible to generate low scale values of masses and mixing angles  
for the normal hierarchical neutrino spectrum. However, this requires non-zero  
values of the Majorana phases. Indeed it is observed that 
threshold effects and Majorana phases can influence the evolution  
of the mixing angles significantly. 
 
We show that even in the case of hierarchical neutrinos, the RG evolution
of neutrino masses and mixing between the thresholds can be substantial
in the Type-III seesaw scenario.
Moreover for quasi-degenerate neutrinos, the large running of masses 
implies that the value of $m_0$ at the high scale can be quite large, 
even if the mass related measurements from the beta decay,
neutrinoless double beta decay, or cosmology, restrict its value
at the low scale.

In conclusion, this work studies threshold effects in the context of the  
Type-III seesaw mechanism. It is crucial for testing the viability of a  
high scale theory with low scale data. Indeed it is 
seen that theories that are excluded by the data  
in the absence of RG running can become viable once these 
effects are included. 
In order to determine the allowed neutrino parameter space  
at the high scale, a detailed exploration of the dependence 
of RG effects on various parameters is necessary. 
This is all the more important in view of the  
onset of the precision era in neutrino physics.

\section*{Acknowledgement} 
J.C. and S.G. thank A. Raychaudhuri for encouragement and discussions.    
S.G. wishes to thank Dilip Ghosh, Anjan Joshipura, Subrata Khan,  
Namit Mahajan and Manimala Mitra for helpful discussions, and 
I. Gogoladze and N. Okada for useful communications.  
S.R. would like to thank R. Loganayagam for useful discussions. 
J.C. acknowledges support from RECAPP project and J.C. and  
S.G. acknowledges support from neutrino project under  
the XI$^{\rm th}$ plan of Harish Chandra Research Institute.  
The work of A.D. and S.R. was partially supported by the 
Max Planck -- India Partnergroup project between Tata Institute  
of Fundamental Research and Max Planck Institute for Physics.

 
\appendix 
 
\section{Feynman rules involving the fermion triplet $\Sigma$} 
\label{App-feyn-Sigma} 
 
In this appendix, we list the Feynmen rules involving the  
fermion triplets $\Sigma$. 
Following \cite{denner}, we introduce the fermion flow arrow for the leptons,  
which is the gray arrow in the diagrams. The black arrows indicate the  
lepton number flow. However interactions involving   
$\Sigma$ may violet lepton numbers and thus the $\Sigma$ line does not  
carry any lepton flow arrow. For the lepton number conserving  
interactions, the two arrows are parallel for particles, and antiparallel  
for the charge-conjugate fields. 
The Feynman rules are also given for the effective operator  
in the low energy limit of the theory obtained by  
integrating out these heavy fermion triplets.

 
\subsection{Propagator} 
 
\begin{center} 
\hspace{-2.0cm} 
  \begin{picture}(300,25) (0,0) 
    \SetWidth{0.5} 
    \SetColor{Black}\Line(100,20)(200,20) 
    \SetColor{Gray}\ArrowLine(130,25)(170,25) 
    \Text(100,2)[lb]{\normalsize{\Black{$\Sigma^{gj}$}}} 
    \Text(190,2)[lb]{\normalsize{\Black{$\Sigma^{fi}$}}}  
    \Text(220,14)[lb]{\large{\Black{$=  
\frac{i (\p + M_{f})}{p^2 - M_{f}^2 + i \epsilon}\; \delta_{fg} \delta_{ij}$}}}   
\end{picture} 
\end{center} 
\vspace{-0.2cm}      

\subsection{Yukawa interactions} 
 
\begin{center} 
\hspace{-3.5cm} 
  \begin{picture}(354,202) (60,-123) 
    \SetWidth{0.5} 
    \SetColor{Black} 
    \Text(61,-118)[lb]{\normalsize{\Black{$\Sigma^{gi}$}}} 
    \Text(61,-52)[lb]{\normalsize{\Black{$l^f_{Lb}$}}} 
    \Text(114,-74)[lb]{\normalsize{\Black{$\phi_a$}}} 
    \Text(131,-86)[lb]{\footnotesize{\Black{$= -i \mu^{\epsilon/2}  
\left( Y_\Sigma \right)_{gf} \left( \varepsilon^T \sigma^i \right)_{ab} P_L$}}} 
    \SetWidth{0.5} 
    \Line(88,-80)(62,-105) 
    \ArrowLine(61,-54)(87,-79) 
    \SetColor{Gray}\ArrowArcn(60,-80)(15,82,-75) 
    \SetColor{Black}\DashArrowLine(121,-79)(87,-80){4} 
    \Vertex(87,-80){2.0} 
    \Text(61,-12)[lb]{\normalsize{\Black{$\Sigma^{gi}$}}} 
    \Text(61,52)[lb]{\normalsize{\Black{$l^f_{Lb}$}}} 
    \Line(61,-1)(87,24) 
    \ArrowLine(86,25)(60,50) 
    \SetColor{Gray}\ArrowArc(60,25)(14.5,-79,80) 
    \SetColor{Black} 
    \DashArrowLine(86,24)(120,25){4} 
    \Vertex(86,24){2.0} 
    \Text(114,30)[lb]{\normalsize{\Black{$\phi_a$}}} 
    \Text(129,13)[lb]{\footnotesize{\Black{$= -i \mu^{\epsilon/2}  
\left( Y_\Sigma^\dagger \right)_{fg} \left( \sigma^i \varepsilon\right)_{ba} P_R$}}} 
    \Line(317,-105)(343,-80) 
    \ArrowLine(342,-79)(316,-54) 
    \SetColor{Gray}\ArrowArcn(314,-80)(15,82,-75) 
    \SetColor{Black} 
    \DashArrowLine(342,-80)(376,-79){4} 
    \Vertex(342,-80){2.0} 
    \Text(317,-118)[lb]{\normalsize{\Black{$\Sigma^{gi}$}}} 
    \Text(373,-74)[lb]{\normalsize{\Black{$\phi_a$}}} 
    \Text(388,-86)[lb]{\footnotesize{\Black{$= -i \mu^{\epsilon/2}  
\left( Y_\Sigma^\ast \right)_{gf} \left( \sigma^i \varepsilon\right)_{ab} P_R$}}} 
    \Text(316,-52)[lb]{\normalsize{\Black{$l^{f}_{Lb}$}}} 
    \Line(343,25)(317,0) 
    \ArrowLine(316,51)(342,26) 
    \SetColor{Gray}\ArrowArc(315,25)(14.5,-79,80) 
    \SetColor{Black} 
    \DashArrowLine(376,26)(342,25){4} 
    \Vertex(342,25){2.0} 
    \Text(317,-12)[lb]{\normalsize{\Black{$\Sigma^{gi}$}}} 
    \Text(315,54)[lb]{\normalsize{\Black{$l^{f}_{Lb}$}}} 
    \Text(373,30)[lb]{\normalsize{\Black{$\phi_a$}}} 
    \Text(388,18)[lb]{\footnotesize{\Black{$= -i \mu^{\epsilon/2}  
\left( Y_\Sigma^T \right)_{fg} \left( \varepsilon^T \sigma^i \right)_{ba} P_L$}}} 
  \end{picture} 
\end{center} 
%
 

\subsection{Gauge boson interactions} 
 
\begin{center} 
\hspace{-2.0cm} 
  \begin{picture}(180,92) (89,-147) 
    \SetWidth{0.5} 
    \SetColor{Black} 
    \Text(191,-109)[lb]{\small{\Black{$= -i \mu^{ \frac{\epsilon}{2} } g_2  
\gamma^\mu \left( i \varepsilon^{jik}\right)$}}} 
    \Text(89,-72)[lb]{\normalsize{\Black{$\Sigma^{fk}$}}} 
    \Text(122,-97)[lb]{\normalsize{\Black{$\mu$}}} 
    \Text(176,-97)[lb]{\normalsize{\Black{$W^i$}}} 
    \SetWidth{0.5} 
    \Line(122,-103)(91,-134) 
    \Line(91,-74)(122,-104) 
    \Vertex(121,-104){2.83} 
    \Photon(122,-105)(184,-102){4}{5} 
    \Text(89,-148)[lb]{\normalsize{\Black{$\Sigma^{gj}$}}} 
  \end{picture} 
\end{center} 
 

\subsection{Counterterms} 
 
\begin{center} 
\hspace{-9.0cm} 
  \begin{picture}(300,50) (0,0) 
    \SetWidth{0.5} 
    \SetColor{Gray} 
    \ArrowLine(140,30)(170,30) 
    \SetColor{Black} 
    \Line(110,20)(148,20) 
    \Line(162,20)(200,20)     
    \COval(155,20)(6,6)(0){Black}{White} 
    \Line(159.5,15.5)(150.0,24.5) 
    \Line(150.5,15.5)(159.5,24.5) 
    \Text(110,2)[lb]{\normalsize{\Black{$\Sigma^{gj}$}}} 
    \Text(190,2)[lb]{\normalsize{\Black{$\Sigma^{fi}$}}}  
    \Text(215,12)[lb]{\small{\Black{$=\;  
i \left[ \p  {\left( \delta Z_{\Sigma} \right)}_{fg}
- \left( \delta Z_{{\mathbbm M}_\Sigma} {\mathbbm M}_\Sigma \right)_{fg} \right] \delta_{ij}$}}}   
\end{picture} 
\end{center} 
 
 
\begin{center} 
\hspace{-2.0cm} 
  \begin{picture}(200,80) (40,-20) 
    \SetWidth{0.5} 
    \SetColor{Black} 
    \COval(25,10)(5.8,5.8)(0){Black}{White} 
    \Line(21,14)(29,6) 
    \Line(21,6)(29,14) 
    \Line(3,-15)(21,5) 
    \ArrowLine(21,15)(1,35) 
    \SetColor{Gray}\ArrowArc(-1,10)(14.5,-79,80) 
    \SetColor{Black} 
    \DashArrowLine(32,10)(60,10){4} 
    \Text(2,36)[lb]{\normalsize{\Black{$l_{Lb}^f$}}} 
    \Text(4,-26)[lb]{\normalsize{\Black{$\Sigma^{gi}$}}} 
    \Text(55,15)[lb]{\normalsize{\Black{$\phi_a$}}} 
    \Text(72,-3)[lb]{\small{\Black{$=\;  
-i \mu^{\epsilon/2} \left(  
\delta Z_{Y_\Sigma}^\dagger Y_\Sigma^\dagger \right)_{fg}  
\left( \sigma^i \varepsilon \right)_{ba} P_R$}}} 
\end{picture} 
\end{center} 

\begin{center} 
\hspace{-2.0cm} 
  \begin{picture}(200,80) (40,-20) 
    \SetWidth{0.5} 
    \SetColor{Black} 
    \COval(25,10)(5.8,5.8)(0){Black}{White} 
    \Line(21,14)(29,6) 
    \Line(21,6)(29,14) 
    \Line(21,5)(3,-15) 
    \ArrowLine(1,35)(21,15) 
    \SetColor{Gray}\ArrowArcn(-1,10)(14.5,80,-79) 
    \SetColor{Black} 
    \DashArrowLine(60,10)(32,10){4} 
    \Text(2,36)[lb]{\normalsize{\Black{$l_{Lb}^f$}}} 
    \Text(4,-26)[lb]{\normalsize{\Black{$\Sigma^{gi}$}}} 
    \Text(55,15)[lb]{\normalsize{\Black{$\phi_a$}}} 
    \Text(72,0)[lb]{\small{\Black{$=\;  
-i \mu^{\epsilon/2} \left(  
\delta Z_{Y_\Sigma} Y_\Sigma \right)_{gf}  
\left( \varepsilon^T \sigma^i \right)_{ab} P_L$}}} 
\end{picture} 
\end{center} 
 
\begin{center} 
\hspace{-2.0cm} 
  \begin{picture}(200,80) (40,-20) 
    \SetWidth{0.5} 
    \SetColor{Black} 
    \COval(25,10)(5.8,5.8)(0){Black}{White} 
    \Line(21,14)(29,6) 
    \Line(21,6)(29,14) 
    \Line(3,-15)(21,5) 
    \ArrowLine(1,35)(21,15) 
    \SetColor{Gray}\ArrowArc(-1,10)(14.5,-79,80) 
    \SetColor{Black} 
    \DashArrowLine(60,10)(32,10){4} 
    \Text(2,36)[lb]{\normalsize{\Black{$l_{Lb}^{f}$}}} 
    \Text(4,-26)[lb]{\normalsize{\Black{$\Sigma^{gi}$}}} 
    \Text(55,15)[lb]{\normalsize{\Black{$\phi_a$}}} 
    \Text(72,-3)[lb]{\small{\Black{$=\;  
-i \mu^{\epsilon/2} \left(  
\delta Z_{Y_\Sigma}^T Y_\Sigma^T \right)_{fg}  
\left( \varepsilon^T \sigma^i \right)_{ba} P_L$}}} 
\end{picture} 
\end{center} 
\begin{center} 
\hspace{-2.0cm} 
  \begin{picture}(200,80) (40,-20) 
    \SetWidth{0.5} 
    \SetColor{Black} 
    \COval(25,10)(5.8,5.8)(0){Black}{White} 
    \Line(21,14)(29,6) 
    \Line(21,6)(29,14) 
    \Line(21,5)(3,-15) 
    \ArrowLine(21,15)(1,35) 
    \SetColor{Gray}\ArrowArcn(-1,10)(14.5,80,-79) 
    \SetColor{Black} 
    \DashArrowLine(32,10)(60,10){4} 
    \Text(2,36)[lb]{\normalsize{\Black{$l_{Lb}^{f}$}}} 
    \Text(4,-26)[lb]{\normalsize{\Black{$\Sigma^{gi}$}}} 
    \Text(55,15)[lb]{\normalsize{\Black{$\phi_a$}}} 
    \Text(72,-2)[lb]{\small{\Black{$=\;  
-i \mu^{\epsilon/2} \left(  
\delta Z_{Y_\Sigma}^\ast Y_\Sigma^\ast \right)_{gf}  
\left( \sigma^i \varepsilon \right)_{ab} P_R$}}} 
\end{picture} 
\end{center}

 
\subsection{Effective vertex $\kappa$} 
 
\begin{center} 
  \begin{picture}(100,96) (201,-149) 
    \SetWidth{0.5} 
    \SetColor{Gray} 
    \ArrowLine(220,-100)(204,-84) 
    \ArrowLine(261,-142)(246,-126) 
    \SetColor{Black} 
    \GBox(226,-119)(246,-99){0.705} 
    \ArrowLine(202,-74)(226,-99) 
    \ArrowLine(271,-145)(247,-120) 
    \DashArrowLine(269,-74)(246,-99){4} 
    \DashArrowLine(203,-144)(226,-119){4} 
    \Text(184,-84)[lb]{\normalsize{\Black{$l^{f}_{Lb}$}}} 
    \Text(274,-82)[lb]{\normalsize{\Black{$\phi_a$}}} 
    \Text(232,-112)[lb]{\Large{\Black{$\kappa$}}} 
    \Text(188,-150)[lb]{\normalsize{\Black{$\phi_c$}}} 
    \Text(276,-148)[lb]{\normalsize{\Black{$l^g_{Ld}$}}} 
    \Text(275,-115)[lb]{\footnotesize{\Black{$=\;  
i \mu^{\epsilon/2} \kappa_{fg} \; \frac{1}{2}  
\left( \varepsilon_{ab} \varepsilon_{cd}  
+ \varepsilon_{ad} \varepsilon_{bc}  \right) P_L$}}} 
 \end{picture} 
\end{center} 
 
 
\begin{center} 
  \begin{picture}(100,96) (201,-149) 
    \SetWidth{0.5} 
    \SetColor{Gray} 
    \ArrowLine(220,-100)(204,-84) 
    \ArrowLine(261,-142)(246,-126) 
    \SetColor{Black} 
    \GBox(226,-119)(246,-99){0.705} 
    \ArrowLine(226,-99)(202,-74) 
    \ArrowLine(247,-120)(271,-145) 
    \DashArrowLine(246,-99)(269,-74){4} 
    \DashArrowLine(226,-119)(203,-144){4} 
    \Text(184,-84)[lb]{\normalsize{\Black{$l^{f}_{Lb}$}}} 
    \Text(274,-82)[lb]{\normalsize{\Black{$\phi_a$}}} 
    \Text(232,-112)[lb]{\Large{\Black{$\kappa$}}} 
    \Text(188,-150)[lb]{\normalsize{\Black{$\phi_c$}}} 
    \Text(276,-148)[lb]{\normalsize{\Black{$l^g_{Ld}$}}} 
    \Text(275,-115)[lb]{\footnotesize{\Black{$=\;  
i \mu^{\epsilon/2} \left(\kappa^\dagger\right)_{fg} \frac{1}{2}  
\left( \varepsilon_{ab} \varepsilon_{cd}  
+ \varepsilon_{ad} \varepsilon_{bc}  \right) P_R$}}} 
 \end{picture} 
\end{center} 
 
\subsection{Counterterms for $\kappa$}

\begin{center} 
  \begin{picture}(100,96) (201,-149) 
    \SetWidth{0.5} 
    \SetColor{Gray} 
    \ArrowLine(220,-100)(204,-84) 
    \ArrowLine(261,-142)(246,-126) 
    \SetColor{Black} 
    \CBox(246,-119)(226,-99){0}{White} 
    \Line(226,-99)(247,-120) 
    \Line(246,-99)(226,-119) 
    \ArrowLine(202,-74)(226,-99) 
    \ArrowLine(271,-145)(247,-120) 
    \DashArrowLine(269,-74)(246,-99){4} 
    \DashArrowLine(203,-144)(226,-119){4} 
    \Text(184,-84)[lb]{\normalsize{\Black{$l^{f}_{Lb}$}}} 
    \Text(274,-82)[lb]{\normalsize{\Black{$\phi_a$}}} 
    \Text(188,-150)[lb]{\normalsize{\Black{$\phi_c$}}} 
    \Text(276,-148)[lb]{\normalsize{\Black{$l^g_{Ld}$}}} 
    \Text(275,-115)[lb]{\footnotesize{\Black{$=\;  
i \mu^{\epsilon/2} \left( \delta \kappa \right)_{fg} \frac{1}{2}  
\left( \varepsilon_{ab} \varepsilon_{cd}  
+ \varepsilon_{ad} \varepsilon_{bc}  \right) P_L$}}} 
 \end{picture} 
\end{center} 
 
\begin{center} 
  \begin{picture}(100,96) (201,-149) 
    \SetWidth{0.5} 
    \SetColor{Gray} 
    \ArrowLine(220,-100)(204,-84) 
    \ArrowLine(261,-142)(246,-126) 
    \SetColor{Black} 
    \CBox(246,-119)(226,-99){0}{White} 
    \Line(226,-99)(247,-120) 
    \Line(246,-99)(226,-119) 
    \ArrowLine(226,-99)(202,-74) 
    \ArrowLine(247,-120)(271,-145) 
    \DashArrowLine(246,-99)(269,-74){4} 
    \DashArrowLine(226,-119)(203,-144){4} 
    \Text(184,-84)[lb]{\normalsize{\Black{$l^{f}_{Lb}$}}} 
    \Text(274,-82)[lb]{\normalsize{\Black{$\phi_a$}}} 
    \Text(188,-150)[lb]{\normalsize{\Black{$\phi_c$}}} 
    \Text(276,-148)[lb]{\normalsize{\Black{$l^g_{Ld}$}}} 
    \Text(275,-115)[lb]{\footnotesize{\Black{$=\;  
i \mu^{\epsilon/2} \left( \delta \kappa^\dagger \right)_{fg} \frac{1}{2}  
\left( \varepsilon_{ab} \varepsilon_{cd}  
+ \varepsilon_{ad} \varepsilon_{bc}  \right) P_R$}}} 
 \end{picture} 
\end{center} 
 
 
\section{Feynman rules for the SM fields} 
\label{App-feyn-SM} 
 
 
In this appendix, we list the Feynmen rules involving the  
SM fields only, also given in \cite{kersten-thesis},  
which are needed for our calculations.The directions of  
the arrows should be interpreted in the  
same way as stated at the beginning of Appendix~\ref{App-feyn-Sigma}.

 
\subsection{Propagators} 
 
\begin{center} 
\hspace{-2.0cm} 
  \begin{picture}(350,25) (90,0) 
    \SetWidth{0.5} 
    \SetColor{Black} 
    \ArrowLine(80,20)(180,20) 
    \Text(80,2)[lb]{\normalsize{\Black{$q^g_{La}$}}} 
    \Text(170,2)[lb]{\normalsize{\Black{$q^f_{Lb}$}}}  
    \Text(190,12)[lb]{\normalsize{\Black{$=  
\frac{i \p }{p^2 + i \epsilon}\; \delta_{fg} \delta_{ab} \quad ;$}}} 
    \ArrowLine(280,20)(380,20) 
    \Text(280,2)[lb]{\normalsize{\Black{$X^g_{R}$}}} 
    \Text(370,2)[lb]{\normalsize{\Black{$X^f_{R}$}}}  
    \Text(390,12)[lb]{\normalsize{\Black{$=  
\frac{i \p }{p^2 + i \epsilon}\; \delta_{fg} \; , \phantom{sp}$  
\small{$X \in \{ u, d \}$}}}}   
\end{picture} 
\end{center} 
 
\begin{center} 
\hspace{-2.0cm} 
  \begin{picture}(350,34) (90,0) 
    \SetWidth{0.5} 
    \SetColor{Gray} 
    \ArrowLine(110,25)(150,25) 
    \SetColor{Black} 
    \ArrowLine(80,20)(180,20) 
    \Text(80,2)[lb]{\normalsize{\Black{$l^g_{La}$}}} 
    \Text(170,2)[lb]{\normalsize{\Black{$l^f_{Lb}$}}}  
    \Text(190,12)[lb]{\normalsize{\Black{$=  
\frac{i \p }{p^2 + i \epsilon}\; \delta_{fg} \delta_{ab} \quad ;$}}} 
    \SetColor{Gray} 
    \ArrowLine(310,25)(350,25) 
    \SetColor{Black} 
    \ArrowLine(280,20)(380,20) 
    \Text(280,2)[lb]{\normalsize{\Black{$e^g_{R}$}}} 
    \Text(370,2)[lb]{\normalsize{\Black{$e^f_{R}$}}}  
    \Text(390,12)[lb]{\normalsize{\Black{$=  
\frac{i \p }{p^2 + i \epsilon}\; \delta_{fg} \; $}}}   
\end{picture} 
\end{center} 
 
\begin{center} 
\hspace{-2.0cm} 
  \begin{picture}(350,34) (90,0) 
    \SetWidth{0.5} 
    \SetColor{Gray} 
    \ArrowLine(150,25)(110,25) 
    \SetColor{Black} 
    \ArrowLine(80,20)(180,20) 
    \Text(80,2)[lb]{\normalsize{\Black{$l^g_{La}$}}} 
    \Text(170,2)[lb]{\normalsize{\Black{$l^f_{Lb}$}}}  
    \Text(190,12)[lb]{\normalsize{\Black{$=  
\frac{-i \p }{p^2 + i \epsilon}\; \delta_{gf} \delta_{ab} \quad ;$}}} 
    \SetColor{Gray} 
    \ArrowLine(350,25)(310,25) 
    \SetColor{Black} 
    \ArrowLine(280,20)(380,20) 
    \Text(280,2)[lb]{\normalsize{\Black{$e^g_{R}$}}} 
    \Text(370,2)[lb]{\normalsize{\Black{$e^f_{R}$}}}  
    \Text(390,12)[lb]{\normalsize{\Black{$=  
\frac{-i \p }{p^2 + i \epsilon}\; \delta_{gf} \; $}}}   
\end{picture} 
\end{center} 

\begin{center} 
\hspace{-2.0cm} 
  \begin{picture}(300,30) (50,0) 
    \SetWidth{0.5} 
    \SetColor{Black} 
    \DashArrowLine(100,20)(200,20){4} 
    \Text(100,2)[lb]{\normalsize{\Black{$\phi_{a}$}}} 
    \Text(190,2)[lb]{\normalsize{\Black{$\phi_{b}$}}}  
    \Text(220,12)[lb]{\normalsize{\Black{$=  
\frac{i}{p^2 - m_\phi^2 + i \epsilon} \; \delta_{ab}$}}}   
\end{picture} 
\end{center} 

\begin{center} 
\hspace{-2.0cm} 
  \begin{picture}(300,30) (50,0) 
    \SetWidth{0.5} 
    \SetColor{Black} 
    \Photon(100,20)(200,20){3}{9} 
    \Text(100,2)[lb]{\normalsize{\Black{$X^\mu$}}} 
    \Text(190,2)[lb]{\normalsize{\Black{$X^\nu$}}}  
    \Text(220,12)[lb]{\normalsize{\Black{$=  
\frac{i \left( -\eta^{\mu \nu} + (1-\xi) p^\mu p^\nu /p^2 \right)} 
{p^2 + i \epsilon} \; ; \quad X \in \{ B, W^i \}$}}} 
\end{picture} 
\end{center} 
where $\xi=\xi_1$ for B boson and $\xi=\xi_2$ for W boson.  
 
\subsection{Yukawa interactions} 
 
\begin{center} 
\hspace{-3.5cm} 
  \begin{picture}(354,202) (60,-123) 
    \SetWidth{0.5} 
    \SetColor{Black} 
    \Text(61,-119)[lb]{\normalsize{\Black{$e^{g}_R$}}} 
    \Text(61,-52)[lb]{\normalsize{\Black{$l^f_{Lb}$}}} 
    \Text(114,-74)[lb]{\normalsize{\Black{$\phi_a$}}} 
    \Text(131,-86)[lb]{\footnotesize{\Black{$= -i \mu^{\epsilon/2}  
\left( Y_e \right)_{gf} \delta_{ab} P_L$}}} 
    \SetWidth{0.5}  
    \SetColor{Gray}\ArrowArcn(60,-80)(15,82,-75) 
    \SetColor{Black} 
    \ArrowLine(88,-80)(62,-105) 
    \ArrowLine(61,-54)(87,-79) 
    \DashArrowLine(87,-80)(121,-80){4} 
    \Vertex(87,-80){2.0} 
    \Text(61,-14)[lb]{\normalsize{\Black{$e^g_R$}}} 
    \Text(61,52)[lb]{\normalsize{\Black{$l^f_{Lb}$}}} 
    \ArrowLine(61,-1)(87,24) 
    \ArrowLine(86,25)(60,50) 
    \SetColor{Gray}\ArrowArc(60,25)(14.5,-79,80) 
    \SetColor{Black} 
    \DashArrowLine(121,24)(87,24){4} 
    \Vertex(86,24){2.0} 
    \Text(114,30)[lb]{\normalsize{\Black{$\phi_a$}}} 
    \Text(129,18)[lb]{\footnotesize{\Black{$= -i \mu^{\epsilon/2}  
\left( Y_e^\dagger \right)_{fg}  \delta_{ab} P_R$}}} 
    \ArrowLine(317,-105)(343,-80) 
    \ArrowLine(342,-79)(316,-54) 
    \SetColor{Gray}\ArrowArcn(314,-80)(15,82,-75) 
    \SetColor{Black} 
    \DashArrowLine(376,-80)(342,-80){4} 
    \Vertex(342,-80){2.0} 
    \Text(317,-119)[lb]{\normalsize{\Black{$ e_R^{g}$}}} 
    \Text(373,-74)[lb]{\normalsize{\Black{$\phi_a$}}} 
    \Text(388,-86)[lb]{\footnotesize{\Black{$= -i \mu^{\epsilon/2}  
\left( Y_e^\ast \right)_{gf} \delta_{ab} P_R$}}} 
    \Text(316,-52)[lb]{\normalsize{\Black{$l^{f}_{Lb}$}}} 
    \ArrowLine(343,25)(317,0) 
    \ArrowLine(316,51)(342,26) 
    \SetColor{Gray}\ArrowArc(315,25)(14.5,-79,80) 
    \SetColor{Black} 
    \DashArrowLine(342,24)(376,24){4} 
    \Vertex(342,25){2.0} 
    \Text(317,-14)[lb]{\normalsize{\Black{$e_R^{g}$}}} 
    \Text(315,54)[lb]{\normalsize{\Black{$l^{f}_{Lb}$}}} 
    \Text(373,30)[lb]{\normalsize{\Black{$\phi_a$}}} 
    \Text(388,18)[lb]{\footnotesize{\Black{$= -i \mu^{\epsilon/2}  
\left( Y_e^T \right)_{fg} \delta_{ab} P_L$}}} 
  \end{picture} 
\end{center} 
Similar Feynman rules, as those in the left panel, are there  
for Yukawa interactions of $q_L$-$u_R$ and $q_L$-$d_R$  
with the Higgs $\phi$ having coefficients $Y_u$ and $Y_d$ respectively. 
 
 
\subsection{Gauge boson -- lepton interactions} 
 
\begin{center} 
\hspace{-2.0cm} 
  \begin{picture}(373,92) (95,-147) 
    \SetWidth{0.5} 
    \SetColor{Black} 
    \Text(191,-109)[lb]{\footnotesize{\Black{ 
$= i \mu^{\frac{\epsilon}{2}} g_1 \gamma^\mu \delta_{gf} P_R$}}} 
    \Text(89,-72)[lb]{\normalsize{\Black{$e^{f}_{R}$}}} 
    \Text(122,-97)[lb]{\normalsize{\Black{$\mu$}}} 
    \Text(176,-97)[lb]{\normalsize{\Black{$B$}}} 
    \SetWidth{0.5} 
    \ArrowLine(122,-103)(91,-134) 
    \ArrowLine(91,-74)(122,-104) 
    \SetColor{Gray}\ArrowArcn(92,-103)(14.5,90,-85) 
    \SetColor{Black} 
    \Vertex(121,-104){2.0} 
    \Photon(122,-105)(184,-102){4}{5} 
    \Text(422,-108)[lb]{\footnotesize{\Black{ 
$= -i \mu^{\frac{\epsilon}{2}} g_1 \gamma^\mu \delta_{fg} P_L$}}} 
    \Text(89,-152)[lb]{\normalsize{\Black{$e^{g}_{R}$}}} 
    \Text(319,-71)[lb]{\normalsize{\Black{$e_R^{f}$}}} 
    \Text(352,-96)[lb]{\normalsize{\Black{$\mu$}}} 
    \Text(406,-96)[lb]{\normalsize{\Black{$B$}}} 
    \ArrowLine(352,-102)(321,-133) 
    \ArrowLine(321,-73)(352,-103) 
    \SetColor{Gray}\ArrowArc(322,-102)(14.5,-85,90) 
    \SetColor{Black} 
    \Vertex(351,-103){2.0} 
    \Photon(352,-104)(414,-101){4}{5} 
    \Text(319,-152)[lb]{\normalsize{\Black{$e^{g}_R$}}} 
  \end{picture} 
\end{center} 
 
 
\begin{center} 
\hspace{-2.0cm} 
  \begin{picture}(373,92) (95,-147) 
    \SetWidth{0.5} 
    \SetColor{Black} 
    \Text(191,-109)[lb]{\footnotesize{\Black{ 
$= -i \mu^{\frac{\epsilon}{2}} g_2 \gamma^\mu \left(\sigma^i\right)_{ba} \delta_{gf} P_L$}}} 
    \Text(89,-72)[lb]{\normalsize{\Black{$l^{f}_{La}$}}} 
    \Text(122,-97)[lb]{\normalsize{\Black{$\mu$}}} 
    \Text(176,-97)[lb]{\normalsize{\Black{$W^i$}}} 
    \SetWidth{0.5} 
    \ArrowLine(122,-103)(91,-134) 
    \ArrowLine(91,-74)(122,-104) 
    \SetColor{Gray}\ArrowArcn(92,-103)(14.5,90,-85) 
    \SetColor{Black} 
    \Vertex(121,-104){2.0} 
    \Photon(122,-105)(184,-102){4}{5} 
    \Text(422,-108)[lb]{\footnotesize{\Black{ 
$= i \mu^{\frac{\epsilon}{2}} g_2 \gamma^\mu \left(\sigma^i\right)_{ba} \delta_{fg} P_R$}}} 
    \Text(89,-152)[lb]{\normalsize{\Black{$l^{g}_{Lb}$}}} 
    \Text(321,-71)[lb]{\normalsize{\Black{$l_{la}^{f}$}}} 
    \Text(351,-96)[lb]{\normalsize{\Black{$\mu$}}} 
    \Text(406,-96)[lb]{\normalsize{\Black{$W^i$}}} 
    \ArrowLine(352,-102)(321,-133) 
    \ArrowLine(321,-73)(352,-103) 
    \SetColor{Gray}\ArrowArc(322,-102)(14.5,-85,90) 
    \SetColor{Black} 
    \Vertex(351,-103){2.0} 
    \Photon(352,-104)(414,-101){4}{5} 
    \Text(319,-152)[lb]{\normalsize{\Black{$l^{g}_{Lb}$}}} 
  \end{picture} 
\end{center} 
 
 
\begin{center} 
\hspace{-2.0cm} 
  \begin{picture}(373,92) (95,-147) 
    \SetWidth{0.5} 
    \SetColor{Black} 
    \Text(191,-109)[lb]{\footnotesize{\Black{ 
$= \frac{i}{2} \mu^{\frac{\epsilon}{2}} g_2 \gamma^\mu \delta_{gf} \delta_{ab} P_L$}}} 
    \Text(89,-72)[lb]{\normalsize{\Black{$l^{f}_{La}$}}} 
    \Text(122,-97)[lb]{\normalsize{\Black{$\mu$}}} 
    \Text(176,-97)[lb]{\normalsize{\Black{$B$}}} 
    \SetWidth{0.5} 
    \ArrowLine(122,-103)(91,-134) 
    \ArrowLine(91,-74)(122,-104) 
    \SetColor{Gray}\ArrowArcn(92,-103)(14.5,90,-85) 
    \SetColor{Black} 
    \Vertex(121,-104){2.0} 
    \Photon(122,-105)(184,-102){4}{5} 
    \Text(422,-108)[lb]{\footnotesize{\Black{ 
$= -\frac{i}{2} \mu^{\frac{\epsilon}{2}} g_2 \gamma^\mu \delta_{fg} \delta_{ab} P_R$}}} 
    \Text(89,-152)[lb]{\normalsize{\Black{$l^{g}_{Lb}$}}} 
    \Text(319,-71)[lb]{\normalsize{\Black{$l_{La}^{f}$}}} 
    \Text(352,-96)[lb]{\normalsize{\Black{$\mu$}}} 
    \Text(406,-96)[lb]{\normalsize{\Black{$B$}}} 
    \ArrowLine(352,-102)(321,-133) 
    \ArrowLine(321,-73)(352,-103) 
    \SetColor{Gray}\ArrowArc(322,-102)(14.5,-85,90) 
    \SetColor{Black} 
    \Vertex(351,-103){2.0} 
    \Photon(352,-104)(414,-101){4}{5} 
    \Text(319,-152)[lb]{\normalsize{\Black{$l^{g}_{Lb}$}}} 
  \end{picture} 
\end{center}

 
\subsection{Gauge boson -- Higgs interactions} 
 
\begin{center} 
\hspace{-2.0cm} 
  \begin{picture}(373,92) (95,-147) 
    \SetWidth{0.5} 
    \SetColor{Black} 
    \Text(186,-109)[lb]{\footnotesize{\Black{ 
$= -\frac{i}{2} \mu^{\frac{\epsilon}{2}} g_1 (p_\mu + q_\mu) \delta_{ab}$}}} 
    \Text(89,-72)[lb]{\normalsize{\Black{$\phi_a$}}} 
    \Text(122,-97)[lb]{\normalsize{\Black{$\mu$}}} 
    \Text(176,-97)[lb]{\normalsize{\Black{$B$}}} 
    \SetWidth{0.5} 
    \DashArrowLine(122,-103)(91,-134){4} 
    \DashArrowLine(91,-74)(122,-104){4} 
    \Vertex(121,-104){2.0} 
    \Photon(122,-105)(180,-102){4}{5} 
    \Text(405,-108)[lb]{\footnotesize{\Black{ 
$= -i \mu^{\frac{\epsilon}{2}} g_2 (p_\mu + q_\mu) \left( \sigma^i \right)_{ba}$}}} 
    \Text(89,-148)[lb]{\normalsize{\Black{$\phi_b$}}} 
    \Text(309,-71)[lb]{\normalsize{\Black{$\phi_a$}}} 
    \Text(342,-96)[lb]{\normalsize{\Black{$\mu$}}} 
    \Text(396,-96)[lb]{\normalsize{\Black{$W^i$}}} 
    \DashArrowLine(342,-102)(311,-133){4} 
    \DashArrowLine(311,-73)(342,-103){4} 
    \Vertex(341,-103){2.0} 
    \Photon(342,-104)(400,-101){4}{5} 
    \Text(309,-148)[lb]{\normalsize{\Black{$\phi_b$}}} 
  \end{picture} 
\end{center} 
The vertices involving two Higgses and two gauge bosons are not shown 
since they do not appear explicitly in our analysis. 

\subsection{Higgs self-interaction}
\begin{center} 
\vspace{-0.4cm} 
\hspace{-2.0cm} 
  \begin{picture}(373,-54) (0,30) 
    \SetWidth{0.5} 
    \SetColor{Black} 
    \Text(156,-9)[lb]{\footnotesize{\Black{ 
$= -i \mu^{\epsilon} \lambda \frac{1}{2} (\delta_{ac} \delta_{bd} + \delta_{bc} \delta_{ad})$}}} 
    \Text(85,28)[lb]{\normalsize{\Black{$\phi_a$}}} 
    \Text(85,-32)[lb]{\normalsize{\Black{$\phi_c$}}} 
    \SetWidth{0.5} 
    \DashArrowLine(118,3)(90,-22){4} 
    \DashArrowLine(90,28)(118,3){4} 
    \Vertex(118,3){2.0} 
    \DashArrowLine(118,3)(146,-22){4} 
    \DashArrowLine(146,28)(118,3){4} 
    \Text(142,28)[lb]{\normalsize{\Black{$\phi_b$}}} 
    \Text(142,-32)[lb]{\normalsize{\Black{$\phi_d$}}} 
  \end{picture} 
\end{center} 
 
\vspace{1.2cm} 
\subsection{Counterterms}
\begin{center} 
\hspace{-2.0cm} 
  \begin{picture}(0,50) (240,0) 
    \SetWidth{0.5} 
    \SetColor{Gray} 
    \ArrowLine(80,30)(110,30) 
    \SetColor{Black} 
    \ArrowLine(52,20)(88,20) 
    \ArrowLine(102,20)(138,20)     
    \COval(95,20)(6,6)(0){Black}{White} 
    \Line(99.5,15.5)(90.0,24.5) 
    \Line(90.5,15.5)(99.5,24.5) 
    \Text(50,2)[lb]{\normalsize{\Black{$l_{La}^{f}$}}} 
    \Text(130,2)[lb]{\normalsize{\Black{$l_{Lb}^{g}$}}}  
    \Text(145,12.5)[lb]{\small{\Black{$=\;  
i \p \left( \delta Z_{l_L} \right)_{gf} P_L \delta_{ba}$ }}} 
    \SetWidth{0.5} 
    \SetColor{Black} 
    \ArrowLine(282,20)(318,20) 
    \ArrowLine(332,20)(368,20) 
    \SetColor{Gray} 
    \ArrowLine(340,30)(310,30)     
    \COval(325,20)(6,6)(0){Black}{White} 
    \Line(329.5,15.5)(320.0,24.5) 
    \Line(320.5,15.5)(329.5,24.5) 
    \Text(280,2)[lb]{\normalsize{\Black{$l_{La}^{f}$}}} 
    \Text(360,2)[lb]{\normalsize{\Black{$l_{Lb}^{g}$}}}  
    \Text(375,12.5)[lb]{\small{\Black{$=\;  
-i \p \left( \delta Z_{l_L} \right)_{fg} P_L \delta_{ba}$ }}} 
\end{picture} 
\end{center} 
\begin{center} 
\vspace{-0.4cm} 
\hspace{-2.0cm} 
  \begin{picture}(0,50) (240,0) 
    \SetWidth{0.5} 
    \SetColor{Gray} 
    \ArrowLine(80,30)(110,30) 
    \SetColor{Black} 
    \ArrowLine(52,20)(88,20) 
    \ArrowLine(102,20)(138,20)     
    \COval(95,20)(6,6)(0){Black}{White} 
    \Line(99.5,15.5)(90.0,24.5) 
    \Line(90.5,15.5)(99.5,24.5) 
    \Text(50,2)[lb]{\normalsize{\Black{$e_{R}^{f}$}}} 
    \Text(130,2)[lb]{\normalsize{\Black{$e_{R}^{g}$}}}  
    \Text(145,12.5)[lb]{\small{\Black{$=\;  
i \p \left( \delta Z_{e_R} \right)_{gf} P_R $ }}} 
    \SetWidth{0.5} 
    \SetColor{Gray} 
    \ArrowLine(340,30)(310,30) 
    \SetColor{Black} 
    \ArrowLine(282,20)(318,20) 
    \ArrowLine(332,20)(368,20)     
    \COval(325,20)(6,6)(0){Black}{White} 
    \Line(329.5,15.5)(320.0,24.5) 
    \Line(320.5,15.5)(329.5,24.5) 
    \Text(280,2)[lb]{\normalsize{\Black{$e_{R}^{f}$}}} 
    \Text(360,2)[lb]{\normalsize{\Black{$e_{R}^{g}$}}}  
    \Text(375,12.5)[lb]{\small{\Black{$=\;  
-i \p \left( \delta Z_{e_R} \right)_{fg} P_R$ }}} 
\end{picture} 
\end{center} 
\begin{center} 
\vspace{-0.4cm} 
\hspace{-2.0cm} 
  \begin{picture}(0,50) (240,0) 
    \SetWidth{0.5} 
    \SetColor{Gray} 
    \ArrowLine(80,30)(110,30) 
    \SetColor{Black} 
    \DashArrowLine(52,20)(88,20){4} 
    \DashArrowLine(102,20)(138,20){4}     
    \COval(95,20)(6,6)(0){Black}{White} 
    \Line(99.5,15.5)(90.0,24.5) 
    \Line(90.5,15.5)(99.5,24.5) 
    \Text(50,2)[lb]{\normalsize{\Black{$\phi_a$}}} 
    \Text(130,2)[lb]{\normalsize{\Black{$\phi_b$}}}  
    \Text(145,12.5)[lb]{\small{\Black{$=\;  
i \left( p^2 \delta Z_\phi - \delta m_\phi^2 \right) \delta_{ba} $ }}} 
\end{picture} 
\end{center} 
%
 
 
\section{Calculation of renormalization constants} 
\label{App-renorm} 
 
Here we show the Feynman diagrams contributing to the renormalization  
constants of different quantities. Note that for particles in the loop,  
we suppress the flavor as well as the SU(2)$_L$ indices. 
 
 
\subsection{Doublet Higgs wavefunction and mass  
$(Z_{\phi}$ and $\delta m_\phi^2)$} 
\label{app-Zphi-dia} 
 
\vspace{-1.6cm} 
 
\begin{center} 
  \begin{picture}(487,276) (15,-11) 
    \SetWidth{0.5} 
    \SetColor{Black} 
    \DashArrowArcn(276.13,215)(12.14,255.19,-64.8){3} 
    \DashArrowLine(277,204)(307,204){3} 
    \DashArrowLine(245,204)(275,204){3} 
    \Vertex(276,204){2.0} 
    \Text(246,192)[lb]{\normalsize{\Black{$\phi$}}} 
    \Text(305,192)[lb]{\normalsize{\Black{$\phi$}}} 
    \Text(276,234)[lb]{\normalsize{\Black{$\phi$}}} 
    \Vertex(359,204){1.8} 
    \Vertex(393,204){1.8} 
    \DashArrowLine(330,204)(360,204){3} 
    \DashArrowLine(392,204)(422,204){3} 
    \ArrowArc(376,204)(17,-0.1,180.1) 
    \ArrowArc(376,205)(17,-172.85,-7.15) 
    \Text(332,192)[lb]{\normalsize{\Black{$\phi$}}} 
    \Text(418,192)[lb]{\normalsize{\Black{$\phi$}}} 
    \Text(376,226)[lb]{\normalsize{\Black{$e_R$}}} 
    \Text(385,178)[lb]{\normalsize{\Black{$l_L$}}} 
    \GOval(90,204)(10,10)(0){0.7} 
    \DashArrowLine(50,204)(80,204){3} 
    \DashArrowLine(100,204)(130,204){3} 
    \Text(52,192)[lb]{\normalsize{\Black{$\phi$}}} 
    \Text(126,192)[lb]{\normalsize{\Black{$\phi$}}} 
    \Text(138,201)[lb]{\normalsize{\Black{$\equiv$}}} 
    \Text(227.5,202)[lb]{\normalsize{\Black{$+$}}} 
    \Text(162,192)[lb]{\normalsize{\Black{$\phi$}}} 
    \DashArrowLine(160,204)(215,204){3} 
    \Text(213,192)[lb]{\normalsize{\Black{$\phi$}}} 
    \Text(316,202)[lb]{\normalsize{\Black{$+$}}} 
    \Text(300,62)[lb]{\normalsize{\Black{$+$}}} 
    \Text(418,58)[lb]{\normalsize{\Black{$\phi$}}} 
    \Text(302,138)[lb]{\normalsize{\Black{$+$}}} 
    \Text(156,137)[lb]{\normalsize{\Black{$+$}}} 
    \Text(156,68)[lb]{\normalsize{\Black{$+$}}} 
    \Text(322,58)[lb]{\normalsize{\Black{$\phi$}}} 
    \Vertex(353,71){1.8} 
    \Vertex(387,71){1.8} 
    \DashArrowLine(320,71)(354,71){3} 
    \DashArrowLine(388,71)(422,71){3} 
    \PhotonArc(369.5,70.5)(17.51,1.64,178.36){-3.5}{5.5} 
    \DashArrowArc(370,74)(17.03,-176.63,-3.37){3} 
    \Text(366,41)[lb]{\normalsize{\Black{$\phi$}}} 
    \Text(376,92)[lb]{\footnotesize{\Black{$B$}}} 
    \Text(326,126)[lb]{\normalsize{\Black{$\phi$}}} 
    \Vertex(354,139){1.8} 
    \Vertex(388,139){1.8} 
    \DashArrowLine(321,139)(355,139){3} 
    \DashArrowLine(389,139)(423,139){3} 
    \ArrowArc(371,139.03)(17,-0.1,180.1) 
    \ArrowArc(371,142.13)(17.13,-172.85,-7.15) 
    \Text(421,128)[lb]{\normalsize{\Black{$\phi$}}} 
    \Text(359,160)[lb]{\normalsize{\Black{$u_R, d_R$}}} 
    \Text(372,112)[lb]{\normalsize{\Black{$q_L$}}} 
    \Vertex(208,140){1.8} 
    \Vertex(242,140){1.8} 
    \DashArrowLine(174,140)(208,140){3} 
    \DashArrowLine(243,140)(277,140){3} 
    \ArrowArc(225,140.03)(17,-0.1,180.1) 
    \CArc(225,143.13)(17.13,-172.85,-7.15) 
    \Text(176,128)[lb]{\normalsize{\Black{$\phi$}}} 
    \Text(277,128)[lb]{\normalsize{\Black{$\phi$}}} 
    \Text(224,163)[lb]{\normalsize{\Black{$l_L$}}} 
    \Text(225,112)[lb]{\normalsize{\Black{$\Sigma$}}} 
    \Text(177,58)[lb]{\normalsize{\Black{$\phi$}}} 
    \Text(277,58)[lb]{\normalsize{\Black{$\phi$}}} 
    \Vertex(208,70){1.8} 
    \Vertex(243,70){1.8} 
    \DashArrowLine(175,70)(209,70){3} 
    \DashArrowLine(244,70)(278,70){3} 
    \PhotonArc(225.5,69.5)(17.51,1.64,178.36){-3.5}{5.5} 
    \DashArrowArc(225.5,73.56)(17.57,-174.9,-5.1){3} 
    \Text(231,90)[lb]{\footnotesize{\Black{$W$}}} 
    \Text(224,42)[lb]{\normalsize{\Black{$\phi$}}} 
    \DashArrowLine(181,20)(215,20){3} 
    \COval(222,20)(6.0,6.0)(45.0){Black}{White} 
    \Line(217.5,24.5)(226.5,15.5) 
    \Line(217.5,15.5)(226.5,24.5) 
    \DashArrowLine(230,20)(264,20){3} 
    \Text(257,8)[lb]{\normalsize{\Black{$\phi$}}} 
    \Text(183,8)[lb]{\normalsize{\Black{$\phi$}}} 
    \Text(156,15)[lb]{\normalsize{\Black{$+$}}} 
    \Text(275,17)[lb]{\normalsize{\Black{=  UV finite}}} 
    \Text(190,220)[lb]{\large{\Blue{A1}}} 
    \Text(295,220)[lb]{\large{\Blue{A2}}} 
    \Text(400,220)[lb]{\large{\Blue{A3}}} 
    \Text(250,150)[lb]{\large{\Blue{A4}}} 
    \Text(400,150)[lb]{\large{\Blue{A5}}} 
    \Text(260,80)[lb]{\large{\Blue{A6}}} 
    \Text(400,80)[lb]{\large{\Blue{A7}}} 
    \Text(240,25)[lb]{\large{\Blue{A8}}} 
    \SetColor{Gray} 
   \end{picture} 
\\ 
\vspace{-1.0cm} 
\beqa 
\Rightarrow \delta Z_{\phi} &=& - \frac{1}{16 \pi^2}  
\Bigl( 2 T - \frac{3}{10} (3 - \xi_1)  g_1^2  
- \frac{3}{2} (3 - \xi_2) g_2^2 \Bigr) \frac{1}{\epsilon} \; , \nn \\ 
{\rm and} \quad  \delta m_\phi^2 &=& \frac{1}{16 \pi^2}  
 \Bigl( 3 \lambda m_\phi^2 - \frac{3}{10} \xi_1 g_1^2 m_\phi^2  
- \frac{3}{2} \xi_2 g_2^2 m_\phi^2 - 4 \; \Tr[3 Y_\Sigma^\dagger Y_\Sigma] \;  
{\mathbbm M}_\Sigma^2 \Bigr) \frac{1}{\epsilon} \; . \nn 
\eeqa 
\end{center}

\subsection{Left-handed lepton wavefunction $(Z_{l_L})$} 
\label{app-Zl-dia} 
 
\vspace{-0.4cm} 
 
\begin{center} 
  \begin{picture}(465,160) (9,-49) 
    \SetWidth{0.5} 
    \SetColor{Black} 
    \Text(329,72)[lb]{\normalsize{\Black{$+$}}} 
    \Text(114,74)[lb]{\normalsize{\Black{$\equiv$}}} 
    \SetWidth{0.5} 
    \ArrowLine(142,76)(204,76) 
    \Text(197,59)[lb]{\normalsize{\Black{$l_L^g$}}} 
    \Text(142,59)[lb]{\normalsize{\Black{$l_L^f$}}} 
    \Text(210,72)[lb]{\normalsize{\Black{$+$}}} 
    \Text(145,18)[lb]{\normalsize{\Black{$+$}}} 
    \Text(287,18)[lb]{\normalsize{\Black{$+$}}} 
    \Text(252,2)[lb]{\normalsize{\Black{$l_L^g$}}} 
    \Text(234,59)[lb]{\normalsize{\Black{$l_L^f$}}} 
    \Text(269,45)[lb]{\normalsize{\Black{$\phi$}}} 
    \Text(270,100)[lb]{\normalsize{\Black{$\Sigma$}}} 
    \Text(310,59)[lb]{\normalsize{\Black{$l_L^g$}}} 
    \Vertex(257,76){1.8}  
    \ArrowLine(232,76)(257,76) 
    \ArrowLine(290,76)(316,76) 
    \DashArrowArcn(273.5,76)(16.32,-11.28,-168.72){3} 
    \CArc(273.5,76)(16,-0.11,180.11) 
    \Vertex(290,76){1.8} 
    \Text(345,59)[lb]{\normalsize{\Black{$l_L^f$}}} 
    \Text(380,103)[lb]{\normalsize{\Black{$e_R$}}} 
    \Text(386,45)[lb]{\normalsize{\Black{$\phi$}}} 
    \Text(420,59)[lb]{\normalsize{\Black{$l_L^g$}}} 
    \Text(314,2)[lb]{\normalsize{\Black{$l_L^f$}}} 
    \Text(395,2)[lb]{\normalsize{\Black{$l_L^g$}}} 
    \Vertex(373,20){1.8} 
    \Text(355,-12)[lb]{\footnotesize{\Black{$l_L$}}} 
    \Text(356,37)[lb]{\footnotesize{\Black{$B$}}} 
    \ArrowArc(357,22)(16.12,-172.87,-7.13) 
    \PhotonArc(357,16.38)(16.66,16.12,163.88){-2.5}{5.5} 
    \Vertex(342,20){1.8} 
    \ArrowLine(313,20)(340,20) 
    \ArrowLine(373,20)(400,20) 
    \Text(145,-40)[lb]{\normalsize{\Black{$+$}}} 
    \ArrowLine(207,-36)(231,-36) 
    \ArrowLine(168,-36)(195,-36) 
    \Text(170,-55)[lb]{\normalsize{\Black{$l_L^f$}}} 
    \Text(220,-55)[lb]{\normalsize{\Black{$l_L^g$}}} 
    \COval(202,-36)(7.0,7.0)(0){Black}{White} 
    \Line(197,-41)(207,-31) 
    \Line(197,-31)(207,-41) 
    \Vertex(230,20){1.8} 
    \Vertex(198,20){1.8} 
    \Text(34,59)[lb]{\normalsize{\Black{$l_L^f$}}} 
    \Text(96,59)[lb]{\normalsize{\Black{$l_L^g$}}} 
    \ArrowLine(80,76)(105,76) 
    \GOval(72,76)(10,10)(0){0.705} 
    \ArrowLine(32,76)(62,76) 
    \Text(168,2)[lb]{\normalsize{\Black{$l_L^f$}}} 
    \Text(212,-13)[lb]{\footnotesize{\Black{$l_L$}}} 
    \ArrowArc(214,22)(16.12,-172.87,-7.13) 
    \PhotonArc(213.5,17.08)(15.99,14.18,165.82){-2.5}{5.5} 
    \ArrowLine(169,20)(196,20) 
    \ArrowLine(229,20)(256,20) 
    \Text(213,37)[lb]{\footnotesize{\Black{$W$}}} 
    \Vertex(370,76){1.8} 
    \Vertex(403,76){1.8} 
    \ArrowLine(345,76)(370,76) 
    \ArrowArcn(386.5,76)(16.51,178.26,1.74) 
    \DashArrowArcn(386.5,76)(16.34,-11.68,-168.32){3} 
    \ArrowLine(403,76)(428,76) 
    \Text(245,-40)[lb]{\normalsize{\Black{=  UV finite}}} 
    \Text(180,88)[lb]{\large{\Blue{B1}}} 
    \Text(295,88)[lb]{\large{\Blue{B2}}} 
    \Text(410,88)[lb]{\large{\Blue{B3}}} 
    \Text(240,28)[lb]{\large{\Blue{B4}}} 
    \Text(385,28)[lb]{\large{\Blue{B5}}} 
    \Text(218,-30)[lb]{\large{\Blue{B6}}} 
    \SetColor{Gray} 
    \ArrowLine(60,90)(85,90) 
    \ArrowLine(155,81)(190,81) 
    \ArrowLine(262,96)(287,96) 
    \ArrowLine(375,98)(400,98) 
    \ArrowLine(202,0)(227,0) 
    \ArrowLine(344,0)(369,0) 
    \ArrowLine(185,-25)(215,-25) 
  \end{picture} 
\\ 
\vspace{-0.5cm} 
\beqa 
\Rightarrow \delta Z_{l_L} = - \frac{1}{16 \pi^2} \Bigl( Y_e^\dagger Y_e  
+ 3 Y_\Sigma^\dagger Y_\Sigma + \frac{3}{10} \xi_1  g_1^2  
+ \frac{3}{2} \xi_2 g_2^2  \Bigr) \frac{1}{\epsilon} \; . \nn 
\eeqa 
\end{center} 
 
\vspace{1.0cm} 
\subsection{Wavefunction and  
mass of fermion triplet $(Z_{\Sigma}$ and $Z_{{\mathbbm M}_\Sigma})$} 
\label{app-Zsigma-dia} 
 
\vspace{-0.2cm} 
 
\begin{center} 
  \begin{picture}(393,127) (52,-34) 
    \SetWidth{0.5} 
    \SetColor{Black} 
    \Text(237,45)[lb]{\normalsize{\Black{$+$}}} 
    \Text(254,35)[lb]{\normalsize{\Black{$\Sigma^{fi}$}}} 
    \Text(335,35)[lb]{\normalsize{\Black{$\Sigma^{gj}$}}} 
    \Text(300,23)[lb]{\normalsize{\Black{$\phi$}}} 
    \Text(298,74)[lb]{\footnotesize{\Black{$l_L$}}} 
    \SetWidth{0.5} 
    \Line(255,50)(282,50) 
    \Line(315,51)(342,51) 
    \SetColor{Gray}  
    \ArrowLine(287,71)(312,71) 
    \SetColor{Black} 
    \ArrowArcn(298.5,49.43)(16.57,174.58,5.42) 
    \DashArrowArc(298.41,53.59)(16.8,-167.64,-8.88){3} 
    \Vertex(282,50){1.8} 
    \Vertex(316,51){1.8} 
    \Line(161,50)(224,50) 
    \SetColor{Gray}  
    \ArrowLine(180,54)(205,54) 
    \SetColor{Black} 
    \Text(159,35)[lb]{\normalsize{\Black{$\Sigma^{fi}$}}} 
    \Text(214,35)[lb]{\normalsize{\Black{$\Sigma^{gj}$}}} 
    \Text(146,46)[lb]{\normalsize{\Black{$\equiv$}}} 
    \Line(103,50)(135,50) 
    \Line(53,50)(85,50) 
    \SetColor{Gray}  
    \ArrowLine(80,65)(110,65) 
    \SetColor{Black} 
    \GOval(94,50)(10,10)(0){0.647} 
    \Text(52,35)[lb]{\normalsize{\Black{$\Sigma^{fi}$}}} 
    \Text(124,35)[lb]{\normalsize{\Black{$\Sigma^{gj}$}}} 
    \Text(154,-12)[lb]{\normalsize{\Black{$+$}}} 
    \Text(300,-22)[lb]{\normalsize{\Black{$\Sigma^{fi}$}}} 
    \Text(365,-22)[lb]{\normalsize{\Black{$\Sigma^{gj}$}}} 
    \Line(304,-6)(333,-6) 
    \COval(340,-6)(6,6)(45.0){Black}{White} 
    \Line(336.46,-9.54)(343.54,-2.46) 
    \Line(336.46,-2.46)(343.54,-9.54) 
    \Line(346,-5)(375,-5) 
    \SetColor{Gray}  
    \ArrowLine(327,5)(352,5) 
    \SetColor{Black} 
    \Text(284,-09)[lb]{\normalsize{\Black{$+$}}} 
    \Text(184,-22)[lb]{\normalsize{\Black{$\Sigma^{fi}$}}} 
    \Text(260,-22)[lb]{\normalsize{\Black{$\Sigma^{gj}$}}} 
    \Text(225,-34)[lb]{\footnotesize{\Black{$W$}}} 
    \Text(230,15)[lb]{\footnotesize{\Black{$\Sigma$}}} 
    \Vertex(212,-7){1.8} 
    \Line(182,-7)(212,-7) 
    \Line(244,-6)(274,-6) 
    \SetColor{Gray}  
    \ArrowLine(215,13)(240,13) 
    \SetColor{Black} 
    \CArc(228,-7.03)(16.03,3.7,176.3) 
    \Vertex(245,-6){1.8} 
    \PhotonArc(227.92,-3.92)(16.21,-169.05,-7.37){-3.5}{5.5} 
    \Text(154,-60)[lb]{\normalsize{\Black{=  UV finite}}} 
    \Text(200,60)[lb]{\large{\Blue{C1}}} 
    \Text(324,61)[lb]{\large{\Blue{C2}}} 
    \Text(255,4)[lb]{\large{\Blue{C3}}} 
    \Text(355,4)[lb]{\large{\Blue{C4}}} 
  \end{picture} \\ 
\vspace{0.2cm} 
\beqa 
\Rightarrow \quad  
\delta Z_{\Sigma} &=& - \frac{1}{16 \pi^2}  
\Biggl[ \Bigl(2 Y_\Sigma Y_\Sigma^\dagger + 4 \xi_2 g_2^2  \Bigr) P_R +  
\Bigl(2 (Y_\Sigma Y_\Sigma^\dagger )^\ast  
+ 4 \xi_2 g_2^2  \Bigr) P_L \Biggr] \frac{1}{\epsilon} \; , \nn \\   
{\rm and} \quad \delta Z_{{\mathbbm M}_\Sigma} &=&  
-\frac{1}{16 \pi^2} \left( 12 + 4 \xi_2 \right)  
g_2^2 \;  \frac{1}{\epsilon}\; . \nn 
\eeqa 
\end{center} 
 

\subsection{Right-handed charged lepton wavefunction $(Z_{e_R})$} 
\label{app-ZeR-dia} 
\vspace{-0.8cm} 
 
\begin{center} 
  \begin{picture}(393,127) (52,-10) 
    \SetWidth{0.5} 
    \SetColor{Black} 
    \Text(237,45)[lb]{\normalsize{\Black{$+$}}} 
    \Text(254,35)[lb]{\normalsize{\Black{$e_R^f$}}} 
    \Text(335,35)[lb]{\normalsize{\Black{$e_R^g$}}} 
    \Text(300,23)[lb]{\footnotesize{\Black{$\phi$}}} 
    \Text(298,74)[lb]{\footnotesize{\Black{$l_L$}}} 
    \SetWidth{0.5} 
    \ArrowLine(255,50)(282,50) 
    \ArrowLine(315,51)(342,51) 
    \SetColor{Gray} 
    \ArrowLine(285,71)(310,71) 
    \ArrowLine(400,71)(425,71) 
    \ArrowLine(180,55)(205,55) 
    \ArrowLine(83,65)(108,65) 
    \ArrowLine(182,5)(207,5) 
    \SetColor{Black} 
    \ArrowArcn(298.5,49.43)(16.57,174.58,5.42) 
    \DashArrowArc(298.41,53.59)(16.8,-167.64,-8.88){3} 
    \Vertex(282,50){1.8} 
    \Vertex(316,51){1.8} 
    \ArrowLine(161,50)(224,50) 
    \Text(159,35)[lb]{\normalsize{\Black{$e_R^{f}$}}} 
    \Text(214,35)[lb]{\normalsize{\Black{$e_R^{g}$}}} 
    \Text(146,46)[lb]{\normalsize{\Black{$\equiv$}}} 
    \ArrowLine(103,50)(135,50) 
    \ArrowLine(53,50)(85,50) 
    \GOval(94,50)(10,10)(0){0.647} 
    \Text(52,35)[lb]{\normalsize{\Black{$e_R^{f}$}}} 
    \Text(124,35)[lb]{\normalsize{\Black{$e_R^{g}$}}} 
    \Text(138,-10)[lb]{\normalsize{\Black{$+$}}} 
    \Text(160,-22)[lb]{\normalsize{\Black{$e_R^{f}$}}} 
    \Text(220,-22)[lb]{\normalsize{\Black{$e_R^{g}$}}} 
    \ArrowLine(160,-6)(190,-6) 
    \COval(193.5,-6)(7.0,7.0)(45.0){Black}{White} 
    \Line(198.0,-1.5)(189.0,-10.5) 
    \Line(198.0,-10.5)(189.0,-1.5) 
    \ArrowLine(200,-6)(227,-6) 
    \Text(352,46)[lb]{\normalsize{\Black{$+$}}} 
    \Text(368,35)[lb]{\normalsize{\Black{$e_R^{f}$}}} 
    \Text(450,35)[lb]{\normalsize{\Black{$e_R^{g}$}}} 
    \Text(416,21)[lb]{\footnotesize{\Black{$B$}}} 
    \Text(410,73.5)[lb]{\footnotesize{\Black{$e_R$}}} 
    \Vertex(394,51){1.8} 
    \ArrowLine(364,51)(394,51) 
    \ArrowLine(427,51)(457,51) 
    \ArrowArcn(411.5,51.05)(16.03,176.3,3.7) 
    \Vertex(427,51){1.8} 
    \PhotonArc(411.5,52.05)(16.21,-169.05,-7.37){-3.5}{5.5} 
    \Text(239,-09)[lb]{\normalsize{\Black{=  UV finite}}} 
    \Text(210,60)[lb]{\large{\Blue{D1}}} 
    \Text(320,60)[lb]{\large{\Blue{D2}}} 
    \Text(435,60)[lb]{\large{\Blue{D3}}} 
    \Text(210,4)[lb]{\large{\Blue{D6}}} 
  \end{picture} \\ 
\beqa 
\Rightarrow \quad  
\delta Z_{e_R} &=& - \frac{1}{16 \pi^2} \Bigl( 2 Y_e Y_e^\dagger  
+ \frac{6}{5} \xi_1 g_1^2  \Bigr) \frac{1}{\epsilon}\; . \nn 
\hspace{0.4cm}  
\eeqa 
\end{center} 
 
\parbox[l]{15.8cm}{ 
\subsection{$l_Le_R\phi$ Yukawa vertex $(Z_{Y_e})$} 
\label{app-ZYe-dia} 
\vspace{-2.0cm} 
 
\begin{center} 
  \begin{picture}(480,365) (0,0) 
    \SetWidth{0.5} 
    \SetColor{Black} 
    \Text(148,255)[lb]{\normalsize{\Black{$\equiv$}}} 
    \SetWidth{0.5} 
    \Vertex(177,258){1.8} 
    \DashArrowLine(181,258)(213,258){3} 
    \ArrowLine(178,258)(153,283) 
    \ArrowLine(150,230)(178,258)  
    \SetColor{Gray} 
    \ArrowArc(57,260)(14.5,-49,50) 
    \ArrowArc(152,258)(14.5,-49,50) 
    \ArrowArc(280,259)(8,-49,50) 
    \ArrowArc(391,257)(8,-46,50) 
    \SetColor{Black} 
    \DashArrowLine(181,258)(213,258){3} 
    \Text(151,217)[lb]{\normalsize{\Black{$l_L^f$}}} 
    \Text(154,288)[lb]{\normalsize{\Black{$e_R^g$}}} 
    \Text(220,254)[lb]{\normalsize{\Black{$\phi$}}} 
    \Text(241,254)[lb]{\normalsize{\Black{$+$}}} 
    \Vertex(297,258){1.8} 
    \DashArrowArc(289,258.5)(17.01,114.3,245.7){3} 
    \Line(283,245)(297,258) 
    \ArrowLine(271,231)(283,245) 
    \ArrowLine(297,258)(283,272) 
    \ArrowLine(283,272)(271,286) 
    \Vertex(282,273){1.8} 
    \Vertex(282,243){1.8} 
    \DashArrowLine(298,258)(334,258){3}    \SetColor{Black} 
    \Text(260,254)[lb]{\normalsize{\Black{$\phi$}}} 
    \Text(288,272)[lb]{\footnotesize{\Black{$l_L$}}} 
    \Text(272,217)[lb]{\normalsize{\Black{$l_L^f$}}} 
    \Text(268,288)[lb]{\normalsize{\Black{$e_R^g$}}} 
    \Text(339,254)[lb]{\normalsize{\Black{$\phi$}}} 
    \Text(291,242)[lb]{\footnotesize{\Black{$\Sigma$}}} 
    \Text(354,254)[lb]{\normalsize{\Black{$+$}}} 
    \Text(380,288)[lb]{\normalsize{\Black{$e_R^g$}}} 
    \Text(382,217)[lb]{\normalsize{\Black{$l_L^f$}}} 
    \Text(402,239)[lb]{\footnotesize{\Black{$l_L$}}} 
    \Text(402,267)[lb]{\footnotesize{\Black{$e_R$}}} 
    \Text(373,257)[lb]{\footnotesize{\Black{$B$}}} 
    \ArrowLine(394,244)(408,257) 
    \ArrowLine(382,230)(394,244) 
    \DashArrowLine(409,257)(445,257){3} 
    \Vertex(408,257){1.8} 
    \ArrowLine(408,257)(394,271) 
    \ArrowLine(394,271)(381,285) 
    \PhotonArc(415,257)(28.02,145.18,214.82){-3}{3.5} 
    \Vertex(392,272){1.8} 
    \Vertex(393,243){1.8} 
    \Text(450,254)[lb]{\normalsize{\Black{$\phi$}}} 
    \Text(139,144)[lb]{\normalsize{\Black{$+$}}} 
    \Text(243,146)[lb]{\normalsize{\Black{$+$}}} 
    \Text(353,144)[lb]{\normalsize{\Black{$+$}}} 
    \Vertex(414,149){1.8} 
    \Vertex(386,135){1.8} 
    \ArrowLine(385,135)(399,149) 
    \ArrowLine(372,122)(385,135) 
    \SetColor{Gray} 
    \ArrowArc(143,150)(17,-40,45) 
    \ArrowArc(249,150)(17,-40,45) 
    \ArrowArc(365,149)(17,-40,45) 
    \ArrowArc(146,36.5)(17,-40,45)  
    \SetColor{Black} 
    \DashArrowLine(400,149)(417,149){3} 
    \DashArrowLine(419,149)(436,149){3} 
    \Vertex(399,149){1.8} 
    \ArrowLine(399,149)(371,177) 
    \PhotonArc(395.68,152.49)(21.6,-125.94,-9.3){-3}{4.5} 
    \Text(373,109)[lb]{\normalsize{\Black{$l_L^f$}}} 
    \Text(371,178)[lb]{\normalsize{\Black{$e_R^g$}}} 
    \Text(443,145)[lb]{\normalsize{\Black{$\phi$}}} 
    \Text(407,154)[lb]{\normalsize{\Black{$\phi$}}} 
    \Text(406,120)[lb]{\footnotesize{\Black{$W$}}} 
    \Text(386,146)[lb]{\footnotesize{\Black{$l_L$}}} 
    \Text(143,37)[lb]{\normalsize{\Black{$+$}}} 
    \COval(175,36)(5,5)(0){Black}{White} 
    \Line(171,32)(179,40) 
    \Line(171,40)(179,32) 
    \ArrowLine(171,41)(155,59) 
    \ArrowLine(155,13)(171,31) 
    \DashArrowLine(182,36)(219,36){3} 
    \Text(154,60)[lb]{\normalsize{\Black{$e_R^g$}}} 
    \Text(150,-4)[lb]{\normalsize{\Black{$l_L^f$}}} 
    \Text(221,34)[lb]{\normalsize{\Black{$\phi$}}} 
    \Text(56,217)[lb]{\normalsize{\Black{$l_L^f$}}} 
    \GOval(85,259)(9,9)(0){0.705} 
    \ArrowLine(58,232)(78,253) 
    \ArrowLine(78,266)(60,284) 
    \DashArrowLine(94,259)(126,259){3} 
    \Text(130,254)[lb]{\normalsize{\Black{$\phi$}}} 
    \Text(60,288)[lb]{\normalsize{\Black{$e_R^g$}}} 
    \Vertex(193,149){1.8} 
    \Vertex(163,135){1.8} 
    \ArrowLine(164,136)(178,149) 
    \ArrowLine(151,122)(164,136) 
    \DashArrowLine(178,149)(195,149){3} 
    \DashArrowLine(195,149)(216,149){3} 
    \Vertex(178,149){1.8} 
    \ArrowLine(178,149)(150,177) 
    \PhotonArc(175.22,152.08)(21.01,-125.58,-8.44){-3}{4.5}    \SetColor{Black} 
    \Text(152,109)[lb]{\normalsize{\Black{$l_L^f$}}} 
    \Text(149,178)[lb]{\normalsize{\Black{$e_R^g$}}} 
    \Text(222,145)[lb]{\normalsize{\Black{$\phi$}}} 
    \Text(186,155)[lb]{\normalsize{\Black{$\phi$}}} 
    \Text(163,145)[lb]{\footnotesize{\Black{$l_L$}}} 
    \Text(184,120)[lb]{\footnotesize{\Black{$B$}}} 
    \Text(332,145)[lb]{\normalsize{\Black{$\phi$}}} 
    \Text(255,109)[lb]{\normalsize{\Black{$l_L^f$}}} 
    \Vertex(268,164){1.8} 
    \Vertex(305,150){1.8} 
    \Text(268,151)[lb]{\footnotesize{\Black{$e_R$}}} 
    \ArrowLine(283,150)(269,164) 
    \ArrowLine(269,164)(255,178) 
    \Text(294,172)[lb]{\footnotesize{\Black{$B$}}} 
    \Text(256,178)[lb]{\normalsize{\Black{$e_R^g$}}} 
    \Text(292,138)[lb]{\normalsize{\Black{$\phi$}}} 
    \PhotonArc(282.3,150.45)(19.71,-1.31,132.42){-3}{4.5} 
    \DashArrowLine(284,150)(301,150){3} 
    \DashArrowLine(303,150)(328,150){3} 
    \Vertex(283,150){1.8} 
    \ArrowLine(256,124)(283,150)    \SetColor{Black} 
    \Text(240,34)[lb]{\normalsize{\Black{=  UV finite}}} 
    \Text(190,270)[lb]{\large{\Blue{E1}}} 
    \Text(325,270)[lb]{\large{\Blue{E2}}} 
    \Text(430,270)[lb]{\large{\Blue{E3}}} 
    \Text(200,160)[lb]{\large{\Blue{E4}}} 
    \Text(315,160)[lb]{\large{\Blue{E5}}} 
    \Text(425,160)[lb]{\large{\Blue{E6}}} 
    \Text(205,45)[lb]{\large{\Blue{E7}}} 
  \end{picture}\\ 
\beqa 
\Rightarrow \quad \delta Z_{Y_e} = - \frac{1}{16 \pi^2}  
\Bigl( -6 Y_\Sigma^\dagger Y_\Sigma 
+ \frac{9}{5} \left( 1 + \frac{1}{2} \xi_1  \right) g_1^2  
+ \frac{3}{2} \xi_2 g_2^2  \Bigr) \frac{1}{\epsilon}\; . \nn 
\eeqa 
\end{center} 
}

\subsection{$l_L \Sigma \phi$ Yukawa vertex $(Z_{Y_\Sigma})$} 
\label{app-ZYsigma-dia} 
 
\begin{center} 
  \begin{picture}(438,294) (63,-56) 
    \SetWidth{0.5} 
    \SetColor{Black} 
    \Vertex(177,187){1.8} 
    \DashArrowLine(213,187)(181,187){3} 
    \Line(178,187)(153,212) 
    \ArrowLine(150,159)(178,187) 
    \Text(151,149)[lb]{\normalsize{\Black{$l_L^f$}}} 
    \Text(152,215)[lb]{\normalsize{\Black{$\Sigma^{gj}$}}} 
    \Text(218,184)[lb]{\normalsize{\Black{$\phi$}}} 
    \Text(231,183)[lb]{\normalsize{\Black{$+$}}} 
    \Text(140,77)[lb]{\normalsize{\Black{$+$}}}
    \Vertex(287,187){1.41} 
    \DashArrowArc(279,187.5)(17.01,114.3,245.7){3} 
    \ArrowLine(274,174)(287,187) 
    \ArrowLine(261,160)(274,174) 
    \ArrowLine(287,187)(273,201) 
    \Line(273,201)(261,215) 
    \SetColor{Gray} 
    \ArrowArc(63,188)(14.5,-49,50) 
    \ArrowArc(152,187)(14.5,-49,50) 
    \ArrowArc(270,187)(8,-49,50) 
    \ArrowArc(356,189)(14.5,-49,50) 
    \SetColor{Black} 
    \Vertex(272,202){1.8} 
    \Vertex(272,172){1.8} 
    \DashArrowLine(324,187)(288,187){3}    \SetColor{Black} 
    \Text(253,185)[lb]{\footnotesize{\Black{$\phi$}}} 
    \Text(281,197)[lb]{\footnotesize{\Black{$l_L$}}} 
    \Text(262,149)[lb]{\normalsize{\Black{$l_L^f$}}} 
    \Text(258,218)[lb]{\normalsize{\Black{$\Sigma^{gj}$}}} 
    \Text(329,185)[lb]{\normalsize{\Black{$\phi$}}} 
    \Text(283,171)[lb]{\footnotesize{\Black{$e_R$}}} 
    \Text(343,182)[lb]{\normalsize{\Black{$+$}}} 
    \Text(354,220)[lb]{\normalsize{\Black{$\Sigma^{gj}$}}} 
    \Text(356,147)[lb]{\normalsize{\Black{$l_L^f$}}} 
    \Vertex(382,188){1.8} 
    \Text(395,195)[lb]{\footnotesize{\Black{$\phi$}}} 
    \Text(400,158)[lb]{\footnotesize{\Black{$B$}}} 
    \Text(381,176)[lb]{\footnotesize{\Black{$l_L$}}} 
    \Text(442,184)[lb]{\normalsize{\Black{$\phi$}}}
    \Vertex(367,174){1.8} 
    \Vertex(409,189){1.8} 
    \ArrowLine(368,175)(382,188) 
    \ArrowLine(353,161)(368,175) 
    \Line(382,188)(355,217) 
    \SetColor{Gray}  
    \ArrowArc(175,84.5)(8,-49,50) 
    \ArrowArc(254,83.5)(17,-40,45) 
    \ArrowArc(347,85)(17,-40,45) 
    \ArrowArc(146,-19.5)(17,-40,45)  
    \SetColor{Black} 
    \DashArrowLine(414,189)(385,189){3} 
    \DashArrowLine(438,189)(417,189){3} 
    \PhotonArc(385.4,190.9)(25.67,-135.79,-4.24){-4}{6.5}
    \Vertex(398,85){1.8} 
    \Vertex(368,71){1.8} 
    \ArrowLine(367,72)(381,85) 
    \ArrowLine(354,58)(367,72) 
    \DashArrowLine(399,85)(382,85){3} 
    \DashArrowLine(418,85)(401,85){3} 
    \Vertex(381,85){1.8} 
    \Line(381,85)(353,113) 
    \PhotonArc(377.68,88.49)(21.6,-125.94,-9.3){-3}{4.5}   \SetColor{Black} 
    \Text(355,45)[lb]{\normalsize{\Black{$l_L^f$}}} 
    \Text(353,118)[lb]{\normalsize{\Black{$\Sigma^{gj}$}}} 
    \Text(425,82)[lb]{\normalsize{\Black{$\phi$}}} 
    \Text(389,91)[lb]{\footnotesize{\Black{$\phi$}}} 
    \Text(388,56)[lb]{\footnotesize{\Black{$W$}}} 
    \Text(368,81)[lb]{\footnotesize{\Black{$l_L$}}} 
    \Text(325,72)[lb]{\normalsize{\Black{$\phi$}}} 
    \Text(257,42)[lb]{\normalsize{\Black{$l_L^f$}}}
    \Vertex(270,97){1.8} 
    \Vertex(307,83){1.8} 
    \Text(280,90)[lb]{\footnotesize{\Black{$\Sigma$}}} 
    \Line(285,83)(271,97) 
    \Line(271,97)(257,111) 
    \Text(296,105)[lb]{\footnotesize{\Black{$W$}}} 
    \Text(258,117)[lb]{\normalsize{\Black{$\Sigma^{gj}$}}} 
    \Text(290,71)[lb]{\footnotesize{\Black{$\phi$}}} 
    \PhotonArc(284.3,83.45)(19.71,-1.31,132.42){-3}{4.5} 
    \DashArrowLine(303,83)(286,83){3} 
    \DashArrowLine(328,83)(305,83){3} 
    \Vertex(285,83){1.8} 
    \ArrowLine(258,57)(285,83) 
    \Vertex(179,70){1.8} 
    \ArrowLine(180,70)(194,84) 
    \ArrowLine(167,57)(180,70) 
    \DashArrowLine(230,84)(194,84){3} 
    \Vertex(194,84){1.8} 
    \Vertex(180,98){1.8} 
    \Line(194,84)(180,98) 
    \Line(180,98)(166,112) 
    \PhotonArc(185.0,83.0)(17.01,114.3,240.7){-2.6}{4.5}  \SetColor{Black} 
    \Text(168,44)[lb]{\normalsize{\Black{$l_L^f$}}} 
    \Text(165,117)[lb]{\normalsize{\Black{$\Sigma^{gj}$}}} 
    \Text(233,82)[lb]{\normalsize{\Black{$\phi$}}} 
    \Text(190,94)[lb]{\footnotesize{\Black{$\Sigma$}}} 
    \Text(187,63)[lb]{\footnotesize{\Black{$l_L$}}} 
    \Text(153,81)[lb]{\footnotesize{\Black{$W$}}} 
    \Text(345,79)[lb]{\normalsize{\Black{$+$}}} 
    \Text(247,81)[lb]{\normalsize{\Black{$+$}}} 
    \COval(175,-20)(5,5)(0){Black}{White} 
    \Line(171,-24)(179,-16)\Line(171,-16)(179,-24) 
    \Line(171,-15)(153,3) 
    \ArrowLine(151,-45)(171,-25) 
    \DashArrowLine(219,-20)(182,-20){3} 
    \Text(154,8)[lb]{\normalsize{\Black{$\Sigma^{gj}$}}} 
    \Text(150,-60)[lb]{\normalsize{\Black{$l_L^f$}}} 
    \Text(221,-22)[lb]{\normalsize{\Black{$\phi$}}} 
    \Text(138,-22)[lb]{\normalsize{\Black{$+$}}} 
    \Text(66,146)[lb]{\normalsize{\Black{$l_L^f$}}} 
    \GOval(92,188)(9,9)(0){0.705} 
    \ArrowLine(65,161)(85,181) 
    \Line(85,195)(67,213) 
    \DashArrowLine(133,188)(101,188){3} 
    \Text(138,186)[lb]{\normalsize{\Black{$\phi$}}} 
    \Text(67,220)[lb]{\normalsize{\Black{$\Sigma^{gj}$}}} 
    \Text(150,183)[lb]{\normalsize{\Black{$\equiv$}}} 
    \Text(240,-25)[lb]{\normalsize{\Black{= UV finite}}} 
    \Text(200,198)[lb]{\large{\Blue{F1}}} 
    \Text(310,198)[lb]{\large{\Blue{F2}}} 
    \Text(420,198)[lb]{\large{\Blue{F3}}} 
    \Text(225,96)[lb]{\large{\Blue{F4}}} 
    \Text(325,96)[lb]{\large{\Blue{F5}}} 
    \Text(405,96)[lb]{\large{\Blue{F6}}} 
    \Text(205,-10)[lb]{\large{\Blue{F7}}} 
  \end{picture} \\ 
\beqa 
\Rightarrow \quad  
\delta Z_{Y_\Sigma} &=& - \frac{1}{16 \pi^2} \Bigl( 2 Y_e^\dagger Y_e  
- \frac{3}{10} \xi_1 g_1^2 - \frac{1}{2}  
\left( 12 + 7 \xi_2  \right) g_2^2\Bigr) \frac{1}{\epsilon} \; . \nn 
\eeqa 
\end{center} 
 
 
\subsection{The extra diagram contributing to $Z_\lambda$} 
\label{app-Zlambda-dia} 
 
\begin{center} 
  \begin{picture}(133,112) (82,-75) 
    \SetWidth{0.5} 
    \SetColor{Black} 
    \ArrowLine(152,-9)(119,-9) 
    \Line(152,-38)(152,-10) 
    \ArrowLine(119,-38)(152,-38) 
    \Line(119,-9)(119,-37) 
    \DashArrowLine(92,19)(120,-9){3} 
    \DashArrowLine(179,-64)(151,-36){3} 
    \DashArrowLine(119,-39)(92,-65){3} 
    \DashArrowLine(153,-8)(180,18){3} 
    \Vertex(119,-9){1.8} 
    \Vertex(119,-38){1.8} 
    \Vertex(152,-38){1.8} 
    \Vertex(152,-9){1.8} 
    \Text(180,-76)[lb]{\normalsize{\Black{$\phi_d$}}} 
    \Text(80,-76)[lb]{\normalsize{\Black{$\phi_c$}}} 
    \Text(85,21)[lb]{\normalsize{\Black{$\phi_a$}}} 
    \Text(183,21)[lb]{\normalsize{\Black{$\phi_b$}}} 
    \Text(100,-30)[lb]{\footnotesize{\Black{$\Sigma$}}} 
    \Text(159,-30)[lb]{\footnotesize{\Black{$\Sigma$}}} 
    \Text(133,-52)[lb]{\footnotesize{\Black{$l_L$}}} 
    \SetColor{Gray} 
    \Text(185,-24)[lb]{\large{\Blue{G1}}} 
    \Text(133,-5)[lb]{\footnotesize{\Black{$l_L$}}} 
  \end{picture} \\ 
\beqa 
= -\frac{5 i}{4 \pi^2}  
\Tr\left[ Y_\Sigma^{\; \dagger} Y_\Sigma Y_\Sigma^{\; \dagger} Y_\Sigma \right] 
\left( \delta_{ab} \delta_{cd} +\delta_{ac} \delta_{bd}  \right)  
\frac{1}{\epsilon} + {\rm UV}\; {\rm finite} \; . \nn 
\label{app-lambda} 
\eeqa 
\end{center} 
 
 
\subsection{Calculation of $Z_\kappa$} 
\label{app-Zkappa-dia} 
 
\begin{center} 
  \begin{picture}(484,513) (54,-32) 
    \SetWidth{0.5} 
    \SetColor{Black} 
    \GOval(98,448)(13,13)(0){0.647} 
    \ArrowLine(64,414)(88,438) 
    \ArrowLine(133,480)(109,456) 
    \SetColor{Black} 
    \DashArrowLine(65,480)(87,458){3} 
    \DashArrowLine(132,411)(109,437){3} 
    \Text(52,483)[lb]{\normalsize{\Black{$\phi_a$}}} 
    \Text(54,404)[lb]{\normalsize{\Black{$l_{Lb}^{g}$}}} 
    \Text(130,404)[lb]{\normalsize{\Black{$\phi_d$}}} 
    \Text(132,481)[lb]{\normalsize{\Black{$l_{Lc}^{f}$}}} 
    \SetColor{Gray} 
    \ArrowLine(63,422)(80,438) 
    \ArrowLine(108,462)(125,480) 
    \ArrowLine(173,422)(190,438) 
    \ArrowLine(213,462)(230,480) 
    \SetColor{Black} 
    \Text(302,253)[lb]{\normalsize{\Black{$l_{Lc}^{f}$}}} 
    \Text(298,163)[lb]{\normalsize{\Black{$l_{Lb}^g$}}} 
    \Text(382,168)[lb]{\normalsize{\Black{$\phi_a$}}} 
    \Text(383,254)[lb]{\normalsize{\Black{$\phi_d$}}} 
    \Text(292,208)[lb]{\footnotesize{\Black{$B$}}} 
    \Text(318,196)[lb]{\footnotesize{\Black{$l_{L}$}}} 
    \Text(318,220)[lb]{\footnotesize{\Black{$l_{L}$}}} 
    \GBox(334,201)(358,223){0.647} 
    \DashArrowLine(382,247)(358,223){3} 
    \DashArrowLine(381,178)(358,201){3} 
    \ArrowLine(320,187)(334,201) 
    \ArrowLine(320,236)(334,223) 
    \ArrowLine(306,173)(320,187) 
    \ArrowLine(307,250)(320,236) 
    \PhotonArc(336.5,211.5)(31.5,125.96,234.04){-3}{5.5} 
    \Vertex(319,236){2.00} 
    \Vertex(319,187){2.00} 
    \Text(170,303)[lb]{\normalsize{\Black{$l_{Lb}^g$}}} 
    \Text(246,365)[lb]{\normalsize{\Black{$l_{Lc}^{f}$}}} 
    \Text(170,365)[lb]{\normalsize{\Black{$\phi_d$}}} 
    \Text(204,358)[lb]{\footnotesize{\Black{$e_R$}}} 
    \Text(226,335)[lb]{\footnotesize{\Black{$\phi$}}} 
    \Text(184,333)[lb]{\footnotesize{\Black{$l_{L}$}}} 
    \Text(242,290)[lb]{\normalsize{\Black{$\phi_a$}}} 
    \SetColor{Black} 
    \Vertex(198,353){2.00} 
    \Vertex(222,353){2.00} 
    \DashArrowLine(247,281)(222,305){3} 
    \ArrowLine(174,281)(198,305) 
    \GBox(198,306)(222,328){0.647} 
    \ArrowLine(198,353)(198,328) 
    \DashArrowLine(222,353)(222,328){3} 
    \ArrowLine(222,353)(198,353) 
    \DashArrowLine(173,377)(198,353){3} 
    \ArrowLine(246,377)(222,353) 
    \SetColor{Black} 
    \Text(152,300)[lb]{\huge{\Black{$\biggl\{$}}} 
    \Text(240,324)[lb]{\normalsize{\Black{$ + \phantom{s} a\leftrightarrow d$}}} 
    \Text(285,300)[lb]{\Huge{\Black{$\biggr\}$}}} 
    \Text(136,324)[lb]{\Large{\Black{$+$}}} 
    \Text(150,75)[lb]{\Huge{\Black{$\biggl\{$}}} 
    \Text(250,100)[lb]{\normalsize{\Black{$ + \phantom{s} a\leftrightarrow d$}}} 
    \Text(295,75)[lb]{\Huge{\Black{$\biggr\}$}}} 
    \Text(335,75)[lb]{\Huge{\Black{$\biggl\{$}}} 
    \Text(450,100)[lb]{\normalsize{\Black{$ + \phantom{s} a\leftrightarrow d$}}} 
    \Text(495,75)[lb]{\Huge{\Black{$\biggr\}$}}} 
    \Text(495,254)[lb]{\normalsize{\Black{$\phi_d$}}} 
    \Text(498,168)[lb]{\normalsize{\Black{$\phi_a$}}} 
    \Text(476,223)[lb]{\footnotesize{\Black{$\phi$}}} 
    \Text(476,196)[lb]{\footnotesize{\Black{$\phi$}}} 
    \Text(501,210)[lb]{\footnotesize{\Black{$B$}}} 
    \Text(412,250)[lb]{\normalsize{\Black{$l_{L}$}}} 
    \Text(410,166)[lb]{\normalsize{\Black{$l_{L}$}}} 
    \GBox(443,201)(467,223){0.647} 
    \ArrowLine(419,247)(443,222) 
    \ArrowLine(419,176)(443,201) 
    \SetColor{Black} 
    \DashArrowLine(481,187)(467,200){3} 
    \DashArrowLine(479,236)(467,222){3} 
    \DashArrowLine(492,251)(480,237){3} 
    \DashArrowLine(495,174)(481,187){3} 
    \PhotonArc(460,213)(34.06,-49.76,49.76){-3}{5.5} 
    \Vertex(479,237){2.00} 
    \Vertex(480,188){2.00} 
    \Text(394,205)[lb]{\Large{\Black{$+$}}} 
    \Text(265,205)[lb]{\Large{\Black{$+$}}} 
    \Text(138,205)[lb]{\Large{\Black{$+$}}} 
    \Text(312,324)[lb]{\Large{\Black{$+$}}} 
    \Text(321,100)[lb]{\Large{\Black{$+$}}} 
    \Text(136,100)[lb]{\Large{\Black{$+$}}} 
    \Text(167,248)[lb]{\normalsize{\Black{$l_{Lc}^{f}$}}} 
    \Text(256,178)[lb]{\normalsize{\Black{$\phi_a$}}} 
    \Text(226,182)[lb]{\footnotesize{\Black{$\phi$}}} 
    \Text(167,160)[lb]{\normalsize{\Black{$l_{Lb}^g$}}} 
    \Text(257,234)[lb]{\normalsize{\Black{$\phi_d$}}} 
    \Text(226,228)[lb]{\footnotesize{\Black{$\phi$}}} 
    \GBox(192,197)(216,219){0.647} 
    \ArrowLine(168,244)(192,219) 
    \ArrowLine(168,173)(192,197) 
    \DashArrowArc(226.05,210.1)(14.11,-8.54,135.41){3} 
    \DashArrowArcn(226.15,206.51)(13.93,6.14,-131.04){3} 
    \DashArrowLine(266,232)(240,208){3} 
    \DashArrowLine(265,185)(240,209){3} 
    \Vertex(242,208){2.00} 
    \SetColor{Gray} 
    \ArrowArc(394,340)(9,120,-120) 
    \ArrowArcn(214,340)(9,-120,120)     
    \ArrowLine(372,306)(355,290) 
    \ArrowLine(420,380)(403,362) 
    \ArrowLine(175,290)(192,306) 
    \ArrowLine(223,362)(241,380) 
    \ArrowLine(184,220)(168,236) 
    \ArrowLine(168,180)(184,196) 
    \ArrowLine(436,222)(420,238) 
    \ArrowLine(420,184)(436,200) 
    \ArrowLine(337,227)(316,247) 
    \ArrowLine(310,170)(334,192) 
    \ArrowLine(362,66)(384,88) 
    \ArrowLine(404,122)(420,140) 
    \ArrowLine(194,66)(216,88) 
    \ArrowLine(236,122)(254,140) 
    \ArrowLine(167,-32)(184,-16) 
    \ArrowLine(208,8)(226,26) 
    \SetColor{Black} 
    \Text(330,300)[lb]{\Huge{\Black{$\biggl\{$}}} 
    \Text(420,324)[lb]{\normalsize{\Black{$ + \phantom{s} a\leftrightarrow d$}}} 
    \Text(465,300)[lb]{\Huge{\Black{$\biggr\}$}}} 
    \Text(352,303)[lb]{\normalsize{\Black{$l_{Lc}^{f}$}}} 
    \Text(427,362)[lb]{\normalsize{\Black{$l_{Lb}^g$}}} 
    \Text(352,362)[lb]{\normalsize{\Black{$\phi_d$}}} 
    \Text(386,358)[lb]{\footnotesize{\Black{$e_R$}}} 
    \Text(408,336)[lb]{\footnotesize{\Black{$\phi$}}} 
    \Text(366,336)[lb]{\footnotesize{\Black{$l_{L}$}}} 
    \Text(422,293)[lb]{\normalsize{\Black{$\phi_a$}}} 
    \Vertex(378,353){2.00} 
    \Vertex(402,353){2.00} 
    \DashArrowLine(428,281)(403,305){3}   
    \ArrowLine(354,281)(378,305) 
    \GBox(378,306)(402,328){0.647} 
    \ArrowLine(378,353)(378,328) 
    \DashArrowLine(402,353)(402,328){3} 
    \ArrowLine(402,353)(378,353) 
    \DashArrowLine(354,377)(379,353){3} 
    \ArrowLine(426,377)(402,353)   
    \Text(180,51)[lb]{\normalsize{\Black{$l_{Lb}^g$}}} 
    \Text(263,56)[lb]{\normalsize{\Black{$\phi_a$}}} 
    \Text(262,143)[lb]{\normalsize{\Black{$l_{Lc}^{f}$}}} 
    \Text(185,144)[lb]{\normalsize{\Black{$\phi_d$}}} 
    \Text(196,90)[lb]{\footnotesize{\Black{$l_{L}$}}} 
    \Text(200,112)[lb]{\footnotesize{\Black{$\phi$}}} 
    \Text(170,101)[lb]{\footnotesize{\Black{$B$}}} 
    \GBox(213,93)(237,115){0.647} 
    \ArrowLine(199,79)(213,93) 
    \ArrowLine(185,65)(199,79) 
    \DashArrowLine(202,127)(213,115){3} 
    \DashArrowLine(190,141)(201,129){3} 
    \Vertex(202,127){2.00} 
    \DashArrowLine(260,70)(237,93){3} 
    \ArrowLine(263,141)(237,115) 
    \SetColor{Black} 
    \Vertex(199,79){2.00} 
    \PhotonArc(205.67,104)(23.69,103.84,256.16){-3}{5.5} 
    \Text(430,52)[lb]{\normalsize{\Black{$\phi_a$}}} 
    \Text(411,88)[lb]{\footnotesize{\Black{$\phi$}}} 
    \Text(435,101)[lb]{\footnotesize{\Black{$B$}}} 
    \Text(430,141)[lb]{\normalsize{\Black{$l_{Lc}^{f}$}}} 
    \Text(412,107)[lb]{\footnotesize{\Black{$l_{L}$}}} 
    \Text(352,142)[lb]{\normalsize{\Black{$\phi_d$}}} 
    \Text(348,52)[lb]{\normalsize{\Black{$l_{Lb}^g$}}} 
    \GBox(379,91)(403,113){0.647} 
    \DashArrowLine(417,78)(403,91){3} 
    \DashArrowLine(431,64)(417,77){3} 
    \PhotonArc(394.83,103.5)(35.2,-48.84,48.84){-3}{5.5} 
    \Vertex(416,79){2.00} 
    \ArrowLine(415,126)(403,113) 
    \ArrowLine(428,140)(416,127) 
    \Vertex(416,126){2.00} 
    \DashArrowLine(354,139)(379,113){3} 
    \ArrowLine(354,66)(379,91) 
    \SetColor{Black} 
    \CBox(190,-16)(208,2){Black}{White} 
    \Line(190,-16)(208,2)\Line(190,2)(208,-16) 
    \ArrowLine(166,-40)(190,-16) 
    \ArrowLine(233,26)(208,2) 
    \DashArrowLine(166,26)(190,2){3} 
    \DashArrowLine(233,-40)(208,-16){3} 
    \Text(136,-12)[lb]{\Large{\Black{$+$}}} 
    \Text(162,28)[lb]{\normalsize{\Black{$\phi_a$}}} 
    \Text(164,-54)[lb]{\normalsize{\Black{$l_{Lb}^{g}$}}} 
    \Text(230,-52)[lb]{\normalsize{\Black{$\phi_d$}}} 
    \Text(232,26)[lb]{\normalsize{\Black{$l_{Lc}^{f}$}}} 
    \Text(240,-12)[lb]{\normalsize{\Black{ = $\quad$ UV finite}}} 
    \GBox(194,435)(217,458){0.647} 
    \ArrowLine(239,481)(217,458) 
    \DashArrowLine(172,479)(193,458){3} 
    \ArrowLine(173,414)(194,435) 
    \DashArrowLine(240,414)(218,435){3} 
    \Text(162,483)[lb]{\normalsize{\Black{$\phi_a$}}} 
    \Text(164,404)[lb]{\normalsize{\Black{$l_{Lb}^{g}$}}} 
    \Text(240,404)[lb]{\normalsize{\Black{$\phi_d$}}} 
    \Text(242,481)[lb]{\normalsize{\Black{$l_{Lc}^{f}$}}} 
    \Text(145,442)[lb]{\Large{\Black{$\equiv$}}} 
    \Text(202,445)[lb]{\normalsize{\Black{$\kappa$}}} 
    \Text(453,211)[lb]{\normalsize{\Black{$\kappa$}}} 
    \Text(344,210)[lb]{\normalsize{\Black{$\kappa$}}} 
    \Text(207,316)[lb]{\normalsize{\Black{$\kappa$}}} 
    \Text(388,316)[lb]{\normalsize{\Black{$\kappa$}}} 
    \Text(202,206)[lb]{\normalsize{\Black{$\kappa$}}} 
    \Text(223,102)[lb]{\normalsize{\Black{$\kappa$}}} 
    \Text(389,101)[lb]{\normalsize{\Black{$\kappa$}}} 
    \Text(240,446)[lb]{\large{\Blue{H1}}} 
    \Text(270,366)[lb]{\large{\Blue{H2}}} 
    \Text(450,366)[lb]{\large{\Blue{H3}}} 
    \Text(230,245)[lb]{\large{\Blue{H4}}} 
    \Text(340,245)[lb]{\large{\Blue{H5}}} 
    \Text(450,245)[lb]{\large{\Blue{H6}}} 
    \Text(280,136)[lb]{\large{\Blue{H7}}} 
    \Text(460,136)[lb]{\large{\Blue{H8}}} 
    \Text(190,25)[lb]{\large{\Blue{H9}}} 
  \end{picture}\\ 
\beqa 
\Rightarrow \quad  
\delta \kappa = - \frac{1}{16 \pi^2}  
\Bigl[ 2 \kappa \left( Y_e^\dagger Y_e \right) 
+ 2 \left( Y_e^\dagger Y_e \right)^T \kappa - \lambda \kappa  
- \left( \frac{3}{2} - \xi_1 \right) g_1^2 \kappa  
- \left( \frac{3}{2} - 3 \xi_2 \right) g_2^2 \kappa \Bigr]  
\frac{1}{\epsilon} \; . 
\nn  
\label{app-kappa} 
\eeqa 
\end{center} 
 

\end{document}